\RequirePackage{lineno}
\documentclass[aps,prl,10pt,twocolumn,superscriptaddress,longbibliography]{revtex4-2}

\usepackage{graphicx}
\usepackage{amssymb,amsmath,amsfonts,gensymb}
\usepackage{bm,hyperref}
\usepackage{color}
\usepackage{ulem}
\usepackage{bbm}
\usepackage{subfigure}
\usepackage{dcolumn}
\usepackage{mathtools}
\usepackage{pifont}

\def\I{\uppercase\expandafter{\romannumeral 1}}
\def\II{\uppercase\expandafter{\romannumeral 2}}
\def\III{{\uppercase\expandafter{\romannumeral 3}}}
\def\IV{{\uppercase\expandafter{\romannumeral 4}}}
\def\V{{\uppercase\expandafter{\romannumeral 5}}}
\def\VI{{\uppercase\expandafter{\romannumeral 6}}}
\def\VII{{\uppercase\expandafter{\romannumeral 7}}}
\def\i{\lowercase\expandafter{\romannumeral 1}}
\def\ii{\lowercase\expandafter{\romannumeral 2}}
\def\iii{{\lowercase\expandafter{\romannumeral 3}}}
\def\iv{{\lowercase\expandafter{\romannumeral 4}}}
\def\v{{\lowercase\expandafter{\romannumeral 5}}}
\def\vi{{\lowercase\expandafter{\romannumeral 6}}}
\def\vii{{\lowercase\expandafter{\romannumeral 7}}}

\def\k{\mathbf{k}}
\def\K{\mathbf{K}}

\def\vr{\mathbf{r}}
\def\kt{\widetilde{\mathbf{k}}}
\def\q{\mathbf{q}}
\def\Q{\mathbf{Q}}

\def\H{\textrm{H}}

\def\hc{\hat{c}}
\def\hcd{\hat{c}^{\dagger}}

\def\lb{\ell_{\rm{B}}}
\def\ha{\hat{a}}
\def\had{\hat{a}^{\dagger}}
\def\hf{\hat{f}}
\def\hfd{\hat{f}^{\dagger}}
\def\degree{^{\circ}}

\newcommand{\ket}[1]{\vert #1 \rangle}
\newcommand{\bra}[1]{\langle #1 \vert}
\newcommand{\bracket}[2]{\langle #1 \vert #2 \rangle}

\begin{document}

\title{Magic momenta and  three dimensional  Landau levels from a three dimensional graphite moir\'e superlattice}
\author{Xin Lu}
\thanks{These  authors contributed equally.}
\affiliation{School of Physical Science and Technology, ShanghaiTech University, Shanghai 201210, China}
\author{Bo Xie}
\thanks{These  authors contributed equally.}
\affiliation{School of Physical Science and Technology, ShanghaiTech University, Shanghai 201210, China}

\author{Yue Yang}
\thanks{These  authors contributed equally.}
\affiliation{School of Physical Science and Technology, ShanghaiTech University, Shanghai 201210, China}

\author{Xiao Kong} 
\affiliation{Institute of Technology for Carbon Neutrality, Shenzhen Institute of Advanced Technology, Chinese Academy of Sciences, Shenzhen, China}
\affiliation{Shanghai Institute of Microsystem and Information Technology, Chinese Academy of Sciences, Shanghai, China}

\author{Jun Li}
\affiliation{School of Physical Science and Technology, ShanghaiTech University, Shanghai 201210, China}
\affiliation{ShanghaiTech Laboratory for Topological Physics, ShanghaiTech University, Shanghai 201210, China}

\author{Feng Ding}
\affiliation{Institute of Technology for Carbon Neutrality, Shenzhen Institute of Advanced Technology, Chinese Academy of Sciences, Shenzhen, China}
\affiliation{Shanghai Institute of Microsystem and Information Technology, Chinese Academy of Sciences, Shanghai, China}

\author{Zhu-Jun Wang}
\email{wangzhj3@shanghaitech.edu.cn}
\affiliation{School of Physical Science and Technology, ShanghaiTech University, Shanghai 201210, China}

\author{Jianpeng Liu}
\email{liujp@shanghaitech.edu.cn}
\affiliation{School of Physical Science and Technology, ShanghaiTech University, Shanghai 201210, China}
\affiliation{ShanghaiTech Laboratory for Topological Physics, ShanghaiTech University, Shanghai 201210, China}
\affiliation{Liaoning Academy of Materials, Shenyang 110167, China}
	
\bibliographystyle{apsrev4-2}

\begin{abstract} 
Twisted bilayer graphene (TBG) and other quasi-two-dimensional moir\'e  superlattices  have attracted significant attention due to the emergence of various  correlated and topological states associated with the flat bands in these systems. In this work, we theoretically explore the  physical properties of a new type of \textit{three dimensional  graphite moir\'e  superlattice}, the bulk alternating twisted graphite (ATG) system with homogeneous twist angle, which is grown by in situ chemical vapor decomposition method. Compared to TBG, the bulk ATG system is bestowed with an additional wavevector degrees of freedom due to the extra dimensionality. As a result, we find that when the twist angle of bulk ATG is smaller than twice of the magic angle of TBG, there always exist ``magic momenta" at which the in-plane Fermi velocities of the moir\'e bands vanish. Moreover, topologically distinct flat bands of TBG at different magic angles can even co-exist at different out-of-plane wavevectors in a single bulk ATG system. Most saliently, when the twist angle is relatively large,  exactly dispersionless three dimensional zeroth Landau level  would emerge in the bulk ATG, which may give rise to robust three dimensional quantum Hall effects  over a large range of twist angles.

\end{abstract} 

\maketitle


Twisted bilayer graphene (TBG) \cite{macdonald-pnas11,castro-neto-prl07,shallcross-tbg-prb-2010} has aroused significant interest in recent years. In TBG, the mismatch between two honeycomb lattices forms the moir\'e pattern with periodic modulations of $AA$ and $AB$ regions in real space \cite{uchida-corrugation-prb14,koshino-prx18}, which enables both  intra- and inter-sublattice interlayer couplings. In particular, when the twist angle is around a discrete set of values known as ``magic angles" \cite{macdonald-pnas11}, the lowest two bands per spin per valley become ultra-flat, and even turn out to be exactly flat in the chiral limit \cite{origin-magic-angle-prl19}. These flat bands are further found to be topologically nontrivial with Landau-level like wavefunctions \cite{origin-magic-angle-prl19,jpliu-prb19,song-tbg-prl19,yang-tbg-prx19,po-tbg-prb19,zaletel-tbg-2019,zaletel-tbg-hf-prx20,wang2021chiral,shi-dai-heavy-prb22}. The unprecedented flatness combined with nontrivial band topology yields plentiful  phenomena in magic-angle TBG such as superconductivity, correlated insulators,  quantum anomalous Hall effect, orbital magnetic states, and so on \cite{cao-nature18-supercond,dean-tbg-science19,marc-tbg-19, efetov-nature19, efetov-nature20, young-tbg-np20, li-tbg-science21, cao-tbg-nematic-science21,cao-nature18-mott,efetov-nature19,tbg-stm-pasupathy19,tbg-stm-andrei19,tbg-stm-yazdani19, tbg-stm-caltech19, young-tbg-science19, efetov-nature20, young-tbg-np20, li-tbg-science21,young-tbg-science19, sharpe-science-19,balents-review-tbg,andrei-review-tbg,jpliu-nrp21,efetov-arxiv20,yazdani-tbg-chern-arxiv20,andrei-tbg-chern-arxiv20,efetov-tbg-chern-arxiv20,pablo-tbg-chern-arxiv21,yang-tbg-cpl21}, which have stimulated tremendous theoretical research \cite{kang-tbg-prl19,Uchoa-ferroMott-prl,xie-tbg-2018,  zaletel-tbg-2019, wu-tbg-collective-prl20, zaletel-tbg-hf-prx20,jpliu-tbghf-prb21,zhang-tbghf-arxiv20,hejazi-tbg-hf,kang-tbg-dmrg-prb20,kang-tbg-topomott,yang-tbg-arxiv20,meng-tbg-arxiv20,Bernevig-tbg3-arxiv20,Lian-tbg4-arxiv20,regnault-tbg-ed,zaletel-dmrg-prb20,macdonald-tbg-ed-arxiv21,meng-tbg-qmc-cpl21,lee-tbg-qmc-arxiv21,bultinck-tbg-strain-prl21,song-heavyfermion-prl22}.

It is well appreciated that tuning a TBG sample to the vicinity of the first magic angle ($\theta^{(1)}_{m}\!\approx\!1.05 \degree$) is extremely challenging, especially if the device were fabricated based on the traditional transferring and stacking technique \cite{dean-hBN_sub-natnano-2010,liu-synthesis_tbg-natmat-2022}. Nevertheless, the  stringent magic-angle condition for the emergence of flat bands in TBG can be somehow released by stacking more  twisted graphene layers in the out-of-plane direction, forming twisted multilayer graphene (TMG) \cite{kim-tdbg-nature20,cao-tdbg-nature20,zhang-tdbg-np20,jpliu-prx19,koshino-tdbg-prb19,ashvin-double-bilayer-nc19,ledwith-prl22,wang-prl22,zhang-tmg-nanolett23} and alternating twisted multilayer graphene (ATMG)  \cite{eslam-tmg-prb19,xie-atmg-npj22,ashvin-atmg-arxiv21,leconte-amtg-2dmat-2022, cea-arxiv19} systems, where  topological  flat bands are robustly present over a finite range of twist angles or at larger values of magic angles. Various exotic states such as unconventional superconductivity \cite{cao2021pauli},  quantum anomalous Hall effect \cite{young-monobi-nature20}, and generalized Wigner crystals \cite{miao-tdbg-nature22} etc., have also been observed in these TMG and ATMG systems. From the perspective of fundamental science, it is then natural to ask what kind of new states and new phenomena would emerge when an infinite number layers of twisted graphene are stacked on top of each other, forming a three dimensional graphite moir\'e superlattice.

Recently, an origami–kirigami approach by chemical vapor deposition (CVD) growth has been developed for the fabrication of graphene spiral \cite{wang-nm2023}. Using such a technique, double-helix structure of hundreds of layers of the intertwined graphene with a uniform twisted interlayer stacking angle has been successfully achieved (see Supplementary Video 1). The structural model of such a graphene spiral, as depicted in Fig.~\ref{fig:structure}(a), showcases this dual-helical configuration. The top view of the dual-helical graphene model ((Fig.~\ref{fig:structure}(b))) indicates a periodic superlattice arrangement. Away from the central dislocation line, the system exhibits an alternating twisted pattern between each of the two adjacent graphene layers, wherein intra-plane moir\'e superlattice vectors ($\mathbf{t}_1, \mathbf{t}_2$ in Fig.~\ref{fig:structure}(b)) can be identified. In principle, it is  feasible to achieve an infinite growth of numerous graphene layers along the vertical direction defined by the helical dislocation, which for the first time realizes a genuine three-dimensional moir\'e graphite superlattice with alternating interlayer twist angle. In Fig.~\ref{fig:structure}(b$_1$), the Transmission Electron Microscopy (TEM) image of the twisted graphene spiral sample is presented. Employing Mask-filtering and Inverse Fast Fourier transform processing in Digital Micrograph, the TEM image distinctly demonstrate a moir\'e periodic pattern characterized by an average superlattice constant of $L_s=4.9\,$nm (as illustrated in the inset of Fig.~\ref{fig:structure}(b$_1$)). The corresponding twisted angle $\theta\!\approx\!2.9\,^{\circ}$ is obtained through the relationship   $L_s=a/(2\sin{\theta/2})$ ($a=2.46$\,\AA\ is graphene’s lattice constant). As depicted in Fig.~\ref{fig:structure}(b$_2$), the electron diffraction patterns from two sets of twisted graphene layers are clearly distinguishable, yielding a measured twist angle of approximately $3\,^{\circ}$. This result is in concordance with the deduction based on the real-space moir\'e superlattice constant. The cross-sectional scanning transmission electron microscopy (STEM) image offers a contrasting feature of the double-helix graphene spiral. As demonstrated in Fig.~\ref{fig:structure}(c$_1$), the helical dislocation nucleus is centrally positioned, giving rise to the growth of two sets of alternatingly twisted graphene layers following the helical dislocation path. Based on the aforementioned details, the double-helix structural model for numerous graphene layers was constructed as schematically shown in Fig.~\ref{fig:structure}(c). To further facilitate a comprehensive understanding of the cross-sectional intricacies of the helical dislocation, simulations were conducted based on the model depicted in Fig.~\ref{fig:structure}(b) and Fig.~\ref{fig:structure}(c), as presented in Fig.~\ref{fig:structure}(d$_1$). Figs.~\ref{fig:structure}(c$_2$) and (d$_2$) respectively illustrate the line profiles within the blue, green, and red rectangle regions in Figs.~\ref{fig:structure}(c$_1$) and (d$_1$). It is evident that the profiles from the two sets exhibit quantitative consistency. In comparison to the regions on both sides (blue and red rectangle), the helical core region (green rectangle) exhibits an approximate vertical displacement of 0.18\,nm (note that the horizontal axes in Figs.~\ref{fig:structure}(c$_2$) and (d$_2$) denote the $z$ coordinate), equivalent to a half of graphene’s interlayer distance. A noteworthy observation is that this screw dislocation itself hosts bound states, elaborated further in the Supplementary Information \cite{supp_info}. The successful realization of this macroscopically extensive bulk alternating twisted graphite (ATG) system effectively extends the realm of intriguing moiré graphene superlattices physics from two dimensions to three dimensions.

\begin{center}
	\begin{figure*}
		\includegraphics[width=0.82\textwidth]{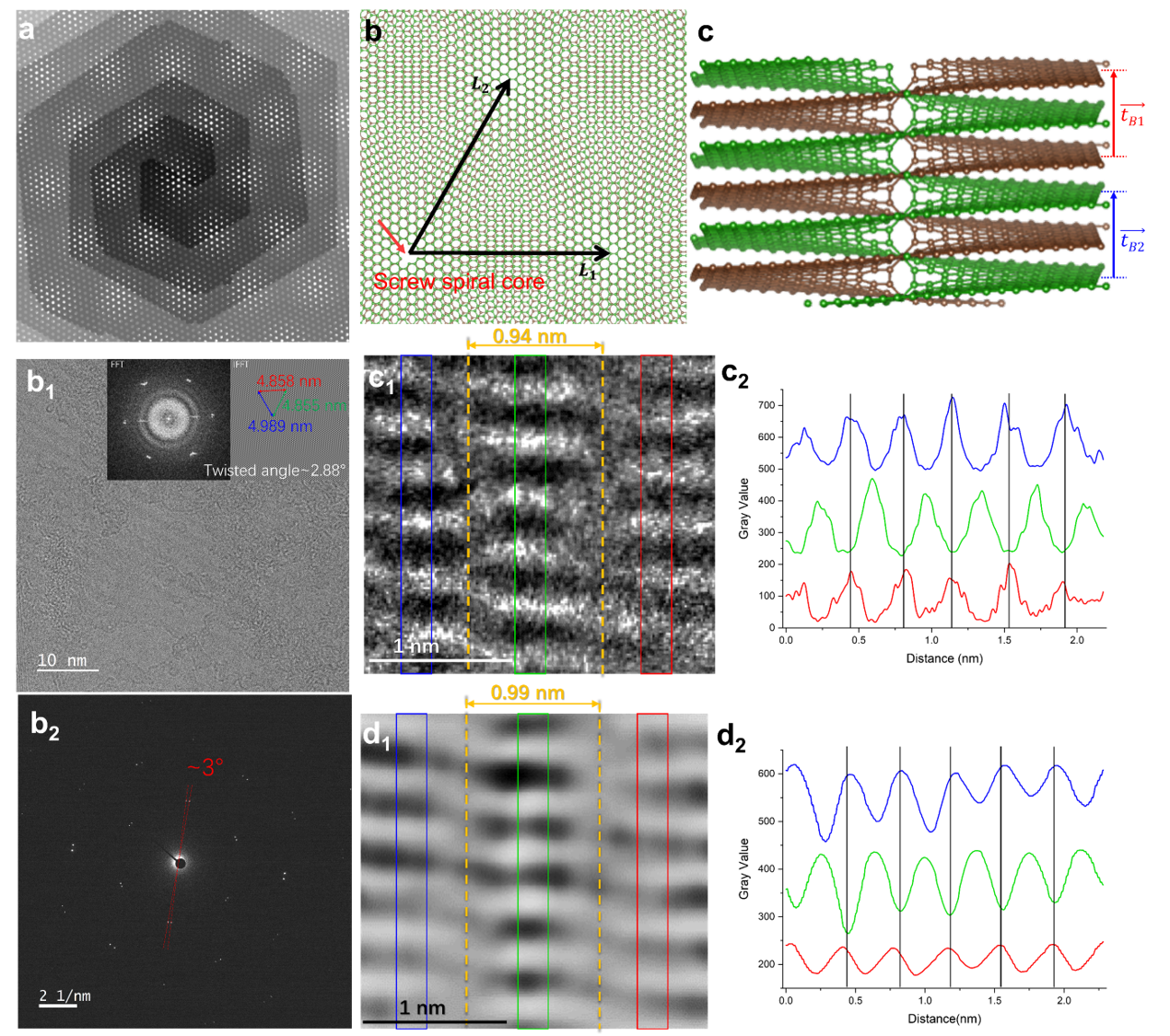}
\caption{Structural characterizations of twisted graphene spiral (bulk alternating twisted graphite). (a) The constructed structural model of a graphene spiral with alternating interlayer twist angle. (b) The top view of the helical graphene which reveals a periodic moir\'e pattern. Away from the central dislocation line, the system exhibits bulk alternating twisted pattern between every two adjacent graphene layers, wherein  in-plane moir\'e superlattice vectors $\mathbf{t}_1$, $\mathbf{t}_2$ are discernible. b$_1$ presents a high-resolution TEM image of the twisted spiral graphene sample, from which the moir\'e period is measured, corresponding to an interlayer twisted angle of $2.9\,^{\circ}$. The angular measurement obtained from electron diffraction in b$_2$ corroborates with the angle depicted in b$_1$. (c) Schematic illustration of the cross-section of dual-helical twisted graphene spiral. The annotations $\mathbf{t}_{B1}$ and $\mathbf{t}_{B2}$ indicate the Burgers vectors of the two inequivalent sets of spiral graphene layers twisted with respect each other by a finite angle. (c$_1$) The STEM image of the twisted graphene spiral sample. (d$_1$) The simulated STEM image of the structural model in (c). Comparative contrast line profile analysis (c$_2$ and d$_2$) reveals that the simulated results are quantitatively consistent with the experimentally measured ones.}
		\label{fig:structure}
	\end{figure*}
\end{center}


In this work, we theoretically study the electronic structures, topological properties, and Landau levels of the bulk ATG system. The additional dimensionality  endows  bulk ATG an additional wavevector degrees of freedom to explore  flat-band physics of magic-angle TBG. In particular, we find that when the twist angle $\theta\leq 2 \theta^{(1)}_m\approx 2.1\,^{\circ}$, one can always find a ``magic quasi momentum" $\hbar k_z^{*}$ at which the 2D band structures and wavefunctions within the $k_x$-$k_y$ plane are exactly the same as that of TBG at the first magic angle. Moreover, when the twist angle $\theta$ is further smaller than twice of the $n$th magic angle $\theta_m^{(n)}$, i.e., $\theta\leq 2\theta^{(n)}_{m}$ (with $n=2, 3, ...$),  then  for each $\theta$, there are \textit{at least $n$ magic momenta $\{ k_z^{(s),*}, s= 1,..., n \}$} at which the band structures and wavefunctions within the $k_x$-$k_y$ planes are exactly the same as those of TBG at the $s$th magic angle $\theta_m^{(s)}$. In other words, when $\theta\leq\theta_m^{(2)}\approx 1\,^{\circ}$, multiple flat bands with distinct topological properties  would co-exist in the same bulk ATG system, which may give rise to unprecedented correlated and topological phases of matter.  On the other hand, for larger twist angles $\theta\!>\!2\theta^{(1)}_m\approx 2.1^{\circ}$, Dirac-cone type  band structures with finite Fermi velocities persist for every $k_z$ in the Brillouin zone of bulk ATG. The Dirac points spectacularly align on a chain along the $k_z$ direction,  forming an exactly straight Weyl nodal chain for each spin and valley species. Such quasi-2D Fermi surface would give rise to weakly dispersive \textit{three dimensional Landau levels} that may lead to 3D quantum Hall effects, especially for twist angle $\theta\gtrapprox 3\,^{\circ}$. Most saliently, given that the zeroth Landau level of graphene is exactly pinned to the Dirac point, and that the Dirac points in bulk ATG are energetically aligned at different $k_z$, there exists a branch of \textit{exactly dispersionless zeroth 3D Landau band} emerging from such 3D bulk ATG system under finite vertical magnetic fields.  Unusual correlated states may emerge at partial fillings of such massively degenerate 3D zeroth Landau levels.


\paragraph{Magic momenta --}
Neglecting the spiral dislocation at the core, we can consider bulk ATG as an infinite aligned stacking of TBG in the out-of-plane direction with lattice constant $2d_0\!=\!6.7\,$\AA. The coupling between two nearest neighbor TBG motifs is just the interlayer moir\'e potential. Such picture leads straightforward to a low-energy effective Hamiltonian for bulk ATG \cite{supp_info} based on the continuum model \cite{macdonald-pnas11,castro-neto-prl07,castro-neto-prb12} of TBG with twist angle $\theta$:
\begin{align}
\H^{\mu}_{\theta}=\hbar v_{F} \left[\begin{array}{cc}
			-(\k-\K_{1}^{\mu})\cdot\bm{\sigma}^{\mu} & T (1+e^{i\phi}) \\
			T^{\dagger}(1+e^{-i\phi}) &-(\k-\K_{2}^{\mu})\cdot\bm{\sigma}^{\mu}
		\end{array}\right]
\label{eq:ham3d}
\end{align}
where the Fermi velocity is $\hbar v_{F}/a\!=\!2.1354\,$eV \cite{moon-tbg-prb13} with graphene's lattice constant $a=2.46\,$\AA. 
Neglecting the intervalley coupling, the diagonal blocks represent the $\k\cdot\textbf{p}$ Hamiltonian of the two layers of graphene in valley $\mu=\pm 1$ near the Dirac points $\K_{1/2}^{\mu}$ with the Pauli matrices $\bm{\sigma}^{\mu}=(\mu\sigma_{x},\sigma_{y})$ defined in the sublattice space, and $\k = (k_x,k_y)$. 
The off-diagonal term stands for the moir\'e potential between two adjacent graphene layers. The phase factor $e^{\pm i\phi}=e^{\pm i2 k_z d_{0}}$ is the only new contribution  arising from the nearest neighbor interlayer couplings along the opposite out-of-plane directions in such 3D bulk system. Here, it is naturally expected that the distance between two adjacent graphene layers is fixed, denoted as $d_0$, since there is no space for out-of-plane atomic corrugations in such 3D bulk moir\'e superlattice. This postulation is verified by direct molecular dynamics simulations of bulk ATG, which gives an equilibrium interlayer distance $d_0=3.35\,$\AA\ \cite{supp_info}.

The moir\'e potential $T$ are characterized by intra- and inter-sublattice interlayer tunnelling amplitudes $u_{0}\!=\!u'_{0}\!=\!0.103\,$eV. They are equal to each other due to the absence of out-of-plane corrugations  in the bulk ATG system \cite{supp_info}. 
Up to an overall bandwidth $\hbar v_F k_\theta$ (with $k_{\theta}=4\pi/(3L_s)$), 
when $u_0=u_0'$ the continuum model of TBG is fully characterized by a dimensionless ratio $\alpha_0(\theta)=u'_0 / (\hbar v_F k_\theta )$. Similarly, since $k_z$ is a good quantum number and the bulk ATG can be seen as a series of decoupled TBG indexed by $k_z$, the non-interacting physics of each of them is totally governed by 
\begin{equation}
\alpha(k_z,\theta) = \alpha_0(\theta) \vert 1+e^{2ik_z d_0}\vert\;,
\end{equation}
where $k_z d_0 \in [0, \pi]$. Only the modulus matters because the phase can be removed by a gauge transformation. 
Therefore, given a  twist angle $\theta$, one can explore a range of $k_z$-dependent effective moir\'e potential, which varies from zero ($k_z=\pi/2 d_0$) to doubled moir\'e potential strength ($k_z=0$) of the corresponding isolated TBG with twist angle $\theta$. 
In the small angle approximation, as long as $\theta/2$ is smaller than $s$th magic angle $\theta_m^{(s)}$ of TBG, one can always find the corresponding $s$th \textit{magic momentum} $k_z^{(s),*}$ such that $\alpha(k_z^{(s),*},\theta)=\alpha_0(\theta_m^{(s)})$,
as shown in Fig.~\ref{fig:magickz}(a). For example, to achieve the flat bands at the first magic angle in TBG, it only suffices to twist the bulk ATG of angle $\theta\!<\!2\theta_{m}^{(1)}\!\approx\!2.1\,\degree$. Furthermore, the bulk ATG can simultaneously host flat bands of TBG at several magic angles if $\theta\!<\!2\theta_{m}^{(n)}$  with $n\!\ge\!2$. Since the $s$th magic angle of TBG approximately equals one $s$th of the first magic angle, the total number of flat bands (denoted by $n$)  found at different out-of-plane wavevectors  in bulk ATG at   twist angle $\theta$ verifies a thumb rule: $n = [2\theta_{m}^{(1)}/\theta]$ ($[\textrm{ } ]$ denotes the round down operation). Accordingly, we can define $n$ magic momenta $\{ k_z^{(s),*}, s= 1,..., n \}$ for bulk ATG at fixed twist angle $\theta$. For example, if $\theta=0.45\,\degree <2\times\theta_{m}^{(4)}$, the flat bands of TBG from first to fourth magic angle emerge concomitantly in the bulk ATG at four magic momenta, as shown in Fig.~\ref{fig:magickz}(b). In Fig.~\ref{fig:magickz}(c), we present the moir\'e band structures of bulk ATG with $\theta=0.8\,\degree$ at several different $k_z$. In such a case, $2\theta_{m}^{(3)}\!<\!\theta\!<\!2\theta^{(2)}_m$, topological flat bands at the first and second magic angles of TBG co-exist at different magnetic momenta $k_z^{(1),*}$ and $k_z^{(2),*}$, as plotted by cyan and magenta lines in Fig.~\ref{fig:magickz}(c).

\paragraph{Band Topology --}
Since bulk ATG integrates a continuum of $k_z$-indexed TBG into a single bulk system, it is intriguing to explore the topological properties  by varying both twist angle $\theta$ and $k_z$.
By introducing a small sublattice mass term ($\sim 0.1\,$meV) into the Hamiltonian Eq.~\eqref{eq:ham3d}, all the degeneracy points are gapped out such  that every band can be associated with a Chern number. At each $k_z$ and $\theta$, we calculate the Chern number of the valence flat band from valley $K$, and 
the calculated topological phase diagram is shown in Fig.~\ref{fig:magickz}(d). First, the Chern number can change only if the twist angle of bulk ATG $\theta$ is smaller than $\theta_{c,0}=2.3 \degree$, which corresponds to the apex point of the rightmost phase boundary. 
From right to left across  $\theta_{c,0}$, the Chern number at $k_z=0$ switches from 1 to 0. Nevertheless, the Chern number shifts back to unity if one continues to cross the apex point of the second rightmost phase boundary, indicating another critical angle $\theta_{c,1}=2.15\,\degree$. If one continues to decrease $\theta$, a series of such phase boundaries are passed, resulting in a rich topological phase diagram in the ($k_z$,$\theta$) space as shown in Fig.~\ref{fig:magickz}(d).

If the twist angle is fixed between $\theta_{c,0}$ and $\theta_{c,1}$, e.g., $\theta=2.2\degree$, the Brillouin zone can be cut into two topologically different parts: the Chern number is zero if $k_z \in [-k_{z,0},k_{z,0}]$ and is one otherwise. A close look on the phase boundary shows that such topological transition is triggered by an accidental crossing at $(0,0,\pm k_{z,0})$ between the valence flat band and the remote band. Therefore, the system behaves as a Weyl semimetal with a pair of Weyl nodes with opposite topological charges at $(0,0,\pm k_{z,0})$ in each valley and spin if spatial inversion symmetry is broken by a slight sublattice mass term. 
Note that in the absence of small mass term, the band touching point is three-fold degenerate forming a three-fold multifermion because the remote band is always twice degenerate dictated by symmetry \cite{hejazi-prb19,supp_info}. 
Furthermore, with smaller twist angle, bulk ATG (with small sublattice mass term) can host eight ($\theta=1.6\degree$), twelve ($\theta=1.2\degree$), and more generally, $4n$ pairs of Weyl nodes with $n=1, 2, 3, 4, \dots$. Most spectacularly, the Weyl nodes in bulk ATG perfectly align on the $k_z$-axis. The topological charge of Weyl node is not limited to unity. For example, one pair of Weyl nodes in bulk ATG of $\theta=0.85\degree$ have topological charges equal to $\pm 2$ since the Chern number of the $k_z$-indexed 2D flat band shifts by 2 across the Weyl nodes. 
\onecolumngrid
\begin{center}
	\begin{figure}
		\includegraphics[width=0.99\textwidth]{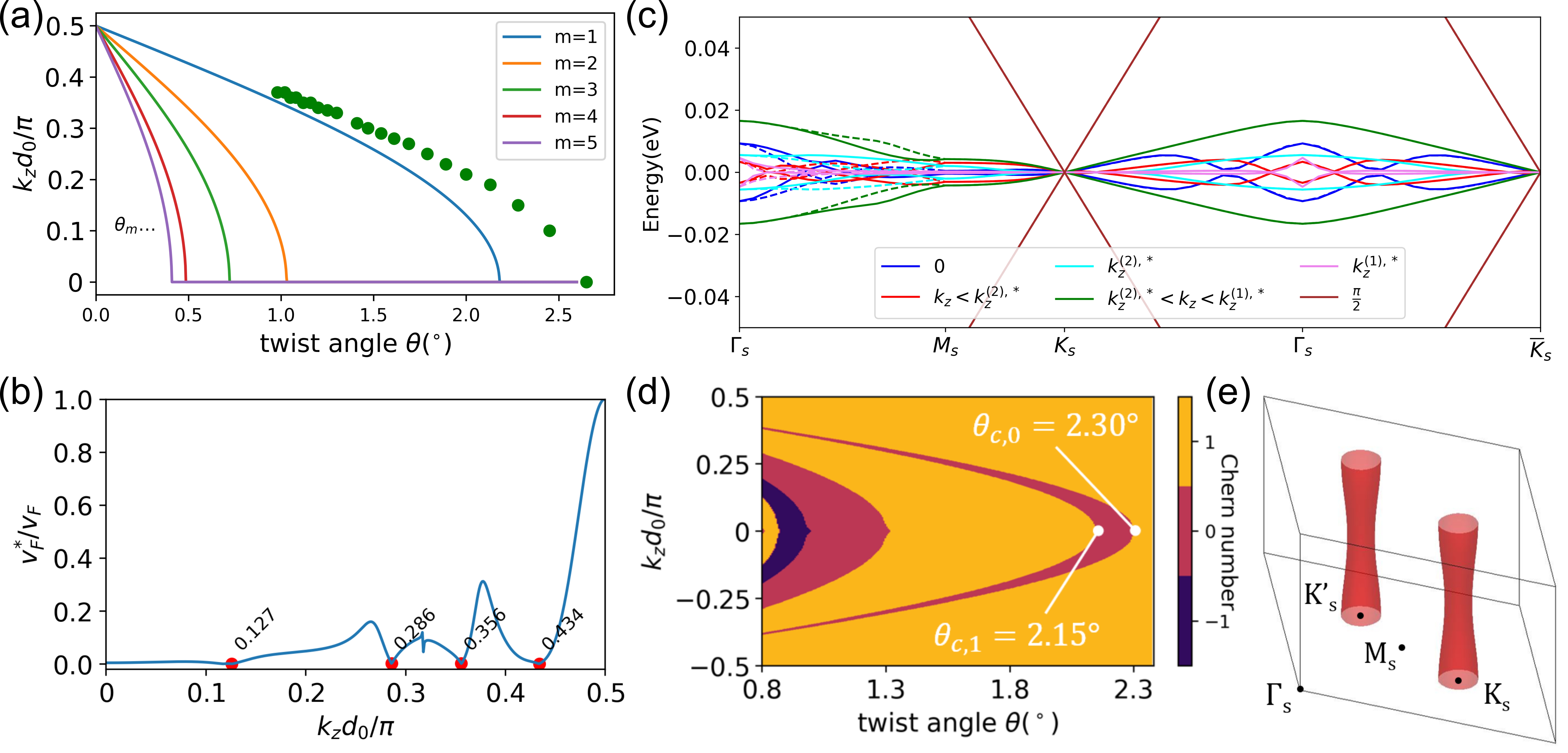}
		\caption{(a) Evolution of magic momenta $k_z^{(m),*}$ as a function of physical twist angle $\theta$. Higher $m$th magic momenta also exist but not shown here for clarity. The green dots are the results from tight-binding calculations \cite{supp_info} (b) Evolution of Fermi velocity near Moir\'e $K$ point as a function of $k_z$ for $\theta=0.45 \degree$. The red dots indicate the four magic momenta $k_z^{*}$, at which the Fermi velocity vanishes exactly and the flat band emerges. (c) Evolution of bulk ATG's 2D band structures in the plane $(k_x,k_y)$ from $k_z d_0 =0$ to $\pi/2$. The twist angle of bulk ATG is chosen to be $0.8 \degree$, at which bulk ATG can host two magic momenta $k_z^{(1),*}$ and $k_z^{(2),*}$, as shown in (a). (d) Topological phase diagram of the flat bands of $k_z$-indexed TBG in bulk ATG at different twist angle. The color indicates the Chern number of the valence flat band in valley $K$. (e) Quasi-1D Fermi surface of ATG of $\theta=7\degree$ for $E_F=0.1$ eV.}
		\label{fig:magickz}
	\end{figure}
\end{center}
\twocolumngrid

\paragraph{3D quantum Hall effect} --

The unusual band dispersions and nontrivial topological properties of bulk ATG imply its unconventional responses to electromagnetic fields.  
It follows from Eq.~\eqref{eq:ham3d} that $k_z$ plays a special role compared to the other two in-plane momenta, leading to a strongly anisotropic, quasi-1D Fermi surface as shown in Fig.~\ref{fig:magickz}(e), which is desirable to realize 3D quantum Hall effect \cite{bernevig-3dqhe_gr-prl-2007,tang-3dqhe-nature-2019}. To this end, a key requirement needs to be fulfilled: the twist angle of the bulk ATG should be sufficiently larger than $2\theta_{m}^{(1)}$, otherwise at some $k_z$ the lowest bands would flatten, which suppresses Landau level spacing thus disfavors quantum Hall effect. In the case of large twist angles, the moir\'e potential can be treated by  perturbation theory, and the bulk ATG around each Dirac point can be described by a Dirac Hamiltonian with renormalized Fermi velocity 
\begin{equation}
	H_{\rm{eff}} = \hbar v_F [1-9 \alpha^2 (k_z,\theta) ] (k_x \sigma_{x} + k_y \sigma_{y})\;,
\label{eq:ham_pert}
\end{equation}
which is valid if $\theta\!\gtrapprox\!6\degree$ based on the criterion $9 \alpha^2 (k_z,\theta)<0.1$. In practice, 3D quantum Hall effect is present at even smaller angles (numerically checked at least for $\theta\!\gtrapprox\!4\,\degree$ at $E_F\!\sim\!20\,$meV where the trigonal warping effect is still weak). Based on the Hamiltonian Eq.~(\ref{eq:ham_pert}), at fixed $k_z$, the Landau level's energy should be exactly the same as graphene but with renormalized Landau level spacing, following $\sqrt{B}$ behavior. Remarkably, for every $k_z$ there exists the zeroth Landau level pinned at the Dirac point, which remains robust even in the presence of slight inhomogeneity and disorder by virtue of Atiyah–Singer index theorem \cite{graphene1}. This gives rise to exactly dispersionless 3D zeroth Landau level in such bulk ATG system.

\onecolumngrid
\begin{center}
	\begin{figure}[h]
		\includegraphics[width=0.99\textwidth]{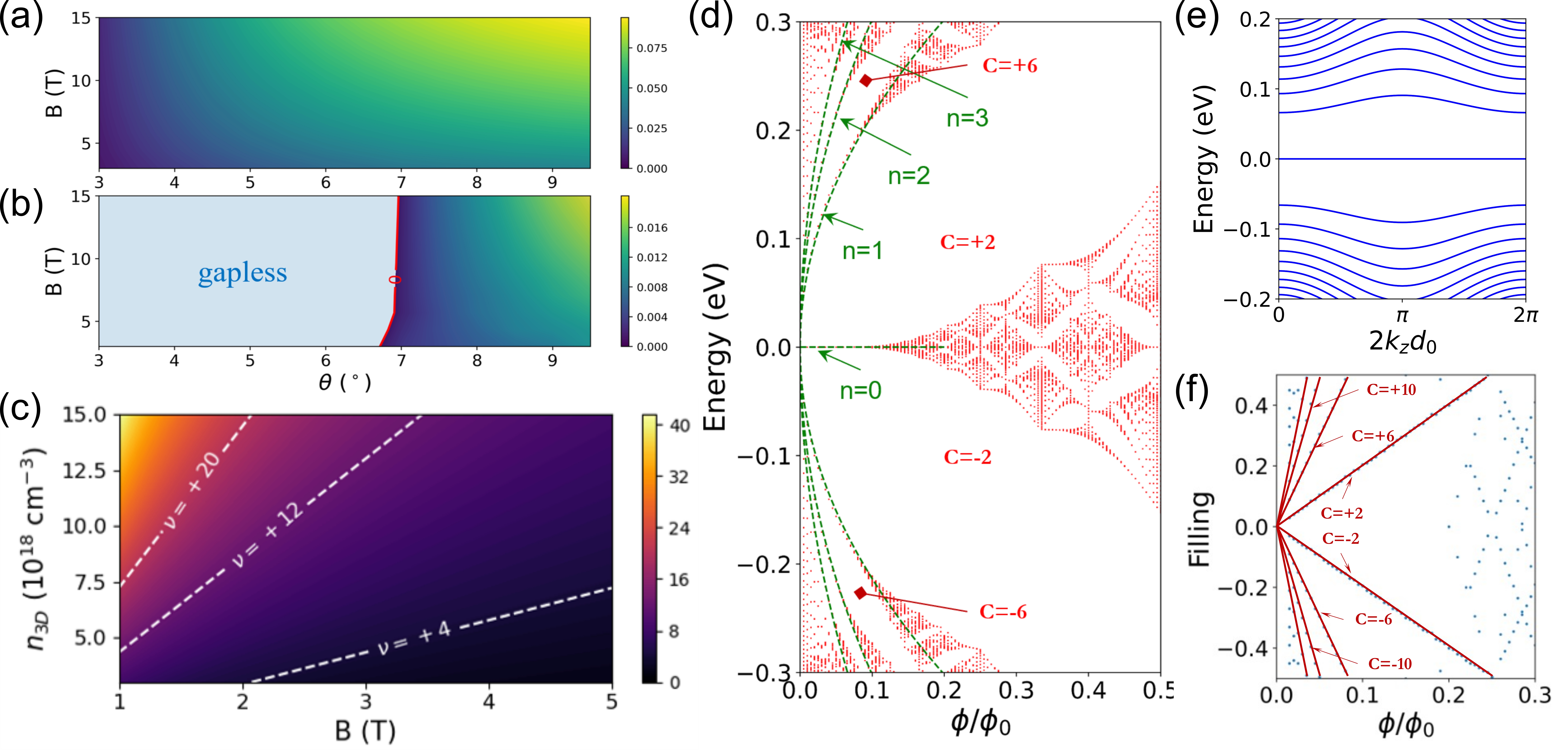}
		\caption{(a) Overall gap between the zeroth and the first 3D Landau levels under magnetic field $B=3$-$15\,$T for twist angle $\theta=3.0$-$9.5\,\degree$. The color coding indicates the amplitude of gap. (b) Overall gap between the first and the second 3D Landau levels in the same parameter space as (a). The gap is closed for $\theta \lessapprox 7\,\degree$. (c) Colormap of the filling factor of 3D Landau levels in bulk ATG as a function of 3D electron density and magnetic field. The white dashed lines indicates the fillings with fully occupied 3D Landau levels. (d) Hofstadter butterfly of $k_z$-indexed TBG in bulk ATG for $k_z=0$ and $\theta=7\degree$ for one spin sector. Green dashed lines are the calculated Laudau level dispersions for $n=1,2,3$ using the approximated Hamiltonian Eq.~\eqref{eq:ham_pert}. The Chern number of gaps are given. (e) Landau level dispersion along the $k_z$-dispersion at $B=9.8\,$T. It turns out that the in-plane dispersion of Hofstadter bands is too small to be visible in the given energy scale. Therefore, here we only show the case for $k_x=k_y=0$ without losing generality.  (f) Wannier plot associated with Hofstadter butterfly in (d). In our convention, the filling is normalized to unity when the first conduction band is filled. }
		\label{fig:3dqhe}
	\end{figure}
\end{center}
\twocolumngrid

Technically, the large moir\'e periodicity makes Landau levels split in a fractal manner forming the well-known Hofstadter butterfly \cite{moire-butterfly-macdonald-prb11,castro-neto-prb11,hejazi-ll_tbg-prb-2019,zhang-ll_tbg-prb-2019,ll-tbg-lian}. Our calculations \cite{supp_info} show that for $\theta\!=\!7\,\degree$, the fractal splitting of the first Landau level starts at $B \sim 100\,$T, which is way beyond the usual experimentally achievable magnitude [see Fig.~\ref{fig:magickz}(d)]. So, in practice, the renormalized Dirac Hamiltonian Eq.~\eqref{eq:ham_pert} suffices to get the qualitative Landau level structure of the bulk ATG if $\theta\!\gtrapprox\!6\,\degree$. Under $B=9.8\,$T, the first and zeroth Landau levels of the bulk ATG at $\theta\!=\!7\,\degree$, extracted from Hofstadter butterfly calculations, are  separated in energy by a large gap $\sim 66\,$meV, and the first and second Landau levels are also separated by an global gap $\sim 2\,$meV (see Fig.~\ref{fig:magickz}(e)), such that several quantized Hall conductivities can be observed. This is guaranteed by the  Dirac dispersion near charge neutrality point which applies to all $k_z$.
Specifically, when the $n$th Landau level is occupied and is separated from the higher one by a global gap, the Hall conductivity should follow the sequence
\begin{equation}
	\sigma_{xy} = 2(4n+2) \frac{e^2}{h} \cdot \frac{1}{2d_0},
	\label{eq:3d_hall_cond}
\end{equation}
where the prefactor of 2 comes from spin degeneracy. This can be shown by Wannier plot Fig.~\ref{fig:magickz}(f), where we only consider one spin sector. According to the Diophantine equation, the slope of each line is precisely the Chern number, or TKNN invariant of the filled Hofstadter bands \cite{niu-rmp10,TKNN,kohmoto-qhe-prb-1989,dana-qhe-jpcssp-1985}. The eight-fold degeneracy is due to two Dirac cones in each of two valleys for two spin sectors.

Unlike the quantum Hall problem in 2D electron gas or in graphene, here the degeneracy of each  zeroth Landau level is: $D=8\,B\,S\,N_z/\phi_0$, where $B$, $S$, $N_z$ denote the magnetic field, total cross-sectional area of the bulk system, and number of primitive cells in the out-of-plane direction, respectively. The factor of $8$ comes from valley, spin, and layer degeneracy. 
The additional $k_z$ degeneracy may lead to a variety of unconventional correlated and topological states for the partially occupied 3D zeroth Landau level. For example, at partial integer fillings a gap may be opened up due to strong $e$-$e$ interaction effects, such that the system may undergo a field-induced metal-to-insulator transition through the spontaneous breaking of flavor (valley, spin, and layer) symmetry. The additional $k_z$ degeneracy may lead to further out-of-plane charge modulation driven by $e$-$e$ and/or electron-phonon couplings, which may give rise to potential 3D topological charge density wave state. Other new states of matter are also anticipated at fractional fillings of the dispersionless 3D zeroth Landau level.

\paragraph{Conclusions and outlook --}
In this work, we theoretically study the electronic structures, topological properties, and Landau levels of a CVD-grown 3D moir\'e graphite superlattice, the bulk ATG system. We have shown that bulk ATG can be seen as a series of decoupled TBG with $k_z$ dependent moir\'e potential. Consequently, the flat bands of TBG at the first magic angle are always present in bulk ATG  at some magic $k_z$ as long as $\theta<2\theta_{m}^{(1)}$. Moreover, multiple topological flat bands of TBG at different magic angles emerge concomitantly in bulk ATG at a set of magic vertical momenta. By introducing a small mass term,a Weyl semimetal phase may anticipated in bulk ATG with tunable Weyl nodal charges. Most saliently, when the twist angle $\theta\gtrapprox 3\rm{-}4\,\degree$, bulk ATG is bestowed with a Dirac-like dispersion and a quasi-2D Fermi surface at low energy, which gives rise to exactly dispersionless 3D zeroth Landau level and weakly dispersive higher  Landau levels. When these 3D Landau levels are fully occupied, the system is expected to realize 3D integer quantum Hall effects.

The physics of bulk ATG has even more to explore than TBG. First, when $\theta <2\theta_{m}^{(1)}$, flat bands of magic-angle TBG occur at some magic $k_z$, which coexist with linearly dispersive Dirac fermions as shown in Fig.~\ref{fig:magickz}(c). The $e$-$e$ interaction effects and electron-phonon coupling effects of such peculiar band structures are open questions. Second, when $\theta <2\theta_{m}^{(2)}$,   topological flat bands of TBG at multiple magic angles are integrated into a single bulk ATG system, and unusual interacting ground states would be anticipated in such multi-flat-band system.  Last, when $\theta$ is sufficiently larger than $2\theta_{m}^{(1)}$, e.g., $\theta\gtrapprox 3$-$4\,\degree$, exactly dispersionless 3D zeroth Landau levels would emerge. 
As a result,  the intriguing quantum Hall physics associated with 2D Landau levels would be bestowed with an additional $k_z$ degrees of freedom.  Unusual quantum states of matter may emerge in the partially occupied 3D Landau levels, which deserves further explorations.

\acknowledgements
This work is supported by the National Key R \& D program of China (grant No. 2020YFA0309601), the National Natural Science Foundation of China (grant No. 12174257), and the start-up grant of ShanghaiTech University. 

\bibliography{tmg}

\begin{thebibliography}{101}%
\makeatletter
\providecommand \@ifxundefined [1]{%
 \@ifx{#1\undefined}
}%
\providecommand \@ifnum [1]{%
 \ifnum #1\expandafter \@firstoftwo
 \else \expandafter \@secondoftwo
 \fi
}%
\providecommand \@ifx [1]{%
 \ifx #1\expandafter \@firstoftwo
 \else \expandafter \@secondoftwo
 \fi
}%
\providecommand \natexlab [1]{#1}%
\providecommand \enquote  [1]{``#1''}%
\providecommand \bibnamefont  [1]{#1}%
\providecommand \bibfnamefont [1]{#1}%
\providecommand \citenamefont [1]{#1}%
\providecommand \href@noop [0]{\@secondoftwo}%
\providecommand \href [0]{\begingroup \@sanitize@url \@href}%
\providecommand \@href[1]{\@@startlink{#1}\@@href}%
\providecommand \@@href[1]{\endgroup#1\@@endlink}%
\providecommand \@sanitize@url [0]{\catcode `\\12\catcode `\$12\catcode
  `\&12\catcode `\#12\catcode `\^12\catcode `\_12\catcode `\%12\relax}%
\providecommand \@@startlink[1]{}%
\providecommand \@@endlink[0]{}%
\providecommand \url  [0]{\begingroup\@sanitize@url \@url }%
\providecommand \@url [1]{\endgroup\@href {#1}{\urlprefix }}%
\providecommand \urlprefix  [0]{URL }%
\providecommand \Eprint [0]{\href }%
\providecommand \doibase [0]{https://doi.org/}%
\providecommand \selectlanguage [0]{\@gobble}%
\providecommand \bibinfo  [0]{\@secondoftwo}%
\providecommand \bibfield  [0]{\@secondoftwo}%
\providecommand \translation [1]{[#1]}%
\providecommand \BibitemOpen [0]{}%
\providecommand \bibitemStop [0]{}%
\providecommand \bibitemNoStop [0]{.\EOS\space}%
\providecommand \EOS [0]{\spacefactor3000\relax}%
\providecommand \BibitemShut  [1]{\csname bibitem#1\endcsname}%
\let\auto@bib@innerbib\@empty
\bibitem [{\citenamefont {Bistritzer}\ and\ \citenamefont
  {MacDonald}(2011{\natexlab{a}})}]{macdonald-pnas11}%
  \BibitemOpen
  \bibfield  {author} {\bibinfo {author} {\bibfnamefont {R.}~\bibnamefont
  {Bistritzer}}\ and\ \bibinfo {author} {\bibfnamefont {A.~H.}\ \bibnamefont
  {MacDonald}},\ }\href@noop {} {\bibfield  {journal} {\bibinfo  {journal}
  {Proceedings of the National Academy of Sciences}\ }\textbf {\bibinfo
  {volume} {108}},\ \bibinfo {pages} {12233} (\bibinfo {year}
  {2011}{\natexlab{a}})}\BibitemShut {NoStop}%
\bibitem [{\citenamefont {Lopes~dos Santos}\ \emph
  {et~al.}(2007{\natexlab{a}})\citenamefont {Lopes~dos Santos}, \citenamefont
  {Peres},\ and\ \citenamefont {Castro~Neto}}]{castro-neto-prl07}%
  \BibitemOpen
  \bibfield  {author} {\bibinfo {author} {\bibfnamefont {J.~M.~B.}\
  \bibnamefont {Lopes~dos Santos}}, \bibinfo {author} {\bibfnamefont
  {N.~M.~R.}\ \bibnamefont {Peres}},\ and\ \bibinfo {author} {\bibfnamefont
  {A.~H.}\ \bibnamefont {Castro~Neto}},\ }\href
  {https://doi.org/10.1103/PhysRevLett.99.256802} {\bibfield  {journal}
  {\bibinfo  {journal} {Phys. Rev. Lett.}\ }\textbf {\bibinfo {volume} {99}},\
  \bibinfo {pages} {256802} (\bibinfo {year} {2007}{\natexlab{a}})}\BibitemShut
  {NoStop}%
\bibitem [{\citenamefont {Shallcross}\ \emph {et~al.}(2010)\citenamefont
  {Shallcross}, \citenamefont {Sharma}, \citenamefont {Kandelaki},\ and\
  \citenamefont {Pankratov}}]{shallcross-tbg-prb-2010}%
  \BibitemOpen
  \bibfield  {author} {\bibinfo {author} {\bibfnamefont {S.}~\bibnamefont
  {Shallcross}}, \bibinfo {author} {\bibfnamefont {S.}~\bibnamefont {Sharma}},
  \bibinfo {author} {\bibfnamefont {E.}~\bibnamefont {Kandelaki}},\ and\
  \bibinfo {author} {\bibfnamefont {O.~A.}\ \bibnamefont {Pankratov}},\ }\href
  {https://doi.org/10.1103/PhysRevB.81.165105} {\bibfield  {journal} {\bibinfo
  {journal} {Phys. Rev. B}\ }\textbf {\bibinfo {volume} {81}},\ \bibinfo
  {pages} {165105} (\bibinfo {year} {2010})}\BibitemShut {NoStop}%
\bibitem [{\citenamefont {Uchida}\ \emph {et~al.}(2014)\citenamefont {Uchida},
  \citenamefont {Furuya}, \citenamefont {Iwata},\ and\ \citenamefont
  {Oshiyama}}]{uchida-corrugation-prb14}%
  \BibitemOpen
  \bibfield  {author} {\bibinfo {author} {\bibfnamefont {K.}~\bibnamefont
  {Uchida}}, \bibinfo {author} {\bibfnamefont {S.}~\bibnamefont {Furuya}},
  \bibinfo {author} {\bibfnamefont {J.-I.}\ \bibnamefont {Iwata}},\ and\
  \bibinfo {author} {\bibfnamefont {A.}~\bibnamefont {Oshiyama}},\ }\href
  {https://doi.org/10.1103/PhysRevB.90.155451} {\bibfield  {journal} {\bibinfo
  {journal} {Phys. Rev. B}\ }\textbf {\bibinfo {volume} {90}},\ \bibinfo
  {pages} {155451} (\bibinfo {year} {2014})}\BibitemShut {NoStop}%
\bibitem [{\citenamefont {Koshino}\ \emph {et~al.}(2018)\citenamefont
  {Koshino}, \citenamefont {Yuan}, \citenamefont {Koretsune}, \citenamefont
  {Ochi}, \citenamefont {Kuroki},\ and\ \citenamefont {Fu}}]{koshino-prx18}%
  \BibitemOpen
  \bibfield  {author} {\bibinfo {author} {\bibfnamefont {M.}~\bibnamefont
  {Koshino}}, \bibinfo {author} {\bibfnamefont {N.~F.~Q.}\ \bibnamefont
  {Yuan}}, \bibinfo {author} {\bibfnamefont {T.}~\bibnamefont {Koretsune}},
  \bibinfo {author} {\bibfnamefont {M.}~\bibnamefont {Ochi}}, \bibinfo {author}
  {\bibfnamefont {K.}~\bibnamefont {Kuroki}},\ and\ \bibinfo {author}
  {\bibfnamefont {L.}~\bibnamefont {Fu}},\ }\href
  {https://doi.org/10.1103/PhysRevX.8.031087} {\bibfield  {journal} {\bibinfo
  {journal} {Phys. Rev. X}\ }\textbf {\bibinfo {volume} {8}},\ \bibinfo {pages}
  {031087} (\bibinfo {year} {2018})}\BibitemShut {NoStop}%
\bibitem [{\citenamefont {Tarnopolsky}\ \emph {et~al.}(2019)\citenamefont
  {Tarnopolsky}, \citenamefont {Kruchkov},\ and\ \citenamefont
  {Vishwanath}}]{origin-magic-angle-prl19}%
  \BibitemOpen
  \bibfield  {author} {\bibinfo {author} {\bibfnamefont {G.}~\bibnamefont
  {Tarnopolsky}}, \bibinfo {author} {\bibfnamefont {A.~J.}\ \bibnamefont
  {Kruchkov}},\ and\ \bibinfo {author} {\bibfnamefont {A.}~\bibnamefont
  {Vishwanath}},\ }\href {https://doi.org/10.1103/PhysRevLett.122.106405}
  {\bibfield  {journal} {\bibinfo  {journal} {Phys. Rev. Lett.}\ }\textbf
  {\bibinfo {volume} {122}},\ \bibinfo {pages} {106405} (\bibinfo {year}
  {2019})}\BibitemShut {NoStop}%
\bibitem [{\citenamefont {Liu}\ \emph {et~al.}(2019{\natexlab{a}})\citenamefont
  {Liu}, \citenamefont {Liu},\ and\ \citenamefont {Dai}}]{jpliu-prb19}%
  \BibitemOpen
  \bibfield  {author} {\bibinfo {author} {\bibfnamefont {J.}~\bibnamefont
  {Liu}}, \bibinfo {author} {\bibfnamefont {J.}~\bibnamefont {Liu}},\ and\
  \bibinfo {author} {\bibfnamefont {X.}~\bibnamefont {Dai}},\ }\href
  {https://doi.org/10.1103/PhysRevB.99.155415} {\bibfield  {journal} {\bibinfo
  {journal} {Phys. Rev. B}\ }\textbf {\bibinfo {volume} {99}},\ \bibinfo
  {pages} {155415} (\bibinfo {year} {2019}{\natexlab{a}})}\BibitemShut
  {NoStop}%
\bibitem [{\citenamefont {Song}\ \emph {et~al.}(2019)\citenamefont {Song},
  \citenamefont {Wang}, \citenamefont {Shi}, \citenamefont {Li}, \citenamefont
  {Fang},\ and\ \citenamefont {Bernevig}}]{song-tbg-prl19}%
  \BibitemOpen
  \bibfield  {author} {\bibinfo {author} {\bibfnamefont {Z.}~\bibnamefont
  {Song}}, \bibinfo {author} {\bibfnamefont {Z.}~\bibnamefont {Wang}}, \bibinfo
  {author} {\bibfnamefont {W.}~\bibnamefont {Shi}}, \bibinfo {author}
  {\bibfnamefont {G.}~\bibnamefont {Li}}, \bibinfo {author} {\bibfnamefont
  {C.}~\bibnamefont {Fang}},\ and\ \bibinfo {author} {\bibfnamefont {B.~A.}\
  \bibnamefont {Bernevig}},\ }\href
  {https://doi.org/10.1103/PhysRevLett.123.036401} {\bibfield  {journal}
  {\bibinfo  {journal} {Phys. Rev. Lett.}\ }\textbf {\bibinfo {volume} {123}},\
  \bibinfo {pages} {036401} (\bibinfo {year} {2019})}\BibitemShut {NoStop}%
\bibitem [{\citenamefont {Ahn}\ \emph {et~al.}(2019)\citenamefont {Ahn},
  \citenamefont {Park},\ and\ \citenamefont {Yang}}]{yang-tbg-prx19}%
  \BibitemOpen
  \bibfield  {author} {\bibinfo {author} {\bibfnamefont {J.}~\bibnamefont
  {Ahn}}, \bibinfo {author} {\bibfnamefont {S.}~\bibnamefont {Park}},\ and\
  \bibinfo {author} {\bibfnamefont {B.-J.}\ \bibnamefont {Yang}},\ }\href
  {https://doi.org/10.1103/PhysRevX.9.021013} {\bibfield  {journal} {\bibinfo
  {journal} {Phys. Rev. X}\ }\textbf {\bibinfo {volume} {9}},\ \bibinfo {pages}
  {021013} (\bibinfo {year} {2019})}\BibitemShut {NoStop}%
\bibitem [{\citenamefont {Po}\ \emph {et~al.}(2019)\citenamefont {Po},
  \citenamefont {Zou}, \citenamefont {Senthil},\ and\ \citenamefont
  {Vishwanath}}]{po-tbg-prb19}%
  \BibitemOpen
  \bibfield  {author} {\bibinfo {author} {\bibfnamefont {H.~C.}\ \bibnamefont
  {Po}}, \bibinfo {author} {\bibfnamefont {L.}~\bibnamefont {Zou}}, \bibinfo
  {author} {\bibfnamefont {T.}~\bibnamefont {Senthil}},\ and\ \bibinfo {author}
  {\bibfnamefont {A.}~\bibnamefont {Vishwanath}},\ }\href
  {https://doi.org/10.1103/PhysRevB.99.195455} {\bibfield  {journal} {\bibinfo
  {journal} {Phys. Rev. B}\ }\textbf {\bibinfo {volume} {99}},\ \bibinfo
  {pages} {195455} (\bibinfo {year} {2019})}\BibitemShut {NoStop}%
\bibitem [{\citenamefont {Bultinck}\ \emph
  {et~al.}(2020{\natexlab{a}})\citenamefont {Bultinck}, \citenamefont
  {Chatterjee},\ and\ \citenamefont {Zaletel}}]{zaletel-tbg-2019}%
  \BibitemOpen
  \bibfield  {author} {\bibinfo {author} {\bibfnamefont {N.}~\bibnamefont
  {Bultinck}}, \bibinfo {author} {\bibfnamefont {S.}~\bibnamefont
  {Chatterjee}},\ and\ \bibinfo {author} {\bibfnamefont {M.~P.}\ \bibnamefont
  {Zaletel}},\ }\href {https://doi.org/10.1103/PhysRevLett.124.166601}
  {\bibfield  {journal} {\bibinfo  {journal} {Phys. Rev. Lett.}\ }\textbf
  {\bibinfo {volume} {124}},\ \bibinfo {pages} {166601} (\bibinfo {year}
  {2020}{\natexlab{a}})}\BibitemShut {NoStop}%
\bibitem [{\citenamefont {Bultinck}\ \emph
  {et~al.}(2020{\natexlab{b}})\citenamefont {Bultinck}, \citenamefont {Khalaf},
  \citenamefont {Liu}, \citenamefont {Chatterjee}, \citenamefont {Vishwanath},\
  and\ \citenamefont {Zaletel}}]{zaletel-tbg-hf-prx20}%
  \BibitemOpen
  \bibfield  {author} {\bibinfo {author} {\bibfnamefont {N.}~\bibnamefont
  {Bultinck}}, \bibinfo {author} {\bibfnamefont {E.}~\bibnamefont {Khalaf}},
  \bibinfo {author} {\bibfnamefont {S.}~\bibnamefont {Liu}}, \bibinfo {author}
  {\bibfnamefont {S.}~\bibnamefont {Chatterjee}}, \bibinfo {author}
  {\bibfnamefont {A.}~\bibnamefont {Vishwanath}},\ and\ \bibinfo {author}
  {\bibfnamefont {M.~P.}\ \bibnamefont {Zaletel}},\ }\href
  {https://doi.org/10.1103/PhysRevX.10.031034} {\bibfield  {journal} {\bibinfo
  {journal} {Phys. Rev. X}\ }\textbf {\bibinfo {volume} {10}},\ \bibinfo
  {pages} {031034} (\bibinfo {year} {2020}{\natexlab{b}})}\BibitemShut
  {NoStop}%
\bibitem [{\citenamefont {Wang}\ \emph {et~al.}(2021)\citenamefont {Wang},
  \citenamefont {Zheng}, \citenamefont {Millis},\ and\ \citenamefont
  {Cano}}]{wang2021chiral}%
  \BibitemOpen
  \bibfield  {author} {\bibinfo {author} {\bibfnamefont {J.}~\bibnamefont
  {Wang}}, \bibinfo {author} {\bibfnamefont {Y.}~\bibnamefont {Zheng}},
  \bibinfo {author} {\bibfnamefont {A.~J.}\ \bibnamefont {Millis}},\ and\
  \bibinfo {author} {\bibfnamefont {J.}~\bibnamefont {Cano}},\ }\href@noop {}
  {\bibfield  {journal} {\bibinfo  {journal} {Physical Review Research}\
  }\textbf {\bibinfo {volume} {3}},\ \bibinfo {pages} {023155} (\bibinfo {year}
  {2021})}\BibitemShut {NoStop}%
\bibitem [{\citenamefont {Shi}\ and\ \citenamefont
  {Dai}(2022)}]{shi-dai-heavy-prb22}%
  \BibitemOpen
  \bibfield  {author} {\bibinfo {author} {\bibfnamefont {H.}~\bibnamefont
  {Shi}}\ and\ \bibinfo {author} {\bibfnamefont {X.}~\bibnamefont {Dai}},\
  }\href {https://doi.org/10.1103/PhysRevB.106.245129} {\bibfield  {journal}
  {\bibinfo  {journal} {Phys. Rev. B}\ }\textbf {\bibinfo {volume} {106}},\
  \bibinfo {pages} {245129} (\bibinfo {year} {2022})}\BibitemShut {NoStop}%
\bibitem [{\citenamefont {Cao}\ \emph {et~al.}(2018{\natexlab{a}})\citenamefont
  {Cao}, \citenamefont {Fatemi}, \citenamefont {Fang}, \citenamefont
  {Watanabe}, \citenamefont {Taniguchi}, \citenamefont {Kaxiras},\ and\
  \citenamefont {Jarillo-Herrero}}]{cao-nature18-supercond}%
  \BibitemOpen
  \bibfield  {author} {\bibinfo {author} {\bibfnamefont {Y.}~\bibnamefont
  {Cao}}, \bibinfo {author} {\bibfnamefont {V.}~\bibnamefont {Fatemi}},
  \bibinfo {author} {\bibfnamefont {S.}~\bibnamefont {Fang}}, \bibinfo {author}
  {\bibfnamefont {K.}~\bibnamefont {Watanabe}}, \bibinfo {author}
  {\bibfnamefont {T.}~\bibnamefont {Taniguchi}}, \bibinfo {author}
  {\bibfnamefont {E.}~\bibnamefont {Kaxiras}},\ and\ \bibinfo {author}
  {\bibfnamefont {P.}~\bibnamefont {Jarillo-Herrero}},\ }\href@noop {}
  {\bibfield  {journal} {\bibinfo  {journal} {Nature}\ }\textbf {\bibinfo
  {volume} {556}},\ \bibinfo {pages} {43} (\bibinfo {year}
  {2018}{\natexlab{a}})}\BibitemShut {NoStop}%
\bibitem [{\citenamefont {Yankowitz}\ \emph {et~al.}(2019)\citenamefont
  {Yankowitz}, \citenamefont {Chen}, \citenamefont {Polshyn}, \citenamefont
  {Zhang}, \citenamefont {Watanabe}, \citenamefont {Taniguchi}, \citenamefont
  {Graf}, \citenamefont {Young},\ and\ \citenamefont
  {Dean}}]{dean-tbg-science19}%
  \BibitemOpen
  \bibfield  {author} {\bibinfo {author} {\bibfnamefont {M.}~\bibnamefont
  {Yankowitz}}, \bibinfo {author} {\bibfnamefont {S.}~\bibnamefont {Chen}},
  \bibinfo {author} {\bibfnamefont {H.}~\bibnamefont {Polshyn}}, \bibinfo
  {author} {\bibfnamefont {Y.}~\bibnamefont {Zhang}}, \bibinfo {author}
  {\bibfnamefont {K.}~\bibnamefont {Watanabe}}, \bibinfo {author}
  {\bibfnamefont {T.}~\bibnamefont {Taniguchi}}, \bibinfo {author}
  {\bibfnamefont {D.}~\bibnamefont {Graf}}, \bibinfo {author} {\bibfnamefont
  {A.~F.}\ \bibnamefont {Young}},\ and\ \bibinfo {author} {\bibfnamefont
  {C.~R.}\ \bibnamefont {Dean}},\ }\href@noop {} {\bibfield  {journal}
  {\bibinfo  {journal} {Science}\ }\textbf {\bibinfo {volume} {363}},\ \bibinfo
  {pages} {1059} (\bibinfo {year} {2019})}\BibitemShut {NoStop}%
\bibitem [{\citenamefont {Codecido}\ \emph {et~al.}(2019)\citenamefont
  {Codecido}, \citenamefont {Wang}, \citenamefont {Koester}, \citenamefont
  {Che}, \citenamefont {Tian}, \citenamefont {Lv}, \citenamefont {Tran},
  \citenamefont {Watanabe}, \citenamefont {Taniguchi}, \citenamefont {Zhang},
  \citenamefont {Bockrath},\ and\ \citenamefont {Lau}}]{marc-tbg-19}%
  \BibitemOpen
  \bibfield  {author} {\bibinfo {author} {\bibfnamefont {E.}~\bibnamefont
  {Codecido}}, \bibinfo {author} {\bibfnamefont {Q.}~\bibnamefont {Wang}},
  \bibinfo {author} {\bibfnamefont {R.}~\bibnamefont {Koester}}, \bibinfo
  {author} {\bibfnamefont {S.}~\bibnamefont {Che}}, \bibinfo {author}
  {\bibfnamefont {H.}~\bibnamefont {Tian}}, \bibinfo {author} {\bibfnamefont
  {R.}~\bibnamefont {Lv}}, \bibinfo {author} {\bibfnamefont {S.}~\bibnamefont
  {Tran}}, \bibinfo {author} {\bibfnamefont {K.}~\bibnamefont {Watanabe}},
  \bibinfo {author} {\bibfnamefont {T.}~\bibnamefont {Taniguchi}}, \bibinfo
  {author} {\bibfnamefont {F.}~\bibnamefont {Zhang}}, \bibinfo {author}
  {\bibfnamefont {M.}~\bibnamefont {Bockrath}},\ and\ \bibinfo {author}
  {\bibfnamefont {C.~N.}\ \bibnamefont {Lau}},\ }\bibfield  {journal} {\bibinfo
   {journal} {Sci. Adv.}\ }\textbf {\bibinfo {volume} {5}},\ \href
  {https://doi.org/10.1126/sciadv.aaw9770} {10.1126/sciadv.aaw9770} (\bibinfo
  {year} {2019})\BibitemShut {NoStop}%
\bibitem [{\citenamefont {Lu}\ \emph {et~al.}(2019)\citenamefont {Lu},
  \citenamefont {Stepanov}, \citenamefont {Yang}, \citenamefont {Xie},
  \citenamefont {Aamir}, \citenamefont {Das}, \citenamefont {Urgell},
  \citenamefont {Watanabe}, \citenamefont {Taniguchi}, \citenamefont {Zhang},
  \citenamefont {Bachtold}, \citenamefont {MacDonald},\ and\ \citenamefont
  {Efetov}}]{efetov-nature19}%
  \BibitemOpen
  \bibfield  {author} {\bibinfo {author} {\bibfnamefont {X.}~\bibnamefont
  {Lu}}, \bibinfo {author} {\bibfnamefont {P.}~\bibnamefont {Stepanov}},
  \bibinfo {author} {\bibfnamefont {W.}~\bibnamefont {Yang}}, \bibinfo {author}
  {\bibfnamefont {M.}~\bibnamefont {Xie}}, \bibinfo {author} {\bibfnamefont
  {M.~A.}\ \bibnamefont {Aamir}}, \bibinfo {author} {\bibfnamefont
  {I.}~\bibnamefont {Das}}, \bibinfo {author} {\bibfnamefont {C.}~\bibnamefont
  {Urgell}}, \bibinfo {author} {\bibfnamefont {K.}~\bibnamefont {Watanabe}},
  \bibinfo {author} {\bibfnamefont {T.}~\bibnamefont {Taniguchi}}, \bibinfo
  {author} {\bibfnamefont {G.}~\bibnamefont {Zhang}}, \bibinfo {author}
  {\bibfnamefont {A.}~\bibnamefont {Bachtold}}, \bibinfo {author}
  {\bibfnamefont {A.~H.}\ \bibnamefont {MacDonald}},\ and\ \bibinfo {author}
  {\bibfnamefont {D.~K.}\ \bibnamefont {Efetov}},\ }\href
  {https://doi.org/10.1038/s41586-019-1695-0} {\bibfield  {journal} {\bibinfo
  {journal} {Nature}\ }\textbf {\bibinfo {volume} {574}},\ \bibinfo {pages}
  {653} (\bibinfo {year} {2019})}\BibitemShut {NoStop}%
\bibitem [{\citenamefont {Stepanov}\ \emph {et~al.}(2020)\citenamefont
  {Stepanov}, \citenamefont {Das}, \citenamefont {Lu}, \citenamefont
  {Fahimniya}, \citenamefont {Watanabe}, \citenamefont {Taniguchi},
  \citenamefont {Koppens}, \citenamefont {Lischner}, \citenamefont {Levitov},\
  and\ \citenamefont {Efetov}}]{efetov-nature20}%
  \BibitemOpen
  \bibfield  {author} {\bibinfo {author} {\bibfnamefont {P.}~\bibnamefont
  {Stepanov}}, \bibinfo {author} {\bibfnamefont {I.}~\bibnamefont {Das}},
  \bibinfo {author} {\bibfnamefont {X.}~\bibnamefont {Lu}}, \bibinfo {author}
  {\bibfnamefont {A.}~\bibnamefont {Fahimniya}}, \bibinfo {author}
  {\bibfnamefont {K.}~\bibnamefont {Watanabe}}, \bibinfo {author}
  {\bibfnamefont {T.}~\bibnamefont {Taniguchi}}, \bibinfo {author}
  {\bibfnamefont {F.~H.~L.}\ \bibnamefont {Koppens}}, \bibinfo {author}
  {\bibfnamefont {J.}~\bibnamefont {Lischner}}, \bibinfo {author}
  {\bibfnamefont {L.}~\bibnamefont {Levitov}},\ and\ \bibinfo {author}
  {\bibfnamefont {D.~K.}\ \bibnamefont {Efetov}},\ }\href
  {https://doi.org/10.1038/s41586-020-2459-6} {\bibfield  {journal} {\bibinfo
  {journal} {Nature}\ }\textbf {\bibinfo {volume} {583}},\ \bibinfo {pages}
  {375} (\bibinfo {year} {2020})}\BibitemShut {NoStop}%
\bibitem [{\citenamefont {Saito}\ \emph {et~al.}(2020)\citenamefont {Saito},
  \citenamefont {Ge}, \citenamefont {Watanabe}, \citenamefont {Taniguchi},\
  and\ \citenamefont {Young}}]{young-tbg-np20}%
  \BibitemOpen
  \bibfield  {author} {\bibinfo {author} {\bibfnamefont {Y.}~\bibnamefont
  {Saito}}, \bibinfo {author} {\bibfnamefont {J.}~\bibnamefont {Ge}}, \bibinfo
  {author} {\bibfnamefont {K.}~\bibnamefont {Watanabe}}, \bibinfo {author}
  {\bibfnamefont {T.}~\bibnamefont {Taniguchi}},\ and\ \bibinfo {author}
  {\bibfnamefont {A.~F.}\ \bibnamefont {Young}},\ }\href
  {https://doi.org/10.1038/s41567-020-0928-3} {\bibfield  {journal} {\bibinfo
  {journal} {Nature Physics}\ }\textbf {\bibinfo {volume} {16}},\ \bibinfo
  {pages} {926} (\bibinfo {year} {2020})}\BibitemShut {NoStop}%
\bibitem [{\citenamefont {Liu}\ \emph {et~al.}(2021)\citenamefont {Liu},
  \citenamefont {Wang}, \citenamefont {Watanabe}, \citenamefont {Taniguchi},
  \citenamefont {Vafek},\ and\ \citenamefont {Li}}]{li-tbg-science21}%
  \BibitemOpen
  \bibfield  {author} {\bibinfo {author} {\bibfnamefont {X.}~\bibnamefont
  {Liu}}, \bibinfo {author} {\bibfnamefont {Z.}~\bibnamefont {Wang}}, \bibinfo
  {author} {\bibfnamefont {K.}~\bibnamefont {Watanabe}}, \bibinfo {author}
  {\bibfnamefont {T.}~\bibnamefont {Taniguchi}}, \bibinfo {author}
  {\bibfnamefont {O.}~\bibnamefont {Vafek}},\ and\ \bibinfo {author}
  {\bibfnamefont {J.}~\bibnamefont {Li}},\ }\href@noop {} {\bibfield  {journal}
  {\bibinfo  {journal} {Science}\ }\textbf {\bibinfo {volume} {371}},\ \bibinfo
  {pages} {1261} (\bibinfo {year} {2021})}\BibitemShut {NoStop}%
\bibitem [{\citenamefont {Cao}\ \emph {et~al.}(2021{\natexlab{a}})\citenamefont
  {Cao}, \citenamefont {Rodan-Legrain}, \citenamefont {Park}, \citenamefont
  {Yuan}, \citenamefont {Watanabe}, \citenamefont {Taniguchi}, \citenamefont
  {Fernandes}, \citenamefont {Fu},\ and\ \citenamefont
  {Jarillo-Herrero}}]{cao-tbg-nematic-science21}%
  \BibitemOpen
  \bibfield  {author} {\bibinfo {author} {\bibfnamefont {Y.}~\bibnamefont
  {Cao}}, \bibinfo {author} {\bibfnamefont {D.}~\bibnamefont {Rodan-Legrain}},
  \bibinfo {author} {\bibfnamefont {J.~M.}\ \bibnamefont {Park}}, \bibinfo
  {author} {\bibfnamefont {N.~F.}\ \bibnamefont {Yuan}}, \bibinfo {author}
  {\bibfnamefont {K.}~\bibnamefont {Watanabe}}, \bibinfo {author}
  {\bibfnamefont {T.}~\bibnamefont {Taniguchi}}, \bibinfo {author}
  {\bibfnamefont {R.~M.}\ \bibnamefont {Fernandes}}, \bibinfo {author}
  {\bibfnamefont {L.}~\bibnamefont {Fu}},\ and\ \bibinfo {author}
  {\bibfnamefont {P.}~\bibnamefont {Jarillo-Herrero}},\ }\href@noop {}
  {\bibfield  {journal} {\bibinfo  {journal} {science}\ }\textbf {\bibinfo
  {volume} {372}},\ \bibinfo {pages} {264} (\bibinfo {year}
  {2021}{\natexlab{a}})}\BibitemShut {NoStop}%
\bibitem [{\citenamefont {Cao}\ \emph {et~al.}(2018{\natexlab{b}})\citenamefont
  {Cao}, \citenamefont {Fatemi}, \citenamefont {Demir}, \citenamefont {Fang},
  \citenamefont {Tomarken}, \citenamefont {Luo}, \citenamefont
  {Sanchez-Yamagishi}, \citenamefont {Watanabe}, \citenamefont {Taniguchi},
  \citenamefont {Kaxiras} \emph {et~al.}}]{cao-nature18-mott}%
  \BibitemOpen
  \bibfield  {author} {\bibinfo {author} {\bibfnamefont {Y.}~\bibnamefont
  {Cao}}, \bibinfo {author} {\bibfnamefont {V.}~\bibnamefont {Fatemi}},
  \bibinfo {author} {\bibfnamefont {A.}~\bibnamefont {Demir}}, \bibinfo
  {author} {\bibfnamefont {S.}~\bibnamefont {Fang}}, \bibinfo {author}
  {\bibfnamefont {S.~L.}\ \bibnamefont {Tomarken}}, \bibinfo {author}
  {\bibfnamefont {J.~Y.}\ \bibnamefont {Luo}}, \bibinfo {author} {\bibfnamefont
  {J.~D.}\ \bibnamefont {Sanchez-Yamagishi}}, \bibinfo {author} {\bibfnamefont
  {K.}~\bibnamefont {Watanabe}}, \bibinfo {author} {\bibfnamefont
  {T.}~\bibnamefont {Taniguchi}}, \bibinfo {author} {\bibfnamefont
  {E.}~\bibnamefont {Kaxiras}}, \emph {et~al.},\ }\href@noop {} {\bibfield
  {journal} {\bibinfo  {journal} {Nature}\ }\textbf {\bibinfo {volume} {556}},\
  \bibinfo {pages} {80} (\bibinfo {year} {2018}{\natexlab{b}})}\BibitemShut
  {NoStop}%
\bibitem [{\citenamefont {Kerelsky}\ \emph {et~al.}(2019)\citenamefont
  {Kerelsky}, \citenamefont {McGilly}, \citenamefont {Kennes}, \citenamefont
  {Xian}, \citenamefont {Yankowitz}, \citenamefont {Chen}, \citenamefont
  {Watanabe}, \citenamefont {Taniguchi}, \citenamefont {Hone}, \citenamefont
  {Dean} \emph {et~al.}}]{tbg-stm-pasupathy19}%
  \BibitemOpen
  \bibfield  {author} {\bibinfo {author} {\bibfnamefont {A.}~\bibnamefont
  {Kerelsky}}, \bibinfo {author} {\bibfnamefont {L.~J.}\ \bibnamefont
  {McGilly}}, \bibinfo {author} {\bibfnamefont {D.~M.}\ \bibnamefont {Kennes}},
  \bibinfo {author} {\bibfnamefont {L.}~\bibnamefont {Xian}}, \bibinfo {author}
  {\bibfnamefont {M.}~\bibnamefont {Yankowitz}}, \bibinfo {author}
  {\bibfnamefont {S.}~\bibnamefont {Chen}}, \bibinfo {author} {\bibfnamefont
  {K.}~\bibnamefont {Watanabe}}, \bibinfo {author} {\bibfnamefont
  {T.}~\bibnamefont {Taniguchi}}, \bibinfo {author} {\bibfnamefont
  {J.}~\bibnamefont {Hone}}, \bibinfo {author} {\bibfnamefont {C.}~\bibnamefont
  {Dean}}, \emph {et~al.},\ }\href@noop {} {\bibfield  {journal} {\bibinfo
  {journal} {Nature}\ }\textbf {\bibinfo {volume} {572}},\ \bibinfo {pages}
  {95} (\bibinfo {year} {2019})}\BibitemShut {NoStop}%
\bibitem [{\citenamefont {{Jiang}}\ \emph {et~al.}(2019)\citenamefont
  {{Jiang}}, \citenamefont {{Lai}}, \citenamefont {{Watanabe}}, \citenamefont
  {{Taniguchi}}, \citenamefont {{Haule}}, \citenamefont {{Mao}},\ and\
  \citenamefont {{Andrei}}}]{tbg-stm-andrei19}%
  \BibitemOpen
  \bibfield  {author} {\bibinfo {author} {\bibfnamefont {Y.}~\bibnamefont
  {{Jiang}}}, \bibinfo {author} {\bibfnamefont {X.}~\bibnamefont {{Lai}}},
  \bibinfo {author} {\bibfnamefont {K.}~\bibnamefont {{Watanabe}}}, \bibinfo
  {author} {\bibfnamefont {T.}~\bibnamefont {{Taniguchi}}}, \bibinfo {author}
  {\bibfnamefont {K.}~\bibnamefont {{Haule}}}, \bibinfo {author} {\bibfnamefont
  {J.}~\bibnamefont {{Mao}}},\ and\ \bibinfo {author} {\bibfnamefont {E.~Y.}\
  \bibnamefont {{Andrei}}},\ }\href {https://doi.org/10.1038/s41586-019-1460-4}
  {\bibfield  {journal} {\bibinfo  {journal} {Nature}\ }\textbf {\bibinfo
  {volume} {573}},\ \bibinfo {pages} {91} (\bibinfo {year} {2019})}\BibitemShut
  {NoStop}%
\bibitem [{\citenamefont {{Xie}}\ \emph {et~al.}(2019)\citenamefont {{Xie}},
  \citenamefont {{Lian}}, \citenamefont {{J{\"a}ck}}, \citenamefont {{Liu}},
  \citenamefont {{Chiu}}, \citenamefont {{Watanabe}}, \citenamefont
  {{Taniguchi}}, \citenamefont {{Bernevig}},\ and\ \citenamefont
  {{Yazdani}}}]{tbg-stm-yazdani19}%
  \BibitemOpen
  \bibfield  {author} {\bibinfo {author} {\bibfnamefont {Y.}~\bibnamefont
  {{Xie}}}, \bibinfo {author} {\bibfnamefont {B.}~\bibnamefont {{Lian}}},
  \bibinfo {author} {\bibfnamefont {B.}~\bibnamefont {{J{\"a}ck}}}, \bibinfo
  {author} {\bibfnamefont {X.}~\bibnamefont {{Liu}}}, \bibinfo {author}
  {\bibfnamefont {C.-L.}\ \bibnamefont {{Chiu}}}, \bibinfo {author}
  {\bibfnamefont {K.}~\bibnamefont {{Watanabe}}}, \bibinfo {author}
  {\bibfnamefont {T.}~\bibnamefont {{Taniguchi}}}, \bibinfo {author}
  {\bibfnamefont {B.~A.}\ \bibnamefont {{Bernevig}}},\ and\ \bibinfo {author}
  {\bibfnamefont {A.}~\bibnamefont {{Yazdani}}},\ }\href
  {https://doi.org/10.1038/s41586-019-1422-x} {\bibfield  {journal} {\bibinfo
  {journal} {Nature}\ }\textbf {\bibinfo {volume} {572}},\ \bibinfo {pages}
  {101} (\bibinfo {year} {2019})}\BibitemShut {NoStop}%
\bibitem [{\citenamefont {Choi}\ \emph {et~al.}(2019)\citenamefont {Choi},
  \citenamefont {Kemmer}, \citenamefont {Peng}, \citenamefont {Thomson},
  \citenamefont {Arora}, \citenamefont {Polski}, \citenamefont {Zhang},
  \citenamefont {Ren}, \citenamefont {Alicea}, \citenamefont {Refael},
  \citenamefont {von Oppen}, \citenamefont {Watanabe}, \citenamefont
  {Taniguchi},\ and\ \citenamefont {Nadj-Perge}}]{tbg-stm-caltech19}%
  \BibitemOpen
  \bibfield  {author} {\bibinfo {author} {\bibfnamefont {Y.}~\bibnamefont
  {Choi}}, \bibinfo {author} {\bibfnamefont {J.}~\bibnamefont {Kemmer}},
  \bibinfo {author} {\bibfnamefont {Y.}~\bibnamefont {Peng}}, \bibinfo {author}
  {\bibfnamefont {A.}~\bibnamefont {Thomson}}, \bibinfo {author} {\bibfnamefont
  {H.}~\bibnamefont {Arora}}, \bibinfo {author} {\bibfnamefont
  {R.}~\bibnamefont {Polski}}, \bibinfo {author} {\bibfnamefont
  {Y.}~\bibnamefont {Zhang}}, \bibinfo {author} {\bibfnamefont
  {H.}~\bibnamefont {Ren}}, \bibinfo {author} {\bibfnamefont {J.}~\bibnamefont
  {Alicea}}, \bibinfo {author} {\bibfnamefont {G.}~\bibnamefont {Refael}},
  \bibinfo {author} {\bibfnamefont {F.}~\bibnamefont {von Oppen}}, \bibinfo
  {author} {\bibfnamefont {K.}~\bibnamefont {Watanabe}}, \bibinfo {author}
  {\bibfnamefont {T.}~\bibnamefont {Taniguchi}},\ and\ \bibinfo {author}
  {\bibfnamefont {S.}~\bibnamefont {Nadj-Perge}},\ }\href
  {https://doi.org/10.1038/s41567-019-0606-5} {\bibfield  {journal} {\bibinfo
  {journal} {Nat. Phys.}\ }\textbf {\bibinfo {volume} {15}},\ \bibinfo {pages}
  {1174} (\bibinfo {year} {2019})}\BibitemShut {NoStop}%
\bibitem [{\citenamefont {Serlin}\ \emph {et~al.}(2019)\citenamefont {Serlin},
  \citenamefont {Tschirhart}, \citenamefont {Polshyn}, \citenamefont {Zhang},
  \citenamefont {Zhu}, \citenamefont {Watanabe}, \citenamefont {Taniguchi},
  \citenamefont {Balents},\ and\ \citenamefont {Young}}]{young-tbg-science19}%
  \BibitemOpen
  \bibfield  {author} {\bibinfo {author} {\bibfnamefont {M.}~\bibnamefont
  {Serlin}}, \bibinfo {author} {\bibfnamefont {C.}~\bibnamefont {Tschirhart}},
  \bibinfo {author} {\bibfnamefont {H.}~\bibnamefont {Polshyn}}, \bibinfo
  {author} {\bibfnamefont {Y.}~\bibnamefont {Zhang}}, \bibinfo {author}
  {\bibfnamefont {J.}~\bibnamefont {Zhu}}, \bibinfo {author} {\bibfnamefont
  {K.}~\bibnamefont {Watanabe}}, \bibinfo {author} {\bibfnamefont
  {T.}~\bibnamefont {Taniguchi}}, \bibinfo {author} {\bibfnamefont
  {L.}~\bibnamefont {Balents}},\ and\ \bibinfo {author} {\bibfnamefont
  {A.}~\bibnamefont {Young}},\ }\href@noop {} {\bibfield  {journal} {\bibinfo
  {journal} {Science}\ } (\bibinfo {year} {2019})}\BibitemShut {NoStop}%
\bibitem [{\citenamefont {{Sharpe}}\ \emph {et~al.}(2019)\citenamefont
  {{Sharpe}}, \citenamefont {{Fox}}, \citenamefont {{Barnard}}, \citenamefont
  {{Finney}}, \citenamefont {{Watanabe}}, \citenamefont {{Taniguchi}},
  \citenamefont {{Kastner}},\ and\ \citenamefont
  {{Goldhaber-Gordon}}}]{sharpe-science-19}%
  \BibitemOpen
  \bibfield  {author} {\bibinfo {author} {\bibfnamefont {A.~L.}\ \bibnamefont
  {{Sharpe}}}, \bibinfo {author} {\bibfnamefont {E.~J.}\ \bibnamefont {{Fox}}},
  \bibinfo {author} {\bibfnamefont {A.~W.}\ \bibnamefont {{Barnard}}}, \bibinfo
  {author} {\bibfnamefont {J.}~\bibnamefont {{Finney}}}, \bibinfo {author}
  {\bibfnamefont {K.}~\bibnamefont {{Watanabe}}}, \bibinfo {author}
  {\bibfnamefont {T.}~\bibnamefont {{Taniguchi}}}, \bibinfo {author}
  {\bibfnamefont {M.~A.}\ \bibnamefont {{Kastner}}},\ and\ \bibinfo {author}
  {\bibfnamefont {D.}~\bibnamefont {{Goldhaber-Gordon}}},\ }\href
  {https://doi.org/10.1126/science.aaw3780} {\bibfield  {journal} {\bibinfo
  {journal} {Science}\ }\textbf {\bibinfo {volume} {365}},\ \bibinfo {pages}
  {605} (\bibinfo {year} {2019})}\BibitemShut {NoStop}%
\bibitem [{\citenamefont {Balents}\ \emph {et~al.}(2020)\citenamefont
  {Balents}, \citenamefont {Dean}, \citenamefont {Efetov},\ and\ \citenamefont
  {Young}}]{balents-review-tbg}%
  \BibitemOpen
  \bibfield  {author} {\bibinfo {author} {\bibfnamefont {L.}~\bibnamefont
  {Balents}}, \bibinfo {author} {\bibfnamefont {C.~R.}\ \bibnamefont {Dean}},
  \bibinfo {author} {\bibfnamefont {D.~K.}\ \bibnamefont {Efetov}},\ and\
  \bibinfo {author} {\bibfnamefont {A.~F.}\ \bibnamefont {Young}},\ }\href
  {https://doi.org/10.1038/s41567-020-0906-9} {\bibfield  {journal} {\bibinfo
  {journal} {Nat. Phys.}\ }\textbf {\bibinfo {volume} {16}},\ \bibinfo {pages}
  {725} (\bibinfo {year} {2020})}\BibitemShut {NoStop}%
\bibitem [{\citenamefont {Andrei}\ \emph {et~al.}(2021)\citenamefont {Andrei},
  \citenamefont {Efetov}, \citenamefont {Jarillo-Herrero}, \citenamefont
  {MacDonald}, \citenamefont {Mak}, \citenamefont {Senthil}, \citenamefont
  {Tutuc}, \citenamefont {Yazdani},\ and\ \citenamefont
  {Young}}]{andrei-review-tbg}%
  \BibitemOpen
  \bibfield  {author} {\bibinfo {author} {\bibfnamefont {E.~Y.}\ \bibnamefont
  {Andrei}}, \bibinfo {author} {\bibfnamefont {D.~K.}\ \bibnamefont {Efetov}},
  \bibinfo {author} {\bibfnamefont {P.}~\bibnamefont {Jarillo-Herrero}},
  \bibinfo {author} {\bibfnamefont {A.~H.}\ \bibnamefont {MacDonald}}, \bibinfo
  {author} {\bibfnamefont {K.~F.}\ \bibnamefont {Mak}}, \bibinfo {author}
  {\bibfnamefont {T.}~\bibnamefont {Senthil}}, \bibinfo {author} {\bibfnamefont
  {E.}~\bibnamefont {Tutuc}}, \bibinfo {author} {\bibfnamefont
  {A.}~\bibnamefont {Yazdani}},\ and\ \bibinfo {author} {\bibfnamefont {A.~F.}\
  \bibnamefont {Young}},\ }\href {https://doi.org/10.1038/s41578-021-00284-1}
  {\bibfield  {journal} {\bibinfo  {journal} {Nat. Rev. Mater.}\ }\textbf
  {\bibinfo {volume} {6}},\ \bibinfo {pages} {201} (\bibinfo {year}
  {2021})}\BibitemShut {NoStop}%
\bibitem [{\citenamefont {Liu}\ and\ \citenamefont
  {Dai}(2021{\natexlab{a}})}]{jpliu-nrp21}%
  \BibitemOpen
  \bibfield  {author} {\bibinfo {author} {\bibfnamefont {J.}~\bibnamefont
  {Liu}}\ and\ \bibinfo {author} {\bibfnamefont {X.}~\bibnamefont {Dai}},\
  }\href {https://doi.org/10.1038/s42254-021-00297-3} {\bibfield  {journal}
  {\bibinfo  {journal} {Nature Reviews Physics}\ }\textbf {\bibinfo {volume}
  {3}},\ \bibinfo {pages} {367} (\bibinfo {year}
  {2021}{\natexlab{a}})}\BibitemShut {NoStop}%
\bibitem [{\citenamefont {Stepanov}\ \emph {et~al.}(2021)\citenamefont
  {Stepanov}, \citenamefont {Xie}, \citenamefont {Taniguchi}, \citenamefont
  {Watanabe}, \citenamefont {Lu}, \citenamefont {MacDonald}, \citenamefont
  {Bernevig},\ and\ \citenamefont {Efetov}}]{efetov-arxiv20}%
  \BibitemOpen
  \bibfield  {author} {\bibinfo {author} {\bibfnamefont {P.}~\bibnamefont
  {Stepanov}}, \bibinfo {author} {\bibfnamefont {M.}~\bibnamefont {Xie}},
  \bibinfo {author} {\bibfnamefont {T.}~\bibnamefont {Taniguchi}}, \bibinfo
  {author} {\bibfnamefont {K.}~\bibnamefont {Watanabe}}, \bibinfo {author}
  {\bibfnamefont {X.}~\bibnamefont {Lu}}, \bibinfo {author} {\bibfnamefont
  {A.~H.}\ \bibnamefont {MacDonald}}, \bibinfo {author} {\bibfnamefont {B.~A.}\
  \bibnamefont {Bernevig}},\ and\ \bibinfo {author} {\bibfnamefont {D.~K.}\
  \bibnamefont {Efetov}},\ }\href
  {https://doi.org/10.1103/PhysRevLett.127.197701} {\bibfield  {journal}
  {\bibinfo  {journal} {Phys. Rev. Lett.}\ }\textbf {\bibinfo {volume} {127}},\
  \bibinfo {pages} {197701} (\bibinfo {year} {2021})}\BibitemShut {NoStop}%
\bibitem [{\citenamefont {Nuckolls}\ \emph {et~al.}(2020)\citenamefont
  {Nuckolls}, \citenamefont {Oh}, \citenamefont {Wong}, \citenamefont {Lian},
  \citenamefont {Watanabe}, \citenamefont {Taniguchi}, \citenamefont
  {Bernevig},\ and\ \citenamefont {Yazdani}}]{yazdani-tbg-chern-arxiv20}%
  \BibitemOpen
  \bibfield  {author} {\bibinfo {author} {\bibfnamefont {K.~P.}\ \bibnamefont
  {Nuckolls}}, \bibinfo {author} {\bibfnamefont {M.}~\bibnamefont {Oh}},
  \bibinfo {author} {\bibfnamefont {D.}~\bibnamefont {Wong}}, \bibinfo {author}
  {\bibfnamefont {B.}~\bibnamefont {Lian}}, \bibinfo {author} {\bibfnamefont
  {K.}~\bibnamefont {Watanabe}}, \bibinfo {author} {\bibfnamefont
  {T.}~\bibnamefont {Taniguchi}}, \bibinfo {author} {\bibfnamefont {B.~A.}\
  \bibnamefont {Bernevig}},\ and\ \bibinfo {author} {\bibfnamefont
  {A.}~\bibnamefont {Yazdani}},\ }\href
  {https://doi.org/10.1038/s41586-020-3028-8} {\bibfield  {journal} {\bibinfo
  {journal} {Nature}\ }\textbf {\bibinfo {volume} {588}},\ \bibinfo {pages}
  {610} (\bibinfo {year} {2020})}\BibitemShut {NoStop}%
\bibitem [{\citenamefont {Wu}\ \emph {et~al.}(2021)\citenamefont {Wu},
  \citenamefont {Zhang}, \citenamefont {Watanabe}, \citenamefont {Taniguchi},\
  and\ \citenamefont {Andrei}}]{andrei-tbg-chern-arxiv20}%
  \BibitemOpen
  \bibfield  {author} {\bibinfo {author} {\bibfnamefont {S.}~\bibnamefont
  {Wu}}, \bibinfo {author} {\bibfnamefont {Z.}~\bibnamefont {Zhang}}, \bibinfo
  {author} {\bibfnamefont {K.}~\bibnamefont {Watanabe}}, \bibinfo {author}
  {\bibfnamefont {T.}~\bibnamefont {Taniguchi}},\ and\ \bibinfo {author}
  {\bibfnamefont {E.~Y.}\ \bibnamefont {Andrei}},\ }\href@noop {} {\bibfield
  {journal} {\bibinfo  {journal} {Nature Materials}\ } (\bibinfo {year}
  {2021})}\BibitemShut {NoStop}%
\bibitem [{\citenamefont {Das}\ \emph {et~al.}(2021)\citenamefont {Das},
  \citenamefont {Lu}, \citenamefont {Herzog-Arbeitman}, \citenamefont {Song},
  \citenamefont {Watanabe}, \citenamefont {Taniguchi}, \citenamefont
  {Bernevig},\ and\ \citenamefont {Efetov}}]{efetov-tbg-chern-arxiv20}%
  \BibitemOpen
  \bibfield  {author} {\bibinfo {author} {\bibfnamefont {I.}~\bibnamefont
  {Das}}, \bibinfo {author} {\bibfnamefont {X.}~\bibnamefont {Lu}}, \bibinfo
  {author} {\bibfnamefont {J.}~\bibnamefont {Herzog-Arbeitman}}, \bibinfo
  {author} {\bibfnamefont {Z.-D.}\ \bibnamefont {Song}}, \bibinfo {author}
  {\bibfnamefont {K.}~\bibnamefont {Watanabe}}, \bibinfo {author}
  {\bibfnamefont {T.}~\bibnamefont {Taniguchi}}, \bibinfo {author}
  {\bibfnamefont {B.~A.}\ \bibnamefont {Bernevig}},\ and\ \bibinfo {author}
  {\bibfnamefont {D.~K.}\ \bibnamefont {Efetov}},\ }\href
  {https://doi.org/10.1038/s41567-021-01186-3} {\bibfield  {journal} {\bibinfo
  {journal} {Nature Physics}\ }\textbf {\bibinfo {volume} {17}},\ \bibinfo
  {pages} {710} (\bibinfo {year} {2021})}\BibitemShut {NoStop}%
\bibitem [{\citenamefont {Pierce}\ \emph {et~al.}(2021)\citenamefont {Pierce},
  \citenamefont {Xie}, \citenamefont {Park}, \citenamefont {Khalaf},
  \citenamefont {Lee}, \citenamefont {Cao}, \citenamefont {Parker},
  \citenamefont {Forrester}, \citenamefont {Chen}, \citenamefont {Watanabe}
  \emph {et~al.}}]{pablo-tbg-chern-arxiv21}%
  \BibitemOpen
  \bibfield  {author} {\bibinfo {author} {\bibfnamefont {A.~T.}\ \bibnamefont
  {Pierce}}, \bibinfo {author} {\bibfnamefont {Y.}~\bibnamefont {Xie}},
  \bibinfo {author} {\bibfnamefont {J.~M.}\ \bibnamefont {Park}}, \bibinfo
  {author} {\bibfnamefont {E.}~\bibnamefont {Khalaf}}, \bibinfo {author}
  {\bibfnamefont {S.~H.}\ \bibnamefont {Lee}}, \bibinfo {author} {\bibfnamefont
  {Y.}~\bibnamefont {Cao}}, \bibinfo {author} {\bibfnamefont {D.~E.}\
  \bibnamefont {Parker}}, \bibinfo {author} {\bibfnamefont {P.~R.}\
  \bibnamefont {Forrester}}, \bibinfo {author} {\bibfnamefont {S.}~\bibnamefont
  {Chen}}, \bibinfo {author} {\bibfnamefont {K.}~\bibnamefont {Watanabe}},
  \emph {et~al.},\ }\href@noop {} {\bibfield  {journal} {\bibinfo  {journal}
  {Nature Physics}\ }\textbf {\bibinfo {volume} {17}},\ \bibinfo {pages} {1210}
  (\bibinfo {year} {2021})}\BibitemShut {NoStop}%
\bibitem [{\citenamefont {Shen}\ \emph {et~al.}(2021)\citenamefont {Shen},
  \citenamefont {Ying}, \citenamefont {Liu}, \citenamefont {Liu}, \citenamefont
  {Li}, \citenamefont {Wang}, \citenamefont {Tang}, \citenamefont {Zhao},
  \citenamefont {Chu}, \citenamefont {Watanabe}, \citenamefont {Taniguchi},
  \citenamefont {Yang}, \citenamefont {Shi}, \citenamefont {Qu}, \citenamefont
  {Lu}, \citenamefont {Yang},\ and\ \citenamefont {Zhang}}]{yang-tbg-cpl21}%
  \BibitemOpen
  \bibfield  {author} {\bibinfo {author} {\bibfnamefont {C.}~\bibnamefont
  {Shen}}, \bibinfo {author} {\bibfnamefont {J.}~\bibnamefont {Ying}}, \bibinfo
  {author} {\bibfnamefont {L.}~\bibnamefont {Liu}}, \bibinfo {author}
  {\bibfnamefont {J.}~\bibnamefont {Liu}}, \bibinfo {author} {\bibfnamefont
  {N.}~\bibnamefont {Li}}, \bibinfo {author} {\bibfnamefont {S.}~\bibnamefont
  {Wang}}, \bibinfo {author} {\bibfnamefont {J.}~\bibnamefont {Tang}}, \bibinfo
  {author} {\bibfnamefont {Y.}~\bibnamefont {Zhao}}, \bibinfo {author}
  {\bibfnamefont {Y.}~\bibnamefont {Chu}}, \bibinfo {author} {\bibfnamefont
  {K.}~\bibnamefont {Watanabe}}, \bibinfo {author} {\bibfnamefont
  {T.}~\bibnamefont {Taniguchi}}, \bibinfo {author} {\bibfnamefont
  {R.}~\bibnamefont {Yang}}, \bibinfo {author} {\bibfnamefont {D.}~\bibnamefont
  {Shi}}, \bibinfo {author} {\bibfnamefont {F.}~\bibnamefont {Qu}}, \bibinfo
  {author} {\bibfnamefont {L.}~\bibnamefont {Lu}}, \bibinfo {author}
  {\bibfnamefont {W.}~\bibnamefont {Yang}},\ and\ \bibinfo {author}
  {\bibfnamefont {G.}~\bibnamefont {Zhang}},\ }\href
  {https://doi.org/10.1088/0256-307X/38/4/047301} {\bibfield  {journal}
  {\bibinfo  {journal} {Chinese Physics Letters}\ }\textbf {\bibinfo {volume}
  {38}},\ \bibinfo {pages} {047301} (\bibinfo {year} {2021})}\BibitemShut
  {NoStop}%
\bibitem [{\citenamefont {Kang}\ and\ \citenamefont
  {Vafek}(2019)}]{kang-tbg-prl19}%
  \BibitemOpen
  \bibfield  {author} {\bibinfo {author} {\bibfnamefont {J.}~\bibnamefont
  {Kang}}\ and\ \bibinfo {author} {\bibfnamefont {O.}~\bibnamefont {Vafek}},\
  }\href@noop {} {\bibfield  {journal} {\bibinfo  {journal} {Phys. Rev. Lett.}\
  }\textbf {\bibinfo {volume} {122}},\ \bibinfo {pages} {246401} (\bibinfo
  {year} {2019})}\BibitemShut {NoStop}%
\bibitem [{\citenamefont {Seo}\ \emph {et~al.}(2019)\citenamefont {Seo},
  \citenamefont {Kotov},\ and\ \citenamefont {Uchoa}}]{Uchoa-ferroMott-prl}%
  \BibitemOpen
  \bibfield  {author} {\bibinfo {author} {\bibfnamefont {K.}~\bibnamefont
  {Seo}}, \bibinfo {author} {\bibfnamefont {V.~N.}\ \bibnamefont {Kotov}},\
  and\ \bibinfo {author} {\bibfnamefont {B.}~\bibnamefont {Uchoa}},\ }\href
  {https://doi.org/10.1103/PhysRevLett.122.246402} {\bibfield  {journal}
  {\bibinfo  {journal} {Phys. Rev. Lett.}\ }\textbf {\bibinfo {volume} {122}},\
  \bibinfo {pages} {246402} (\bibinfo {year} {2019})}\BibitemShut {NoStop}%
\bibitem [{\citenamefont {Xie}\ and\ \citenamefont
  {MacDonald}(2020)}]{xie-tbg-2018}%
  \BibitemOpen
  \bibfield  {author} {\bibinfo {author} {\bibfnamefont {M.}~\bibnamefont
  {Xie}}\ and\ \bibinfo {author} {\bibfnamefont {A.~H.}\ \bibnamefont
  {MacDonald}},\ }\href {https://doi.org/10.1103/PhysRevLett.124.097601}
  {\bibfield  {journal} {\bibinfo  {journal} {Phys. Rev. Lett.}\ }\textbf
  {\bibinfo {volume} {124}},\ \bibinfo {pages} {097601} (\bibinfo {year}
  {2020})}\BibitemShut {NoStop}%
\bibitem [{\citenamefont {Wu}\ and\ \citenamefont
  {Das~Sarma}(2020)}]{wu-tbg-collective-prl20}%
  \BibitemOpen
  \bibfield  {author} {\bibinfo {author} {\bibfnamefont {F.}~\bibnamefont
  {Wu}}\ and\ \bibinfo {author} {\bibfnamefont {S.}~\bibnamefont {Das~Sarma}},\
  }\href {https://doi.org/10.1103/PhysRevLett.124.046403} {\bibfield  {journal}
  {\bibinfo  {journal} {Phys. Rev. Lett.}\ }\textbf {\bibinfo {volume} {124}},\
  \bibinfo {pages} {046403} (\bibinfo {year} {2020})}\BibitemShut {NoStop}%
\bibitem [{\citenamefont {Liu}\ and\ \citenamefont
  {Dai}(2021{\natexlab{b}})}]{jpliu-tbghf-prb21}%
  \BibitemOpen
  \bibfield  {author} {\bibinfo {author} {\bibfnamefont {J.}~\bibnamefont
  {Liu}}\ and\ \bibinfo {author} {\bibfnamefont {X.}~\bibnamefont {Dai}},\
  }\href {https://doi.org/10.1103/PhysRevB.103.035427} {\bibfield  {journal}
  {\bibinfo  {journal} {Phys. Rev. B}\ }\textbf {\bibinfo {volume} {103}},\
  \bibinfo {pages} {035427} (\bibinfo {year} {2021}{\natexlab{b}})}\BibitemShut
  {NoStop}%
\bibitem [{\citenamefont {Zhang}\ \emph {et~al.}(2020)\citenamefont {Zhang},
  \citenamefont {Jiang}, \citenamefont {Wang},\ and\ \citenamefont
  {Zhang}}]{zhang-tbghf-arxiv20}%
  \BibitemOpen
  \bibfield  {author} {\bibinfo {author} {\bibfnamefont {Y.}~\bibnamefont
  {Zhang}}, \bibinfo {author} {\bibfnamefont {K.}~\bibnamefont {Jiang}},
  \bibinfo {author} {\bibfnamefont {Z.}~\bibnamefont {Wang}},\ and\ \bibinfo
  {author} {\bibfnamefont {F.}~\bibnamefont {Zhang}},\ }\href
  {https://doi.org/10.1103/PhysRevB.102.035136} {\bibfield  {journal} {\bibinfo
   {journal} {Phys. Rev. B}\ }\textbf {\bibinfo {volume} {102}},\ \bibinfo
  {pages} {035136} (\bibinfo {year} {2020})}\BibitemShut {NoStop}%
\bibitem [{\citenamefont {Hejazi}\ \emph {et~al.}(2021)\citenamefont {Hejazi},
  \citenamefont {Chen},\ and\ \citenamefont {Balents}}]{hejazi-tbg-hf}%
  \BibitemOpen
  \bibfield  {author} {\bibinfo {author} {\bibfnamefont {K.}~\bibnamefont
  {Hejazi}}, \bibinfo {author} {\bibfnamefont {X.}~\bibnamefont {Chen}},\ and\
  \bibinfo {author} {\bibfnamefont {L.}~\bibnamefont {Balents}},\ }\href
  {https://doi.org/10.1103/PhysRevResearch.3.013242} {\bibfield  {journal}
  {\bibinfo  {journal} {Phys. Rev. Research}\ }\textbf {\bibinfo {volume}
  {3}},\ \bibinfo {pages} {013242} (\bibinfo {year} {2021})}\BibitemShut
  {NoStop}%
\bibitem [{\citenamefont {Kang}\ and\ \citenamefont
  {Vafek}(2020)}]{kang-tbg-dmrg-prb20}%
  \BibitemOpen
  \bibfield  {author} {\bibinfo {author} {\bibfnamefont {J.}~\bibnamefont
  {Kang}}\ and\ \bibinfo {author} {\bibfnamefont {O.}~\bibnamefont {Vafek}},\
  }\href {https://doi.org/10.1103/PhysRevB.102.035161} {\bibfield  {journal}
  {\bibinfo  {journal} {Phys. Rev. B}\ }\textbf {\bibinfo {volume} {102}},\
  \bibinfo {pages} {035161} (\bibinfo {year} {2020})}\BibitemShut {NoStop}%
\bibitem [{\citenamefont {Chen}\ \emph {et~al.}(2021)\citenamefont {Chen},
  \citenamefont {Liao}, \citenamefont {Chen}, \citenamefont {Vafek},
  \citenamefont {Kang}, \citenamefont {Li},\ and\ \citenamefont
  {Meng}}]{kang-tbg-topomott}%
  \BibitemOpen
  \bibfield  {author} {\bibinfo {author} {\bibfnamefont {B.-B.}\ \bibnamefont
  {Chen}}, \bibinfo {author} {\bibfnamefont {Y.~D.}\ \bibnamefont {Liao}},
  \bibinfo {author} {\bibfnamefont {Z.}~\bibnamefont {Chen}}, \bibinfo {author}
  {\bibfnamefont {O.}~\bibnamefont {Vafek}}, \bibinfo {author} {\bibfnamefont
  {J.}~\bibnamefont {Kang}}, \bibinfo {author} {\bibfnamefont {W.}~\bibnamefont
  {Li}},\ and\ \bibinfo {author} {\bibfnamefont {Z.~Y.}\ \bibnamefont {Meng}},\
  }\href {https://doi.org/10.1038/s41467-021-25438-1} {\bibfield  {journal}
  {\bibinfo  {journal} {Nat. Commun.}\ }\textbf {\bibinfo {volume} {12}},\
  \bibinfo {pages} {5480} (\bibinfo {year} {2021})}\BibitemShut {NoStop}%
\bibitem [{\citenamefont {Lu}\ \emph {et~al.}(2022)\citenamefont {Lu},
  \citenamefont {Zhang}, \citenamefont {Zhang}, \citenamefont {Zhang},
  \citenamefont {Liu}, \citenamefont {Wang}, \citenamefont {Gu}, \citenamefont
  {Chen},\ and\ \citenamefont {Yang}}]{yang-tbg-arxiv20}%
  \BibitemOpen
  \bibfield  {author} {\bibinfo {author} {\bibfnamefont {C.}~\bibnamefont
  {Lu}}, \bibinfo {author} {\bibfnamefont {Y.}~\bibnamefont {Zhang}}, \bibinfo
  {author} {\bibfnamefont {Y.}~\bibnamefont {Zhang}}, \bibinfo {author}
  {\bibfnamefont {M.}~\bibnamefont {Zhang}}, \bibinfo {author} {\bibfnamefont
  {C.-C.}\ \bibnamefont {Liu}}, \bibinfo {author} {\bibfnamefont
  {Y.}~\bibnamefont {Wang}}, \bibinfo {author} {\bibfnamefont {Z.-C.}\
  \bibnamefont {Gu}}, \bibinfo {author} {\bibfnamefont {W.-Q.}\ \bibnamefont
  {Chen}},\ and\ \bibinfo {author} {\bibfnamefont {F.}~\bibnamefont {Yang}},\
  }\href {https://doi.org/10.1103/PhysRevB.106.024518} {\bibfield  {journal}
  {\bibinfo  {journal} {Phys. Rev. B}\ }\textbf {\bibinfo {volume} {106}},\
  \bibinfo {pages} {024518} (\bibinfo {year} {2022})}\BibitemShut {NoStop}%
\bibitem [{\citenamefont {Da~Liao}\ \emph {et~al.}(2021)\citenamefont
  {Da~Liao}, \citenamefont {Kang}, \citenamefont {Brei\o{}}, \citenamefont
  {Xu}, \citenamefont {Wu}, \citenamefont {Andersen}, \citenamefont
  {Fernandes},\ and\ \citenamefont {Meng}}]{meng-tbg-arxiv20}%
  \BibitemOpen
  \bibfield  {author} {\bibinfo {author} {\bibfnamefont {Y.}~\bibnamefont
  {Da~Liao}}, \bibinfo {author} {\bibfnamefont {J.}~\bibnamefont {Kang}},
  \bibinfo {author} {\bibfnamefont {C.~N.}\ \bibnamefont {Brei\o{}}}, \bibinfo
  {author} {\bibfnamefont {X.~Y.}\ \bibnamefont {Xu}}, \bibinfo {author}
  {\bibfnamefont {H.-Q.}\ \bibnamefont {Wu}}, \bibinfo {author} {\bibfnamefont
  {B.~M.}\ \bibnamefont {Andersen}}, \bibinfo {author} {\bibfnamefont {R.~M.}\
  \bibnamefont {Fernandes}},\ and\ \bibinfo {author} {\bibfnamefont {Z.~Y.}\
  \bibnamefont {Meng}},\ }\href {https://doi.org/10.1103/PhysRevX.11.011014}
  {\bibfield  {journal} {\bibinfo  {journal} {Phys. Rev. X}\ }\textbf {\bibinfo
  {volume} {11}},\ \bibinfo {pages} {011014} (\bibinfo {year}
  {2021})}\BibitemShut {NoStop}%
\bibitem [{\citenamefont {Bernevig}\ \emph {et~al.}(2021)\citenamefont
  {Bernevig}, \citenamefont {Song}, \citenamefont {Regnault},\ and\
  \citenamefont {Lian}}]{Bernevig-tbg3-arxiv20}%
  \BibitemOpen
  \bibfield  {author} {\bibinfo {author} {\bibfnamefont {B.~A.}\ \bibnamefont
  {Bernevig}}, \bibinfo {author} {\bibfnamefont {Z.-D.}\ \bibnamefont {Song}},
  \bibinfo {author} {\bibfnamefont {N.}~\bibnamefont {Regnault}},\ and\
  \bibinfo {author} {\bibfnamefont {B.}~\bibnamefont {Lian}},\ }\href
  {https://doi.org/10.1103/PhysRevB.103.205413} {\bibfield  {journal} {\bibinfo
   {journal} {Phys. Rev. B}\ }\textbf {\bibinfo {volume} {103}},\ \bibinfo
  {pages} {205413} (\bibinfo {year} {2021})}\BibitemShut {NoStop}%
\bibitem [{\citenamefont {Lian}\ \emph {et~al.}(2021)\citenamefont {Lian},
  \citenamefont {Song}, \citenamefont {Regnault}, \citenamefont {Efetov},
  \citenamefont {Yazdani},\ and\ \citenamefont {Bernevig}}]{Lian-tbg4-arxiv20}%
  \BibitemOpen
  \bibfield  {author} {\bibinfo {author} {\bibfnamefont {B.}~\bibnamefont
  {Lian}}, \bibinfo {author} {\bibfnamefont {Z.-D.}\ \bibnamefont {Song}},
  \bibinfo {author} {\bibfnamefont {N.}~\bibnamefont {Regnault}}, \bibinfo
  {author} {\bibfnamefont {D.~K.}\ \bibnamefont {Efetov}}, \bibinfo {author}
  {\bibfnamefont {A.}~\bibnamefont {Yazdani}},\ and\ \bibinfo {author}
  {\bibfnamefont {B.~A.}\ \bibnamefont {Bernevig}},\ }\href
  {https://doi.org/10.1103/PhysRevB.103.205414} {\bibfield  {journal} {\bibinfo
   {journal} {Phys. Rev. B}\ }\textbf {\bibinfo {volume} {103}},\ \bibinfo
  {pages} {205414} (\bibinfo {year} {2021})}\BibitemShut {NoStop}%
\bibitem [{\citenamefont {Xie}\ \emph {et~al.}(2021)\citenamefont {Xie},
  \citenamefont {Cowsik}, \citenamefont {Song}, \citenamefont {Lian},
  \citenamefont {Bernevig},\ and\ \citenamefont {Regnault}}]{regnault-tbg-ed}%
  \BibitemOpen
  \bibfield  {author} {\bibinfo {author} {\bibfnamefont {F.}~\bibnamefont
  {Xie}}, \bibinfo {author} {\bibfnamefont {A.}~\bibnamefont {Cowsik}},
  \bibinfo {author} {\bibfnamefont {Z.-D.}\ \bibnamefont {Song}}, \bibinfo
  {author} {\bibfnamefont {B.}~\bibnamefont {Lian}}, \bibinfo {author}
  {\bibfnamefont {B.~A.}\ \bibnamefont {Bernevig}},\ and\ \bibinfo {author}
  {\bibfnamefont {N.}~\bibnamefont {Regnault}},\ }\href
  {https://doi.org/10.1103/PhysRevB.103.205416} {\bibfield  {journal} {\bibinfo
   {journal} {Phys. Rev. B}\ }\textbf {\bibinfo {volume} {103}},\ \bibinfo
  {pages} {205416} (\bibinfo {year} {2021})}\BibitemShut {NoStop}%
\bibitem [{\citenamefont {Soejima}\ \emph {et~al.}(2020)\citenamefont
  {Soejima}, \citenamefont {Parker}, \citenamefont {Bultinck}, \citenamefont
  {Hauschild},\ and\ \citenamefont {Zaletel}}]{zaletel-dmrg-prb20}%
  \BibitemOpen
  \bibfield  {author} {\bibinfo {author} {\bibfnamefont {T.}~\bibnamefont
  {Soejima}}, \bibinfo {author} {\bibfnamefont {D.~E.}\ \bibnamefont {Parker}},
  \bibinfo {author} {\bibfnamefont {N.}~\bibnamefont {Bultinck}}, \bibinfo
  {author} {\bibfnamefont {J.}~\bibnamefont {Hauschild}},\ and\ \bibinfo
  {author} {\bibfnamefont {M.~P.}\ \bibnamefont {Zaletel}},\ }\href
  {https://doi.org/10.1103/PhysRevB.102.205111} {\bibfield  {journal} {\bibinfo
   {journal} {Phys. Rev. B}\ }\textbf {\bibinfo {volume} {102}},\ \bibinfo
  {pages} {205111} (\bibinfo {year} {2020})}\BibitemShut {NoStop}%
\bibitem [{\citenamefont {Potasz}\ \emph {et~al.}(2021)\citenamefont {Potasz},
  \citenamefont {Xie},\ and\ \citenamefont
  {MacDonald}}]{macdonald-tbg-ed-arxiv21}%
  \BibitemOpen
  \bibfield  {author} {\bibinfo {author} {\bibfnamefont {P.}~\bibnamefont
  {Potasz}}, \bibinfo {author} {\bibfnamefont {M.}~\bibnamefont {Xie}},\ and\
  \bibinfo {author} {\bibfnamefont {A.~H.}\ \bibnamefont {MacDonald}},\ }\href
  {https://doi.org/10.1103/PhysRevLett.127.147203} {\bibfield  {journal}
  {\bibinfo  {journal} {Phys. Rev. Lett.}\ }\textbf {\bibinfo {volume} {127}},\
  \bibinfo {pages} {147203} (\bibinfo {year} {2021})}\BibitemShut {NoStop}%
\bibitem [{\citenamefont {Zhang}\ \emph {et~al.}(2021)\citenamefont {Zhang},
  \citenamefont {Pan}, \citenamefont {Zhang}, \citenamefont {Kang},\ and\
  \citenamefont {Meng}}]{meng-tbg-qmc-cpl21}%
  \BibitemOpen
  \bibfield  {author} {\bibinfo {author} {\bibfnamefont {X.}~\bibnamefont
  {Zhang}}, \bibinfo {author} {\bibfnamefont {G.}~\bibnamefont {Pan}}, \bibinfo
  {author} {\bibfnamefont {Y.}~\bibnamefont {Zhang}}, \bibinfo {author}
  {\bibfnamefont {J.}~\bibnamefont {Kang}},\ and\ \bibinfo {author}
  {\bibfnamefont {Z.~Y.}\ \bibnamefont {Meng}},\ }\href
  {https://doi.org/10.1088/0256-307x/38/7/077305} {\bibfield  {journal}
  {\bibinfo  {journal} {Chinese Physics Letters}\ }\textbf {\bibinfo {volume}
  {38}},\ \bibinfo {pages} {077305} (\bibinfo {year} {2021})}\BibitemShut
  {NoStop}%
\bibitem [{\citenamefont {Hofmann}\ \emph {et~al.}(2022)\citenamefont
  {Hofmann}, \citenamefont {Khalaf}, \citenamefont {Vishwanath}, \citenamefont
  {Berg},\ and\ \citenamefont {Lee}}]{lee-tbg-qmc-arxiv21}%
  \BibitemOpen
  \bibfield  {author} {\bibinfo {author} {\bibfnamefont {J.~S.}\ \bibnamefont
  {Hofmann}}, \bibinfo {author} {\bibfnamefont {E.}~\bibnamefont {Khalaf}},
  \bibinfo {author} {\bibfnamefont {A.}~\bibnamefont {Vishwanath}}, \bibinfo
  {author} {\bibfnamefont {E.}~\bibnamefont {Berg}},\ and\ \bibinfo {author}
  {\bibfnamefont {J.~Y.}\ \bibnamefont {Lee}},\ }\href
  {https://doi.org/10.1103/PhysRevX.12.011061} {\bibfield  {journal} {\bibinfo
  {journal} {Phys. Rev. X}\ }\textbf {\bibinfo {volume} {12}},\ \bibinfo
  {pages} {011061} (\bibinfo {year} {2022})}\BibitemShut {NoStop}%
\bibitem [{\citenamefont {Parker}\ \emph {et~al.}(2021)\citenamefont {Parker},
  \citenamefont {Soejima}, \citenamefont {Hauschild}, \citenamefont {Zaletel},\
  and\ \citenamefont {Bultinck}}]{bultinck-tbg-strain-prl21}%
  \BibitemOpen
  \bibfield  {author} {\bibinfo {author} {\bibfnamefont {D.~E.}\ \bibnamefont
  {Parker}}, \bibinfo {author} {\bibfnamefont {T.}~\bibnamefont {Soejima}},
  \bibinfo {author} {\bibfnamefont {J.}~\bibnamefont {Hauschild}}, \bibinfo
  {author} {\bibfnamefont {M.~P.}\ \bibnamefont {Zaletel}},\ and\ \bibinfo
  {author} {\bibfnamefont {N.}~\bibnamefont {Bultinck}},\ }\href
  {https://doi.org/10.1103/PhysRevLett.127.027601} {\bibfield  {journal}
  {\bibinfo  {journal} {Phys. Rev. Lett.}\ }\textbf {\bibinfo {volume} {127}},\
  \bibinfo {pages} {027601} (\bibinfo {year} {2021})}\BibitemShut {NoStop}%
\bibitem [{\citenamefont {Song}\ and\ \citenamefont
  {Bernevig}(2022)}]{song-heavyfermion-prl22}%
  \BibitemOpen
  \bibfield  {author} {\bibinfo {author} {\bibfnamefont {Z.-D.}\ \bibnamefont
  {Song}}\ and\ \bibinfo {author} {\bibfnamefont {B.~A.}\ \bibnamefont
  {Bernevig}},\ }\href {https://doi.org/10.1103/PhysRevLett.129.047601}
  {\bibfield  {journal} {\bibinfo  {journal} {Phys. Rev. Lett.}\ }\textbf
  {\bibinfo {volume} {129}},\ \bibinfo {pages} {047601} (\bibinfo {year}
  {2022})}\BibitemShut {NoStop}%
\bibitem [{\citenamefont {Dean}\ \emph {et~al.}(2010)\citenamefont {Dean},
  \citenamefont {Young}, \citenamefont {Meric}, \citenamefont {Lee},
  \citenamefont {Wang}, \citenamefont {Sorgenfrei}, \citenamefont {Watanabe},
  \citenamefont {Taniguchi}, \citenamefont {Kim}, \citenamefont {Shepard} \emph
  {et~al.}}]{dean-hBN_sub-natnano-2010}%
  \BibitemOpen
  \bibfield  {author} {\bibinfo {author} {\bibfnamefont {C.~R.}\ \bibnamefont
  {Dean}}, \bibinfo {author} {\bibfnamefont {A.~F.}\ \bibnamefont {Young}},
  \bibinfo {author} {\bibfnamefont {I.}~\bibnamefont {Meric}}, \bibinfo
  {author} {\bibfnamefont {C.}~\bibnamefont {Lee}}, \bibinfo {author}
  {\bibfnamefont {L.}~\bibnamefont {Wang}}, \bibinfo {author} {\bibfnamefont
  {S.}~\bibnamefont {Sorgenfrei}}, \bibinfo {author} {\bibfnamefont
  {K.}~\bibnamefont {Watanabe}}, \bibinfo {author} {\bibfnamefont
  {T.}~\bibnamefont {Taniguchi}}, \bibinfo {author} {\bibfnamefont
  {P.}~\bibnamefont {Kim}}, \bibinfo {author} {\bibfnamefont {K.~L.}\
  \bibnamefont {Shepard}}, \emph {et~al.},\ }\href@noop {} {\bibfield
  {journal} {\bibinfo  {journal} {Nature nanotechnology}\ }\textbf {\bibinfo
  {volume} {5}},\ \bibinfo {pages} {722} (\bibinfo {year} {2010})}\BibitemShut
  {NoStop}%
\bibitem [{\citenamefont {Liu}\ \emph {et~al.}(2022)\citenamefont {Liu},
  \citenamefont {Li}, \citenamefont {Qiao}, \citenamefont {Wang}, \citenamefont
  {Zhang}, \citenamefont {Liu}, \citenamefont {Zhou}, \citenamefont {Shang},
  \citenamefont {Fang}, \citenamefont {Wang} \emph
  {et~al.}}]{liu-synthesis_tbg-natmat-2022}%
  \BibitemOpen
  \bibfield  {author} {\bibinfo {author} {\bibfnamefont {C.}~\bibnamefont
  {Liu}}, \bibinfo {author} {\bibfnamefont {Z.}~\bibnamefont {Li}}, \bibinfo
  {author} {\bibfnamefont {R.}~\bibnamefont {Qiao}}, \bibinfo {author}
  {\bibfnamefont {Q.}~\bibnamefont {Wang}}, \bibinfo {author} {\bibfnamefont
  {Z.}~\bibnamefont {Zhang}}, \bibinfo {author} {\bibfnamefont
  {F.}~\bibnamefont {Liu}}, \bibinfo {author} {\bibfnamefont {Z.}~\bibnamefont
  {Zhou}}, \bibinfo {author} {\bibfnamefont {N.}~\bibnamefont {Shang}},
  \bibinfo {author} {\bibfnamefont {H.}~\bibnamefont {Fang}}, \bibinfo {author}
  {\bibfnamefont {M.}~\bibnamefont {Wang}}, \emph {et~al.},\ }\href@noop {}
  {\bibfield  {journal} {\bibinfo  {journal} {Nature Materials}\ }\textbf
  {\bibinfo {volume} {21}},\ \bibinfo {pages} {1263} (\bibinfo {year}
  {2022})}\BibitemShut {NoStop}%
\bibitem [{\citenamefont {Liu}\ \emph {et~al.}(2020)\citenamefont {Liu},
  \citenamefont {Hao}, \citenamefont {Khalaf}, \citenamefont {Lee},
  \citenamefont {Ronen}, \citenamefont {Yoo}, \citenamefont {Haei~Najafabadi},
  \citenamefont {Watanabe}, \citenamefont {Taniguchi}, \citenamefont
  {Vishwanath},\ and\ \citenamefont {Kim}}]{kim-tdbg-nature20}%
  \BibitemOpen
  \bibfield  {author} {\bibinfo {author} {\bibfnamefont {X.}~\bibnamefont
  {Liu}}, \bibinfo {author} {\bibfnamefont {Z.}~\bibnamefont {Hao}}, \bibinfo
  {author} {\bibfnamefont {E.}~\bibnamefont {Khalaf}}, \bibinfo {author}
  {\bibfnamefont {J.~Y.}\ \bibnamefont {Lee}}, \bibinfo {author} {\bibfnamefont
  {Y.}~\bibnamefont {Ronen}}, \bibinfo {author} {\bibfnamefont
  {H.}~\bibnamefont {Yoo}}, \bibinfo {author} {\bibfnamefont {D.}~\bibnamefont
  {Haei~Najafabadi}}, \bibinfo {author} {\bibfnamefont {K.}~\bibnamefont
  {Watanabe}}, \bibinfo {author} {\bibfnamefont {T.}~\bibnamefont {Taniguchi}},
  \bibinfo {author} {\bibfnamefont {A.}~\bibnamefont {Vishwanath}},\ and\
  \bibinfo {author} {\bibfnamefont {P.}~\bibnamefont {Kim}},\ }\href
  {https://doi.org/10.1038/s41586-020-2458-7} {\bibfield  {journal} {\bibinfo
  {journal} {Nature}\ }\textbf {\bibinfo {volume} {583}},\ \bibinfo {pages}
  {221} (\bibinfo {year} {2020})}\BibitemShut {NoStop}%
\bibitem [{\citenamefont {Cao}\ \emph {et~al.}(2020)\citenamefont {Cao},
  \citenamefont {Rodan-Legrain}, \citenamefont {Rubies-Bigorda}, \citenamefont
  {Park}, \citenamefont {Watanabe}, \citenamefont {Taniguchi},\ and\
  \citenamefont {Jarillo-Herrero}}]{cao-tdbg-nature20}%
  \BibitemOpen
  \bibfield  {author} {\bibinfo {author} {\bibfnamefont {Y.}~\bibnamefont
  {Cao}}, \bibinfo {author} {\bibfnamefont {D.}~\bibnamefont {Rodan-Legrain}},
  \bibinfo {author} {\bibfnamefont {O.}~\bibnamefont {Rubies-Bigorda}},
  \bibinfo {author} {\bibfnamefont {J.~M.}\ \bibnamefont {Park}}, \bibinfo
  {author} {\bibfnamefont {K.}~\bibnamefont {Watanabe}}, \bibinfo {author}
  {\bibfnamefont {T.}~\bibnamefont {Taniguchi}},\ and\ \bibinfo {author}
  {\bibfnamefont {P.}~\bibnamefont {Jarillo-Herrero}},\ }\bibfield  {journal}
  {\bibinfo  {journal} {Nature}\ }\href
  {https://doi.org/10.1038/s41586-020-2260-6} {10.1038/s41586-020-2260-6}
  (\bibinfo {year} {2020})\BibitemShut {NoStop}%
\bibitem [{\citenamefont {Shen}\ \emph {et~al.}(2020)\citenamefont {Shen},
  \citenamefont {Chu}, \citenamefont {Wu}, \citenamefont {Li}, \citenamefont
  {Wang}, \citenamefont {Zhao}, \citenamefont {Tang}, \citenamefont {Liu},
  \citenamefont {Tian}, \citenamefont {Watanabe}, \citenamefont {Taniguchi},
  \citenamefont {Yang}, \citenamefont {Meng}, \citenamefont {Shi},
  \citenamefont {Yazyev},\ and\ \citenamefont {Zhang}}]{zhang-tdbg-np20}%
  \BibitemOpen
  \bibfield  {author} {\bibinfo {author} {\bibfnamefont {C.}~\bibnamefont
  {Shen}}, \bibinfo {author} {\bibfnamefont {Y.}~\bibnamefont {Chu}}, \bibinfo
  {author} {\bibfnamefont {Q.}~\bibnamefont {Wu}}, \bibinfo {author}
  {\bibfnamefont {N.}~\bibnamefont {Li}}, \bibinfo {author} {\bibfnamefont
  {S.}~\bibnamefont {Wang}}, \bibinfo {author} {\bibfnamefont {Y.}~\bibnamefont
  {Zhao}}, \bibinfo {author} {\bibfnamefont {J.}~\bibnamefont {Tang}}, \bibinfo
  {author} {\bibfnamefont {J.}~\bibnamefont {Liu}}, \bibinfo {author}
  {\bibfnamefont {J.}~\bibnamefont {Tian}}, \bibinfo {author} {\bibfnamefont
  {K.}~\bibnamefont {Watanabe}}, \bibinfo {author} {\bibfnamefont
  {T.}~\bibnamefont {Taniguchi}}, \bibinfo {author} {\bibfnamefont
  {R.}~\bibnamefont {Yang}}, \bibinfo {author} {\bibfnamefont {Z.~Y.}\
  \bibnamefont {Meng}}, \bibinfo {author} {\bibfnamefont {D.}~\bibnamefont
  {Shi}}, \bibinfo {author} {\bibfnamefont {O.~V.}\ \bibnamefont {Yazyev}},\
  and\ \bibinfo {author} {\bibfnamefont {G.}~\bibnamefont {Zhang}},\ }\href
  {https://doi.org/10.1038/s41567-020-0825-9} {\bibfield  {journal} {\bibinfo
  {journal} {Nature Physics}\ }\textbf {\bibinfo {volume} {16}},\ \bibinfo
  {pages} {520} (\bibinfo {year} {2020})}\BibitemShut {NoStop}%
\bibitem [{\citenamefont {Liu}\ \emph {et~al.}(2019{\natexlab{b}})\citenamefont
  {Liu}, \citenamefont {Ma}, \citenamefont {Gao},\ and\ \citenamefont
  {Dai}}]{jpliu-prx19}%
  \BibitemOpen
  \bibfield  {author} {\bibinfo {author} {\bibfnamefont {J.}~\bibnamefont
  {Liu}}, \bibinfo {author} {\bibfnamefont {Z.}~\bibnamefont {Ma}}, \bibinfo
  {author} {\bibfnamefont {J.}~\bibnamefont {Gao}},\ and\ \bibinfo {author}
  {\bibfnamefont {X.}~\bibnamefont {Dai}},\ }\href
  {https://doi.org/10.1103/PhysRevX.9.031021} {\bibfield  {journal} {\bibinfo
  {journal} {Phys. Rev. X}\ }\textbf {\bibinfo {volume} {9}},\ \bibinfo {pages}
  {031021} (\bibinfo {year} {2019}{\natexlab{b}})}\BibitemShut {NoStop}%
\bibitem [{\citenamefont {Koshino}(2019)}]{koshino-tdbg-prb19}%
  \BibitemOpen
  \bibfield  {author} {\bibinfo {author} {\bibfnamefont {M.}~\bibnamefont
  {Koshino}},\ }\href {https://doi.org/10.1103/PhysRevB.99.235406} {\bibfield
  {journal} {\bibinfo  {journal} {Phys. Rev. B}\ }\textbf {\bibinfo {volume}
  {99}},\ \bibinfo {pages} {235406} (\bibinfo {year} {2019})}\BibitemShut
  {NoStop}%
\bibitem [{\citenamefont {Lee}\ \emph {et~al.}(2019)\citenamefont {Lee},
  \citenamefont {Khalaf}, \citenamefont {Liu}, \citenamefont {Liu},
  \citenamefont {Hao}, \citenamefont {Kim},\ and\ \citenamefont
  {Vishwanath}}]{ashvin-double-bilayer-nc19}%
  \BibitemOpen
  \bibfield  {author} {\bibinfo {author} {\bibfnamefont {J.~Y.}\ \bibnamefont
  {Lee}}, \bibinfo {author} {\bibfnamefont {E.}~\bibnamefont {Khalaf}},
  \bibinfo {author} {\bibfnamefont {S.}~\bibnamefont {Liu}}, \bibinfo {author}
  {\bibfnamefont {X.}~\bibnamefont {Liu}}, \bibinfo {author} {\bibfnamefont
  {Z.}~\bibnamefont {Hao}}, \bibinfo {author} {\bibfnamefont {P.}~\bibnamefont
  {Kim}},\ and\ \bibinfo {author} {\bibfnamefont {A.}~\bibnamefont
  {Vishwanath}},\ }\href {https://doi.org/10.1038/s41467-019-12981-1}
  {\bibfield  {journal} {\bibinfo  {journal} {Nature Communications}\ }\textbf
  {\bibinfo {volume} {10}},\ \bibinfo {pages} {5333} (\bibinfo {year}
  {2019})}\BibitemShut {NoStop}%
\bibitem [{\citenamefont {Ledwith}\ \emph {et~al.}(2022)\citenamefont
  {Ledwith}, \citenamefont {Vishwanath},\ and\ \citenamefont
  {Khalaf}}]{ledwith-prl22}%
  \BibitemOpen
  \bibfield  {author} {\bibinfo {author} {\bibfnamefont {P.~J.}\ \bibnamefont
  {Ledwith}}, \bibinfo {author} {\bibfnamefont {A.}~\bibnamefont
  {Vishwanath}},\ and\ \bibinfo {author} {\bibfnamefont {E.}~\bibnamefont
  {Khalaf}},\ }\href {https://doi.org/10.1103/PhysRevLett.128.176404}
  {\bibfield  {journal} {\bibinfo  {journal} {Phys. Rev. Lett.}\ }\textbf
  {\bibinfo {volume} {128}},\ \bibinfo {pages} {176404} (\bibinfo {year}
  {2022})}\BibitemShut {NoStop}%
\bibitem [{\citenamefont {Wang}\ and\ \citenamefont {Liu}(2022)}]{wang-prl22}%
  \BibitemOpen
  \bibfield  {author} {\bibinfo {author} {\bibfnamefont {J.}~\bibnamefont
  {Wang}}\ and\ \bibinfo {author} {\bibfnamefont {Z.}~\bibnamefont {Liu}},\
  }\href {https://doi.org/10.1103/PhysRevLett.128.176403} {\bibfield  {journal}
  {\bibinfo  {journal} {Phys. Rev. Lett.}\ }\textbf {\bibinfo {volume} {128}},\
  \bibinfo {pages} {176403} (\bibinfo {year} {2022})}\BibitemShut {NoStop}%
\bibitem [{\citenamefont {Zhang}\ \emph {et~al.}(2023)\citenamefont {Zhang},
  \citenamefont {Xie}, \citenamefont {Wu}, \citenamefont {Liu},\ and\
  \citenamefont {Yazyev}}]{zhang-tmg-nanolett23}%
  \BibitemOpen
  \bibfield  {author} {\bibinfo {author} {\bibfnamefont {S.}~\bibnamefont
  {Zhang}}, \bibinfo {author} {\bibfnamefont {B.}~\bibnamefont {Xie}}, \bibinfo
  {author} {\bibfnamefont {Q.}~\bibnamefont {Wu}}, \bibinfo {author}
  {\bibfnamefont {J.}~\bibnamefont {Liu}},\ and\ \bibinfo {author}
  {\bibfnamefont {O.~V.}\ \bibnamefont {Yazyev}},\ }\href@noop {} {\bibfield
  {journal} {\bibinfo  {journal} {Nano Letters}\ }\textbf {\bibinfo {volume}
  {23}},\ \bibinfo {pages} {2921} (\bibinfo {year} {2023})}\BibitemShut
  {NoStop}%
\bibitem [{\citenamefont {Khalaf}\ \emph {et~al.}(2019)\citenamefont {Khalaf},
  \citenamefont {Kruchkov}, \citenamefont {Tarnopolsky},\ and\ \citenamefont
  {Vishwanath}}]{eslam-tmg-prb19}%
  \BibitemOpen
  \bibfield  {author} {\bibinfo {author} {\bibfnamefont {E.}~\bibnamefont
  {Khalaf}}, \bibinfo {author} {\bibfnamefont {A.~J.}\ \bibnamefont
  {Kruchkov}}, \bibinfo {author} {\bibfnamefont {G.}~\bibnamefont
  {Tarnopolsky}},\ and\ \bibinfo {author} {\bibfnamefont {A.}~\bibnamefont
  {Vishwanath}},\ }\href {https://doi.org/10.1103/PhysRevB.100.085109}
  {\bibfield  {journal} {\bibinfo  {journal} {Phys. Rev. B}\ }\textbf {\bibinfo
  {volume} {100}},\ \bibinfo {pages} {085109} (\bibinfo {year}
  {2019})}\BibitemShut {NoStop}%
\bibitem [{\citenamefont {Xie}\ \emph {et~al.}(2022)\citenamefont {Xie},
  \citenamefont {Peng}, \citenamefont {Zhang},\ and\ \citenamefont
  {Liu}}]{xie-atmg-npj22}%
  \BibitemOpen
  \bibfield  {author} {\bibinfo {author} {\bibfnamefont {B.}~\bibnamefont
  {Xie}}, \bibinfo {author} {\bibfnamefont {R.}~\bibnamefont {Peng}}, \bibinfo
  {author} {\bibfnamefont {S.}~\bibnamefont {Zhang}},\ and\ \bibinfo {author}
  {\bibfnamefont {J.}~\bibnamefont {Liu}},\ }\href
  {https://doi.org/10.1038/s41524-022-00789-5} {\bibfield  {journal} {\bibinfo
  {journal} {npj Computational Materials}\ }\textbf {\bibinfo {volume} {8}},\
  \bibinfo {pages} {110} (\bibinfo {year} {2022})}\BibitemShut {NoStop}%
\bibitem [{\citenamefont {Ledwith}\ \emph {et~al.}(2021)\citenamefont
  {Ledwith}, \citenamefont {Khalaf}, \citenamefont {Zhu}, \citenamefont {Carr},
  \citenamefont {Kaxiras},\ and\ \citenamefont
  {Vishwanath}}]{ashvin-atmg-arxiv21}%
  \BibitemOpen
  \bibfield  {author} {\bibinfo {author} {\bibfnamefont {P.~J.}\ \bibnamefont
  {Ledwith}}, \bibinfo {author} {\bibfnamefont {E.}~\bibnamefont {Khalaf}},
  \bibinfo {author} {\bibfnamefont {Z.}~\bibnamefont {Zhu}}, \bibinfo {author}
  {\bibfnamefont {S.}~\bibnamefont {Carr}}, \bibinfo {author} {\bibfnamefont
  {E.}~\bibnamefont {Kaxiras}},\ and\ \bibinfo {author} {\bibfnamefont
  {A.}~\bibnamefont {Vishwanath}},\ }\href
  {https://doi.org/10.48550/ARXIV.2111.11060} {\bibinfo {title} {Tb or not tb?
  contrasting properties of twisted bilayer graphene and the alternating twist
  $n$-layer structures ($n=3, 4, 5, \dots$)}} (\bibinfo {year}
  {2021})\BibitemShut {NoStop}%
\bibitem [{\citenamefont {Leconte}\ \emph {et~al.}(2022)\citenamefont
  {Leconte}, \citenamefont {Park}, \citenamefont {An}, \citenamefont
  {Samudrala},\ and\ \citenamefont {Jung}}]{leconte-amtg-2dmat-2022}%
  \BibitemOpen
  \bibfield  {author} {\bibinfo {author} {\bibfnamefont {N.}~\bibnamefont
  {Leconte}}, \bibinfo {author} {\bibfnamefont {Y.}~\bibnamefont {Park}},
  \bibinfo {author} {\bibfnamefont {J.}~\bibnamefont {An}}, \bibinfo {author}
  {\bibfnamefont {A.}~\bibnamefont {Samudrala}},\ and\ \bibinfo {author}
  {\bibfnamefont {J.}~\bibnamefont {Jung}},\ }\href@noop {} {\bibfield
  {journal} {\bibinfo  {journal} {2D Materials}\ }\textbf {\bibinfo {volume}
  {9}},\ \bibinfo {pages} {044002} (\bibinfo {year} {2022})}\BibitemShut
  {NoStop}%
\bibitem [{\citenamefont {Cea}\ \emph {et~al.}(2019)\citenamefont {Cea},
  \citenamefont {Walet},\ and\ \citenamefont {Guinea}}]{cea-arxiv19}%
  \BibitemOpen
  \bibfield  {author} {\bibinfo {author} {\bibfnamefont {T.}~\bibnamefont
  {Cea}}, \bibinfo {author} {\bibfnamefont {N.~R.}\ \bibnamefont {Walet}},\
  and\ \bibinfo {author} {\bibfnamefont {F.}~\bibnamefont {Guinea}},\
  }\href@noop {} {\bibfield  {journal} {\bibinfo  {journal} {Nano letters}\
  }\textbf {\bibinfo {volume} {19}},\ \bibinfo {pages} {8683} (\bibinfo {year}
  {2019})}\BibitemShut {NoStop}%
\bibitem [{\citenamefont {Cao}\ \emph {et~al.}(2021{\natexlab{b}})\citenamefont
  {Cao}, \citenamefont {Park}, \citenamefont {Watanabe}, \citenamefont
  {Taniguchi},\ and\ \citenamefont {Jarillo-Herrero}}]{cao2021pauli}%
  \BibitemOpen
  \bibfield  {author} {\bibinfo {author} {\bibfnamefont {Y.}~\bibnamefont
  {Cao}}, \bibinfo {author} {\bibfnamefont {J.~M.}\ \bibnamefont {Park}},
  \bibinfo {author} {\bibfnamefont {K.}~\bibnamefont {Watanabe}}, \bibinfo
  {author} {\bibfnamefont {T.}~\bibnamefont {Taniguchi}},\ and\ \bibinfo
  {author} {\bibfnamefont {P.}~\bibnamefont {Jarillo-Herrero}},\ }\href@noop {}
  {\bibfield  {journal} {\bibinfo  {journal} {Nature}\ }\textbf {\bibinfo
  {volume} {595}},\ \bibinfo {pages} {526} (\bibinfo {year}
  {2021}{\natexlab{b}})}\BibitemShut {NoStop}%
\bibitem [{\citenamefont {Polshyn}\ \emph {et~al.}(2020)\citenamefont
  {Polshyn}, \citenamefont {Zhu}, \citenamefont {Kumar}, \citenamefont {Zhang},
  \citenamefont {Yang}, \citenamefont {Tschirhart}, \citenamefont {Serlin},
  \citenamefont {Watanabe}, \citenamefont {Taniguchi}, \citenamefont
  {MacDonald} \emph {et~al.}}]{young-monobi-nature20}%
  \BibitemOpen
  \bibfield  {author} {\bibinfo {author} {\bibfnamefont {H.}~\bibnamefont
  {Polshyn}}, \bibinfo {author} {\bibfnamefont {J.}~\bibnamefont {Zhu}},
  \bibinfo {author} {\bibfnamefont {M.}~\bibnamefont {Kumar}}, \bibinfo
  {author} {\bibfnamefont {Y.}~\bibnamefont {Zhang}}, \bibinfo {author}
  {\bibfnamefont {F.}~\bibnamefont {Yang}}, \bibinfo {author} {\bibfnamefont
  {C.}~\bibnamefont {Tschirhart}}, \bibinfo {author} {\bibfnamefont
  {M.}~\bibnamefont {Serlin}}, \bibinfo {author} {\bibfnamefont
  {K.}~\bibnamefont {Watanabe}}, \bibinfo {author} {\bibfnamefont
  {T.}~\bibnamefont {Taniguchi}}, \bibinfo {author} {\bibfnamefont
  {A.}~\bibnamefont {MacDonald}}, \emph {et~al.},\ }\href@noop {} {\bibfield
  {journal} {\bibinfo  {journal} {Nature}\ ,\ \bibinfo {pages} {1}} (\bibinfo
  {year} {2020})}\BibitemShut {NoStop}%
\bibitem [{\citenamefont {Li}\ \emph {et~al.}(2022)\citenamefont {Li},
  \citenamefont {Cheng}, \citenamefont {Chen}, \citenamefont {Xie},
  \citenamefont {Xie}, \citenamefont {Wang}, \citenamefont {Chen},
  \citenamefont {Liu}, \citenamefont {Watanabe}, \citenamefont {Taniguchi},
  \citenamefont {Liang}, \citenamefont {Wang}, \citenamefont {Wang},
  \citenamefont {Wang}, \citenamefont {Liu},\ and\ \citenamefont
  {Miao}}]{miao-tdbg-nature22}%
  \BibitemOpen
  \bibfield  {author} {\bibinfo {author} {\bibfnamefont {Q.}~\bibnamefont
  {Li}}, \bibinfo {author} {\bibfnamefont {B.}~\bibnamefont {Cheng}}, \bibinfo
  {author} {\bibfnamefont {M.}~\bibnamefont {Chen}}, \bibinfo {author}
  {\bibfnamefont {B.}~\bibnamefont {Xie}}, \bibinfo {author} {\bibfnamefont
  {Y.}~\bibnamefont {Xie}}, \bibinfo {author} {\bibfnamefont {P.}~\bibnamefont
  {Wang}}, \bibinfo {author} {\bibfnamefont {F.}~\bibnamefont {Chen}}, \bibinfo
  {author} {\bibfnamefont {Z.}~\bibnamefont {Liu}}, \bibinfo {author}
  {\bibfnamefont {K.}~\bibnamefont {Watanabe}}, \bibinfo {author}
  {\bibfnamefont {T.}~\bibnamefont {Taniguchi}}, \bibinfo {author}
  {\bibfnamefont {S.-J.}\ \bibnamefont {Liang}}, \bibinfo {author}
  {\bibfnamefont {D.}~\bibnamefont {Wang}}, \bibinfo {author} {\bibfnamefont
  {C.}~\bibnamefont {Wang}}, \bibinfo {author} {\bibfnamefont {Q.-H.}\
  \bibnamefont {Wang}}, \bibinfo {author} {\bibfnamefont {J.}~\bibnamefont
  {Liu}},\ and\ \bibinfo {author} {\bibfnamefont {F.}~\bibnamefont {Miao}},\
  }\href {https://doi.org/10.1038/s41586-022-05106-0} {\bibfield  {journal}
  {\bibinfo  {journal} {Nature}\ }\textbf {\bibinfo {volume} {609}},\ \bibinfo
  {pages} {479} (\bibinfo {year} {2022})}\BibitemShut {NoStop}%
\bibitem [{\citenamefont {Wang}\ \emph {et~al.}(2023)\citenamefont {Wang},
  \citenamefont {Kong}, \citenamefont {Huang}, \citenamefont {Li},
  \citenamefont {Bao}, \citenamefont {Cao}, \citenamefont {Hu}, \citenamefont
  {Cai}, \citenamefont {Wang}, \citenamefont {Chen}, \citenamefont {Wu},
  \citenamefont {Zhang}, \citenamefont {Pang}, \citenamefont {Cheng},
  \citenamefont {Babor}, \citenamefont {Kolibal}, \citenamefont {Liu},
  \citenamefont {Chen}, \citenamefont {Zhang}, \citenamefont {Cui},
  \citenamefont {Liu}, \citenamefont {Yang}, \citenamefont {Bao}, \citenamefont
  {Gao}, \citenamefont {Liu}, \citenamefont {Ji}, \citenamefont {Ding},\ and\
  \citenamefont {Willinger}}]{wang-nm2023}%
  \BibitemOpen
  \bibfield  {author} {\bibinfo {author} {\bibfnamefont {Z.-J.}\ \bibnamefont
  {Wang}}, \bibinfo {author} {\bibfnamefont {X.}~\bibnamefont {Kong}}, \bibinfo
  {author} {\bibfnamefont {Y.}~\bibnamefont {Huang}}, \bibinfo {author}
  {\bibfnamefont {J.}~\bibnamefont {Li}}, \bibinfo {author} {\bibfnamefont
  {L.}~\bibnamefont {Bao}}, \bibinfo {author} {\bibfnamefont {K.}~\bibnamefont
  {Cao}}, \bibinfo {author} {\bibfnamefont {Y.}~\bibnamefont {Hu}}, \bibinfo
  {author} {\bibfnamefont {J.}~\bibnamefont {Cai}}, \bibinfo {author}
  {\bibfnamefont {L.}~\bibnamefont {Wang}}, \bibinfo {author} {\bibfnamefont
  {H.}~\bibnamefont {Chen}}, \bibinfo {author} {\bibfnamefont {Y.}~\bibnamefont
  {Wu}}, \bibinfo {author} {\bibfnamefont {Y.}~\bibnamefont {Zhang}}, \bibinfo
  {author} {\bibfnamefont {F.}~\bibnamefont {Pang}}, \bibinfo {author}
  {\bibfnamefont {Z.}~\bibnamefont {Cheng}}, \bibinfo {author} {\bibfnamefont
  {P.}~\bibnamefont {Babor}}, \bibinfo {author} {\bibfnamefont
  {M.}~\bibnamefont {Kolibal}}, \bibinfo {author} {\bibfnamefont
  {Z.}~\bibnamefont {Liu}}, \bibinfo {author} {\bibfnamefont {Y.}~\bibnamefont
  {Chen}}, \bibinfo {author} {\bibfnamefont {Q.}~\bibnamefont {Zhang}},
  \bibinfo {author} {\bibfnamefont {Y.}~\bibnamefont {Cui}}, \bibinfo {author}
  {\bibfnamefont {K.}~\bibnamefont {Liu}}, \bibinfo {author} {\bibfnamefont
  {H.}~\bibnamefont {Yang}}, \bibinfo {author} {\bibfnamefont {X.}~\bibnamefont
  {Bao}}, \bibinfo {author} {\bibfnamefont {H.-J.}\ \bibnamefont {Gao}},
  \bibinfo {author} {\bibfnamefont {Z.}~\bibnamefont {Liu}}, \bibinfo {author}
  {\bibfnamefont {W.}~\bibnamefont {Ji}}, \bibinfo {author} {\bibfnamefont
  {F.}~\bibnamefont {Ding}},\ and\ \bibinfo {author} {\bibfnamefont {M.-G.}\
  \bibnamefont {Willinger}},\ }\bibfield  {journal} {\bibinfo  {journal}
  {Nature Materials}\ }\href {https://doi.org/10.1038/s41563-023-01632-y}
  {10.1038/s41563-023-01632-y} (\bibinfo {year} {2023})\BibitemShut {NoStop}%
\bibitem [{sup()}]{supp_info}%
  \BibitemOpen
  \href@noop {} {}\bibinfo {note} {See Supplemental Information for: (a)
  details of the continuum model for bulk alternating twisted graphite, (b)
  detailed band structures and topological transitions in bulk alternating
  twisted graphite, (c) lattice relaxations and atomistic tight-binding model
  for bulk alternating twisted graphite, (d) detailed formalism of calculating
  Hofstadter butterfly spectra based on the continuum model, (e) more results
  about the Hofstadter butterfly spectra, (f) bound states associated with the
  spiral dislocation line.}\BibitemShut {Stop}%
\bibitem [{\citenamefont {Lopes~dos Santos}\ \emph {et~al.}(2012)\citenamefont
  {Lopes~dos Santos}, \citenamefont {Peres},\ and\ \citenamefont
  {Castro~Neto}}]{castro-neto-prb12}%
  \BibitemOpen
  \bibfield  {author} {\bibinfo {author} {\bibfnamefont {J.~M.~B.}\
  \bibnamefont {Lopes~dos Santos}}, \bibinfo {author} {\bibfnamefont
  {N.~M.~R.}\ \bibnamefont {Peres}},\ and\ \bibinfo {author} {\bibfnamefont
  {A.~H.}\ \bibnamefont {Castro~Neto}},\ }\href
  {https://doi.org/10.1103/PhysRevB.86.155449} {\bibfield  {journal} {\bibinfo
  {journal} {Phys. Rev. B}\ }\textbf {\bibinfo {volume} {86}},\ \bibinfo
  {pages} {155449} (\bibinfo {year} {2012})}\BibitemShut {NoStop}%
\bibitem [{\citenamefont {Moon}\ and\ \citenamefont
  {Koshino}(2013)}]{moon-tbg-prb13}%
  \BibitemOpen
  \bibfield  {author} {\bibinfo {author} {\bibfnamefont {P.}~\bibnamefont
  {Moon}}\ and\ \bibinfo {author} {\bibfnamefont {M.}~\bibnamefont {Koshino}},\
  }\href@noop {} {\bibfield  {journal} {\bibinfo  {journal} {Physical Review
  B}\ }\textbf {\bibinfo {volume} {87}},\ \bibinfo {pages} {205404} (\bibinfo
  {year} {2013})}\BibitemShut {NoStop}%
\bibitem [{\citenamefont {Hejazi}\ \emph
  {et~al.}(2019{\natexlab{a}})\citenamefont {Hejazi}, \citenamefont {Liu},
  \citenamefont {Shapourian}, \citenamefont {Chen},\ and\ \citenamefont
  {Balents}}]{hejazi-prb19}%
  \BibitemOpen
  \bibfield  {author} {\bibinfo {author} {\bibfnamefont {K.}~\bibnamefont
  {Hejazi}}, \bibinfo {author} {\bibfnamefont {C.}~\bibnamefont {Liu}},
  \bibinfo {author} {\bibfnamefont {H.}~\bibnamefont {Shapourian}}, \bibinfo
  {author} {\bibfnamefont {X.}~\bibnamefont {Chen}},\ and\ \bibinfo {author}
  {\bibfnamefont {L.}~\bibnamefont {Balents}},\ }\href
  {https://doi.org/10.1103/PhysRevB.99.035111} {\bibfield  {journal} {\bibinfo
  {journal} {Phys. Rev. B}\ }\textbf {\bibinfo {volume} {99}},\ \bibinfo
  {pages} {035111} (\bibinfo {year} {2019}{\natexlab{a}})}\BibitemShut
  {NoStop}%
\bibitem [{\citenamefont {Bernevig}\ \emph {et~al.}(2007)\citenamefont
  {Bernevig}, \citenamefont {Hughes}, \citenamefont {Raghu},\ and\
  \citenamefont {Arovas}}]{bernevig-3dqhe_gr-prl-2007}%
  \BibitemOpen
  \bibfield  {author} {\bibinfo {author} {\bibfnamefont {B.~A.}\ \bibnamefont
  {Bernevig}}, \bibinfo {author} {\bibfnamefont {T.~L.}\ \bibnamefont
  {Hughes}}, \bibinfo {author} {\bibfnamefont {S.}~\bibnamefont {Raghu}},\ and\
  \bibinfo {author} {\bibfnamefont {D.~P.}\ \bibnamefont {Arovas}},\ }\href
  {https://doi.org/10.1103/PhysRevLett.99.146804} {\bibfield  {journal}
  {\bibinfo  {journal} {Phys. Rev. Lett.}\ }\textbf {\bibinfo {volume} {99}},\
  \bibinfo {pages} {146804} (\bibinfo {year} {2007})}\BibitemShut {NoStop}%
\bibitem [{\citenamefont {Tang}\ \emph {et~al.}(2019)\citenamefont {Tang},
  \citenamefont {Ren}, \citenamefont {Wang}, \citenamefont {Zhong},
  \citenamefont {Schneeloch}, \citenamefont {Yang}, \citenamefont {Yang},
  \citenamefont {Lee}, \citenamefont {Gu}, \citenamefont {Qiao} \emph
  {et~al.}}]{tang-3dqhe-nature-2019}%
  \BibitemOpen
  \bibfield  {author} {\bibinfo {author} {\bibfnamefont {F.}~\bibnamefont
  {Tang}}, \bibinfo {author} {\bibfnamefont {Y.}~\bibnamefont {Ren}}, \bibinfo
  {author} {\bibfnamefont {P.}~\bibnamefont {Wang}}, \bibinfo {author}
  {\bibfnamefont {R.}~\bibnamefont {Zhong}}, \bibinfo {author} {\bibfnamefont
  {J.}~\bibnamefont {Schneeloch}}, \bibinfo {author} {\bibfnamefont {S.~A.}\
  \bibnamefont {Yang}}, \bibinfo {author} {\bibfnamefont {K.}~\bibnamefont
  {Yang}}, \bibinfo {author} {\bibfnamefont {P.~A.}\ \bibnamefont {Lee}},
  \bibinfo {author} {\bibfnamefont {G.}~\bibnamefont {Gu}}, \bibinfo {author}
  {\bibfnamefont {Z.}~\bibnamefont {Qiao}}, \emph {et~al.},\ }\href@noop {}
  {\bibfield  {journal} {\bibinfo  {journal} {Nature}\ }\textbf {\bibinfo
  {volume} {569}},\ \bibinfo {pages} {537} (\bibinfo {year}
  {2019})}\BibitemShut {NoStop}%
\bibitem [{\citenamefont {Novoselov}\ \emph {et~al.}(2005)\citenamefont
  {Novoselov}, \citenamefont {Geim}, \citenamefont {Morozov}, \citenamefont
  {Jiang}, \citenamefont {Katsnelson}, \citenamefont {Grigorieva},
  \citenamefont {Dubonos},\ and\ \citenamefont {Firsov}}]{graphene1}%
  \BibitemOpen
  \bibfield  {author} {\bibinfo {author} {\bibfnamefont {K.}~\bibnamefont
  {Novoselov}}, \bibinfo {author} {\bibfnamefont {A.~K.}\ \bibnamefont {Geim}},
  \bibinfo {author} {\bibfnamefont {S.}~\bibnamefont {Morozov}}, \bibinfo
  {author} {\bibfnamefont {D.}~\bibnamefont {Jiang}}, \bibinfo {author}
  {\bibfnamefont {M.}~\bibnamefont {Katsnelson}}, \bibinfo {author}
  {\bibfnamefont {I.}~\bibnamefont {Grigorieva}}, \bibinfo {author}
  {\bibfnamefont {S.}~\bibnamefont {Dubonos}},\ and\ \bibinfo {author}
  {\bibfnamefont {A.}~\bibnamefont {Firsov}},\ }\href@noop {} {\bibfield
  {journal} {\bibinfo  {journal} {nature}\ }\textbf {\bibinfo {volume} {438}},\
  \bibinfo {pages} {197} (\bibinfo {year} {2005})}\BibitemShut {NoStop}%
\bibitem [{\citenamefont {Bistritzer}\ and\ \citenamefont
  {MacDonald}(2011{\natexlab{b}})}]{moire-butterfly-macdonald-prb11}%
  \BibitemOpen
  \bibfield  {author} {\bibinfo {author} {\bibfnamefont {R.}~\bibnamefont
  {Bistritzer}}\ and\ \bibinfo {author} {\bibfnamefont {A.~H.}\ \bibnamefont
  {MacDonald}},\ }\href {https://doi.org/10.1103/PhysRevB.84.035440} {\bibfield
   {journal} {\bibinfo  {journal} {Phys. Rev. B}\ }\textbf {\bibinfo {volume}
  {84}},\ \bibinfo {pages} {035440} (\bibinfo {year}
  {2011}{\natexlab{b}})}\BibitemShut {NoStop}%
\bibitem [{\citenamefont {de~Gail}\ \emph {et~al.}(2011)\citenamefont
  {de~Gail}, \citenamefont {Goerbig}, \citenamefont {Guinea}, \citenamefont
  {Montambaux},\ and\ \citenamefont {Castro~Neto}}]{castro-neto-prb11}%
  \BibitemOpen
  \bibfield  {author} {\bibinfo {author} {\bibfnamefont {R.}~\bibnamefont
  {de~Gail}}, \bibinfo {author} {\bibfnamefont {M.~O.}\ \bibnamefont
  {Goerbig}}, \bibinfo {author} {\bibfnamefont {F.}~\bibnamefont {Guinea}},
  \bibinfo {author} {\bibfnamefont {G.}~\bibnamefont {Montambaux}},\ and\
  \bibinfo {author} {\bibfnamefont {A.~H.}\ \bibnamefont {Castro~Neto}},\
  }\href {https://doi.org/10.1103/PhysRevB.84.045436} {\bibfield  {journal}
  {\bibinfo  {journal} {Phys. Rev. B}\ }\textbf {\bibinfo {volume} {84}},\
  \bibinfo {pages} {045436} (\bibinfo {year} {2011})}\BibitemShut {NoStop}%
\bibitem [{\citenamefont {Hejazi}\ \emph
  {et~al.}(2019{\natexlab{b}})\citenamefont {Hejazi}, \citenamefont {Liu},\
  and\ \citenamefont {Balents}}]{hejazi-ll_tbg-prb-2019}%
  \BibitemOpen
  \bibfield  {author} {\bibinfo {author} {\bibfnamefont {K.}~\bibnamefont
  {Hejazi}}, \bibinfo {author} {\bibfnamefont {C.}~\bibnamefont {Liu}},\ and\
  \bibinfo {author} {\bibfnamefont {L.}~\bibnamefont {Balents}},\ }\href
  {https://doi.org/10.1103/PhysRevB.100.035115} {\bibfield  {journal} {\bibinfo
   {journal} {Phys. Rev. B}\ }\textbf {\bibinfo {volume} {100}},\ \bibinfo
  {pages} {035115} (\bibinfo {year} {2019}{\natexlab{b}})}\BibitemShut
  {NoStop}%
\bibitem [{\citenamefont {Zhang}\ \emph {et~al.}(2019)\citenamefont {Zhang},
  \citenamefont {Po},\ and\ \citenamefont {Senthil}}]{zhang-ll_tbg-prb-2019}%
  \BibitemOpen
  \bibfield  {author} {\bibinfo {author} {\bibfnamefont {Y.-H.}\ \bibnamefont
  {Zhang}}, \bibinfo {author} {\bibfnamefont {H.~C.}\ \bibnamefont {Po}},\ and\
  \bibinfo {author} {\bibfnamefont {T.}~\bibnamefont {Senthil}},\ }\href
  {https://doi.org/10.1103/PhysRevB.100.125104} {\bibfield  {journal} {\bibinfo
   {journal} {Phys. Rev. B}\ }\textbf {\bibinfo {volume} {100}},\ \bibinfo
  {pages} {125104} (\bibinfo {year} {2019})}\BibitemShut {NoStop}%
\bibitem [{\citenamefont {Lian}\ \emph {et~al.}(2020)\citenamefont {Lian},
  \citenamefont {Xie},\ and\ \citenamefont {Bernevig}}]{ll-tbg-lian}%
  \BibitemOpen
  \bibfield  {author} {\bibinfo {author} {\bibfnamefont {B.}~\bibnamefont
  {Lian}}, \bibinfo {author} {\bibfnamefont {F.}~\bibnamefont {Xie}},\ and\
  \bibinfo {author} {\bibfnamefont {B.~A.}\ \bibnamefont {Bernevig}},\ }\href
  {https://doi.org/10.1103/PhysRevB.102.041402} {\bibfield  {journal} {\bibinfo
   {journal} {Phys. Rev. B}\ }\textbf {\bibinfo {volume} {102}},\ \bibinfo
  {pages} {041402} (\bibinfo {year} {2020})}\BibitemShut {NoStop}%
\bibitem [{\citenamefont {Xiao}\ \emph {et~al.}(2010)\citenamefont {Xiao},
  \citenamefont {Chang},\ and\ \citenamefont {Niu}}]{niu-rmp10}%
  \BibitemOpen
  \bibfield  {author} {\bibinfo {author} {\bibfnamefont {D.}~\bibnamefont
  {Xiao}}, \bibinfo {author} {\bibfnamefont {M.-C.}\ \bibnamefont {Chang}},\
  and\ \bibinfo {author} {\bibfnamefont {Q.}~\bibnamefont {Niu}},\ }\href@noop
  {} {\bibfield  {journal} {\bibinfo  {journal} {Reviews of modern physics}\
  }\textbf {\bibinfo {volume} {82}},\ \bibinfo {pages} {1959} (\bibinfo {year}
  {2010})}\BibitemShut {NoStop}%
\bibitem [{\citenamefont {Thouless}\ \emph {et~al.}(1982)\citenamefont
  {Thouless}, \citenamefont {Kohmoto}, \citenamefont {Nightingale},\ and\
  \citenamefont {den Nijs}}]{TKNN}%
  \BibitemOpen
  \bibfield  {author} {\bibinfo {author} {\bibfnamefont {D.~J.}\ \bibnamefont
  {Thouless}}, \bibinfo {author} {\bibfnamefont {M.}~\bibnamefont {Kohmoto}},
  \bibinfo {author} {\bibfnamefont {M.~P.}\ \bibnamefont {Nightingale}},\ and\
  \bibinfo {author} {\bibfnamefont {M.}~\bibnamefont {den Nijs}},\ }\href
  {https://doi.org/10.1103/PhysRevLett.49.405} {\bibfield  {journal} {\bibinfo
  {journal} {Phys. Rev. Lett.}\ }\textbf {\bibinfo {volume} {49}},\ \bibinfo
  {pages} {405} (\bibinfo {year} {1982})}\BibitemShut {NoStop}%
\bibitem [{\citenamefont {Kohmoto}(1989)}]{kohmoto-qhe-prb-1989}%
  \BibitemOpen
  \bibfield  {author} {\bibinfo {author} {\bibfnamefont {M.}~\bibnamefont
  {Kohmoto}},\ }\href {https://doi.org/10.1103/PhysRevB.39.11943} {\bibfield
  {journal} {\bibinfo  {journal} {Phys. Rev. B}\ }\textbf {\bibinfo {volume}
  {39}},\ \bibinfo {pages} {11943} (\bibinfo {year} {1989})}\BibitemShut
  {NoStop}%
\bibitem [{\citenamefont {Dana}\ \emph {et~al.}(1985)\citenamefont {Dana},
  \citenamefont {Avron},\ and\ \citenamefont {Zak}}]{dana-qhe-jpcssp-1985}%
  \BibitemOpen
  \bibfield  {author} {\bibinfo {author} {\bibfnamefont {I.}~\bibnamefont
  {Dana}}, \bibinfo {author} {\bibfnamefont {Y.}~\bibnamefont {Avron}},\ and\
  \bibinfo {author} {\bibfnamefont {J.}~\bibnamefont {Zak}},\ }\href
  {https://doi.org/10.1088/0022-3719/18/22/004} {\bibfield  {journal} {\bibinfo
   {journal} {Journal of Physics C: Solid State Physics}\ }\textbf {\bibinfo
  {volume} {18}},\ \bibinfo {pages} {L679} (\bibinfo {year}
  {1985})}\BibitemShut {NoStop}%
\bibitem [{\citenamefont {Lopes~dos Santos}\ \emph
  {et~al.}(2007{\natexlab{b}})\citenamefont {Lopes~dos Santos}, \citenamefont
  {Peres},\ and\ \citenamefont {Castro~Neto}}]{santos-tbg-prl07}%
  \BibitemOpen
  \bibfield  {author} {\bibinfo {author} {\bibfnamefont {J.~M.~B.}\
  \bibnamefont {Lopes~dos Santos}}, \bibinfo {author} {\bibfnamefont
  {N.~M.~R.}\ \bibnamefont {Peres}},\ and\ \bibinfo {author} {\bibfnamefont
  {A.~H.}\ \bibnamefont {Castro~Neto}},\ }\href
  {https://doi.org/10.1103/PhysRevLett.99.256802} {\bibfield  {journal}
  {\bibinfo  {journal} {Phys. Rev. Lett.}\ }\textbf {\bibinfo {volume} {99}},\
  \bibinfo {pages} {256802} (\bibinfo {year} {2007}{\natexlab{b}})}\BibitemShut
  {NoStop}%
\bibitem [{\citenamefont {Thompson}\ \emph {et~al.}(2022)\citenamefont
  {Thompson}, \citenamefont {Aktulga}, \citenamefont {Berger}, \citenamefont
  {Bolintineanu}, \citenamefont {Brown}, \citenamefont {Crozier}, \citenamefont
  {in~'t Veld}, \citenamefont {Kohlmeyer}, \citenamefont {Moore}, \citenamefont
  {Nguyen}, \citenamefont {Shan}, \citenamefont {Stevens}, \citenamefont
  {Tranchida}, \citenamefont {Trott},\ and\ \citenamefont {Plimpton}}]{LAMMPS}%
  \BibitemOpen
  \bibfield  {author} {\bibinfo {author} {\bibfnamefont {A.~P.}\ \bibnamefont
  {Thompson}}, \bibinfo {author} {\bibfnamefont {H.~M.}\ \bibnamefont
  {Aktulga}}, \bibinfo {author} {\bibfnamefont {R.}~\bibnamefont {Berger}},
  \bibinfo {author} {\bibfnamefont {D.~S.}\ \bibnamefont {Bolintineanu}},
  \bibinfo {author} {\bibfnamefont {W.~M.}\ \bibnamefont {Brown}}, \bibinfo
  {author} {\bibfnamefont {P.~S.}\ \bibnamefont {Crozier}}, \bibinfo {author}
  {\bibfnamefont {P.~J.}\ \bibnamefont {in~'t Veld}}, \bibinfo {author}
  {\bibfnamefont {A.}~\bibnamefont {Kohlmeyer}}, \bibinfo {author}
  {\bibfnamefont {S.~G.}\ \bibnamefont {Moore}}, \bibinfo {author}
  {\bibfnamefont {T.~D.}\ \bibnamefont {Nguyen}}, \bibinfo {author}
  {\bibfnamefont {R.}~\bibnamefont {Shan}}, \bibinfo {author} {\bibfnamefont
  {M.~J.}\ \bibnamefont {Stevens}}, \bibinfo {author} {\bibfnamefont
  {J.}~\bibnamefont {Tranchida}}, \bibinfo {author} {\bibfnamefont
  {C.}~\bibnamefont {Trott}},\ and\ \bibinfo {author} {\bibfnamefont {S.~J.}\
  \bibnamefont {Plimpton}},\ }\href {https://doi.org/10.1016/j.cpc.2021.108171}
  {\bibfield  {journal} {\bibinfo  {journal} {Comp. Phys. Comm.}\ }\textbf
  {\bibinfo {volume} {271}},\ \bibinfo {pages} {108171} (\bibinfo {year}
  {2022})}\BibitemShut {NoStop}%
\bibitem [{\citenamefont {Los}\ and\ \citenamefont {Fasolino}(2003)}]{LCBOP}%
  \BibitemOpen
  \bibfield  {author} {\bibinfo {author} {\bibfnamefont {J.~H.}\ \bibnamefont
  {Los}}\ and\ \bibinfo {author} {\bibfnamefont {A.}~\bibnamefont {Fasolino}},\
  }\href {https://doi.org/10.1103/PhysRevB.68.024107} {\bibfield  {journal}
  {\bibinfo  {journal} {Phys. Rev. B}\ }\textbf {\bibinfo {volume} {68}},\
  \bibinfo {pages} {024107} (\bibinfo {year} {2003})}\BibitemShut {NoStop}%
\bibitem [{\citenamefont {Nam}\ and\ \citenamefont
  {Koshino}(2017)}]{koshino-tbg-prb17}%
  \BibitemOpen
  \bibfield  {author} {\bibinfo {author} {\bibfnamefont {N.~N.~T.}\
  \bibnamefont {Nam}}\ and\ \bibinfo {author} {\bibfnamefont {M.}~\bibnamefont
  {Koshino}},\ }\href {https://doi.org/10.1103/PhysRevB.96.075311} {\bibfield
  {journal} {\bibinfo  {journal} {Phys. Rev. B}\ }\textbf {\bibinfo {volume}
  {96}},\ \bibinfo {pages} {075311} (\bibinfo {year} {2017})}\BibitemShut
  {NoStop}%
\bibitem [{\citenamefont {Angeli}\ \emph {et~al.}(2018)\citenamefont {Angeli},
  \citenamefont {Mandelli}, \citenamefont {Valli}, \citenamefont {Amaricci},
  \citenamefont {Capone}, \citenamefont {Tosatti},\ and\ \citenamefont
  {Fabrizio}}]{angeli-tbg-prb18}%
  \BibitemOpen
  \bibfield  {author} {\bibinfo {author} {\bibfnamefont {M.}~\bibnamefont
  {Angeli}}, \bibinfo {author} {\bibfnamefont {D.}~\bibnamefont {Mandelli}},
  \bibinfo {author} {\bibfnamefont {A.}~\bibnamefont {Valli}}, \bibinfo
  {author} {\bibfnamefont {A.}~\bibnamefont {Amaricci}}, \bibinfo {author}
  {\bibfnamefont {M.}~\bibnamefont {Capone}}, \bibinfo {author} {\bibfnamefont
  {E.}~\bibnamefont {Tosatti}},\ and\ \bibinfo {author} {\bibfnamefont
  {M.}~\bibnamefont {Fabrizio}},\ }\href
  {https://doi.org/10.1103/PhysRevB.98.235137} {\bibfield  {journal} {\bibinfo
  {journal} {Phys. Rev. B}\ }\textbf {\bibinfo {volume} {98}},\ \bibinfo
  {pages} {235137} (\bibinfo {year} {2018})}\BibitemShut {NoStop}%
\bibitem [{\citenamefont {Cahill}\ and\ \citenamefont
  {Glauber}(1969)}]{cahill-glauber-pr-1969}%
  \BibitemOpen
  \bibfield  {author} {\bibinfo {author} {\bibfnamefont {K.~E.}\ \bibnamefont
  {Cahill}}\ and\ \bibinfo {author} {\bibfnamefont {R.~J.}\ \bibnamefont
  {Glauber}},\ }\href {https://doi.org/10.1103/PhysRev.177.1857} {\bibfield
  {journal} {\bibinfo  {journal} {Phys. Rev.}\ }\textbf {\bibinfo {volume}
  {177}},\ \bibinfo {pages} {1857} (\bibinfo {year} {1969})}\BibitemShut
  {NoStop}%
\bibitem [{\citenamefont {Streda}(1982)}]{streda-1982}%
  \BibitemOpen
  \bibfield  {author} {\bibinfo {author} {\bibfnamefont {P.}~\bibnamefont
  {Streda}},\ }\href {https://doi.org/10.1088/0022-3719/15/22/005} {\bibfield
  {journal} {\bibinfo  {journal} {Journal of Physics C: Solid State Physics}\
  }\textbf {\bibinfo {volume} {15}},\ \bibinfo {pages} {L717} (\bibinfo {year}
  {1982})}\BibitemShut {NoStop}%
\end{thebibliography}%

\widetext
\clearpage

\begin{center}
\textbf{\large Supplementary Information for “Magic momenta and three dimensional  Landau levels from a three dimensional graphite moir\'e superlattice"}
\end{center}
\maketitle

\tableofcontents

\section{S1. Continuum model for bulk alternating twisted graphite}
\label{sec:continuum_model}
Away from the dislocation line, graphite spiral (GS) can be thought as bulk \textit{alternating twisted multilayer graphene} (ATG), as  shown in Fig.~1 of the main text. Bulk ATG consists of an infinitely large number of graphene layers stacked with an alternating twist angle $\pm\theta$ between each pair of adjacent graphene layers. Equivalently, bulk ATG can be viewed as a periodic stacking of twisted bilayer graphene (TBG) along the out-of-plane direction with lattice constant $2 d_0$, where $d_0=3.35\,$\AA\ is the distance between two adjacent graphene layers. Within each stacking motif, TBG forms a moir\'e pattern with spatial periodicity $L_{s}=a/(2\sin(\theta/2))$, where $a=2.46\,$\AA\  is the lattice constant of monolayer graphene.

To obtain the low-energy electronic structure of bulk ATG, we start from the continuum model of TBG for valley $\mu$ at twist angle $\theta$ \cite{santos-tbg-prl07,macdonald-pnas11}:
\begin{align}
\H^{\mu}_{\theta}= \left[\begin{array}{cc}
			-\hbar v_{F}(\k-\K_{1}^{\mu})\cdot\bm{\sigma}^{\mu} & T \\
			T^{\dagger} &-\hbar v_{F}(\k-\K_{2}^{\mu})\cdot\bm{\sigma}^{\mu}
		\end{array}\right]
	\label{eq:hamtbg}
\end{align}
where the Fermi velocity is $\hbar v_{F}/a\!=\!2.1354\,$eV \cite{moon-tbg-prb13}. The diagonal blocks represent the low-energy Hamiltonian of the two layers of graphene in valley $\mu=\pm1$ near its Dirac points $\K_{1/2,\mu}$ with the Pauli matrices $\bm{\sigma}^{\mu}=(\mp\sigma_{x},\sigma_{y})$ and $\k = (k_x,k_y)$. The intervalley coupling is omitted for the low-energy states on the moir\'e length scale so that $\mu$ is a good quantum number. The off-diagonal term is the moir\'e potential. 
 $T=U(\vr)\exp (-i\Delta\K_\mu \cdot\vr)$ with the shift between two Dirac points in the same valley from two adjacent layer $\Delta\K_\mu = \K_{2,\mu} - \K_{1,\mu}=\mu(0,k_{\theta})$ and  
\begin{align}
	U(\textbf{r})=u_0' \left[\begin{array}{cc}
			\kappa g(\vr) & g(\vr-\vr_{\rm{AB}}) \\
			g(\vr-\vr_{\rm{AB}}) & \kappa g(\vr)
		\end{array}\right]
	\label{eq:moire_pot}
\end{align}
where $\kappa = u_{0}/u'_{0}$ with intra-sublattice and inter-sublattice interlayer tunneling amplitudes denoted by $u_0$ and $u_0'$, respectively. Here $k_\theta =4 \pi / 3 L_s$. If out-of-plane corrugations are taken into account, one finds that $u_0'=0.0975\,$eV and $u_0=0.0797\,$eV \cite{koshino-prx18}.
The phase factor is defined as $g(\vr)\!=\!\sum^{3}_{j=1}e^{i\q_{j}\cdot\vr}$, $\q_{1}=k_\theta (0,1)$, $\q_{2}=k_\theta(-\sqrt{3}/2,-1/2)$ and $\q_{3}=k_\theta(\sqrt{3}/2,-1/2)$ with $\vr_{\rm{AB}} = (\sqrt{3} L_s/3,0)$. Up to a multiplicative factor $\hbar v_F k_\theta$, the continuum model of TBG is fully characterized by the dimensionless constants $\alpha_0 = u'_0 / (\hbar v_F k_\theta )$ and $\kappa=u_0'/u_0$.

Bulk ATG is formed by stacking vertically TBG in the $z$-direction. The tunnelling between two adjacent TBGs is  given by a $k_z$-dependent moir\'e potential $T'$ with possibly different dimensionless ratio $\alpha_0'$ and $\kappa'$. In the second quantization, the Hamiltonian is written:
\begin{align}
	\hat{H}^{\mu}= \sum_{z} \sum_{l=1,2} \left( H^{\mu}_{l} \right)_{\alpha \alpha'} \hcd_{l,\alpha}(z) \hc_{l,\alpha'}(z)  + \sum_{z} \left( T \right)_{\alpha \alpha'} \hcd_{1,\alpha}(z) \hc_{2,\alpha'}(z) + \left( T' \right)_{\alpha \alpha'} \hcd_{1,\alpha}(z) \hc_{2,\alpha'}(z+d_0+d_0') + \rm{H.c.}
	\label{eq:ham3d_2ndq}
\end{align}
where $l=1, 2$ and $\alpha=A, B$ represent the layer and the sublattice indices within TBG unit, and $H^{\mu}_{l}$ is the diagonal term of Eq.~\eqref{eq:hamtbg}.  $T_{\alpha,\alpha'}$ in Eq.~\eqref{eq:ham3d_2ndq} is just the off-diagonal moir\'e potential term within a TBG unit as expressed in Eqs.~\eqref{eq:hamtbg}-\eqref{eq:moire_pot}. 
Without $T'$, the Hamiltonian Eq.~\eqref{eq:ham3d_2ndq} is just a sum of decoupled TBG at different $z$. This form is reminiscent of the tight-binding model of quasi-1D chain, where the hopping between nearest primitive cells is precisely given by $T'$ and the lattice constant in the out-of-plane direction is $d_0 + d_0'$. Here $d_0$ denotes the interlayer distance within a TBG unit, and $d_0'$ denotes the interlayer distance between two neighboring TBG units. 
By virtue of lattice translational symmetry in the $z$ direction, we use the Fourier transform
\begin{align}
	\hc_{l,\alpha} (z) = \frac{1}{\sqrt{N_z}} \sum_{k_z} e^{i k_z z} \hc_{l,\alpha} (k_z)
	\label{eq:FT_c}
\end{align}
to rewrite the Hamiltonian such that
\begin{align}
	\hat{H}^{\mu}= \sum_{k_z} \sum_{l=1,2} \left( H^{\mu}_{l} \right)_{\alpha \alpha'} \hcd_{l,\alpha}(k_z) \hcd_{l,\alpha'}(k_z)  + \sum_{k_z} \left( T \right)_{\alpha \alpha'} \hcd_{1,\alpha}(k_z) \hc_{2,\alpha'}(k_z) + \left( T' \right)_{\alpha \alpha'} \hcd_{1,\alpha}(k_z) \hc_{2,\alpha'}(k_z) e^{i k_z (d_0+d_0')} + \rm{H.c.} \ .
	\label{eq:ham3d_2ndq_FT}
\end{align}
In principal, dimerization could occur such that $d_0\neq d_0'$, such that the system at each in-plane moir\'e wavevector resembles a Su-Schrieffer-Hegger chain. However, our transmission electron microscopy measurements indicate that dimerization is absent in the bulk ATG system. Therefore,
hereafter we take $d_0=d_0'=3.35\,$\AA, and $T=T'$, thus we obtain the Hamiltonian Eq.~(1) in the main text.

\begin{figure}
\centering
\includegraphics[width=16cm]{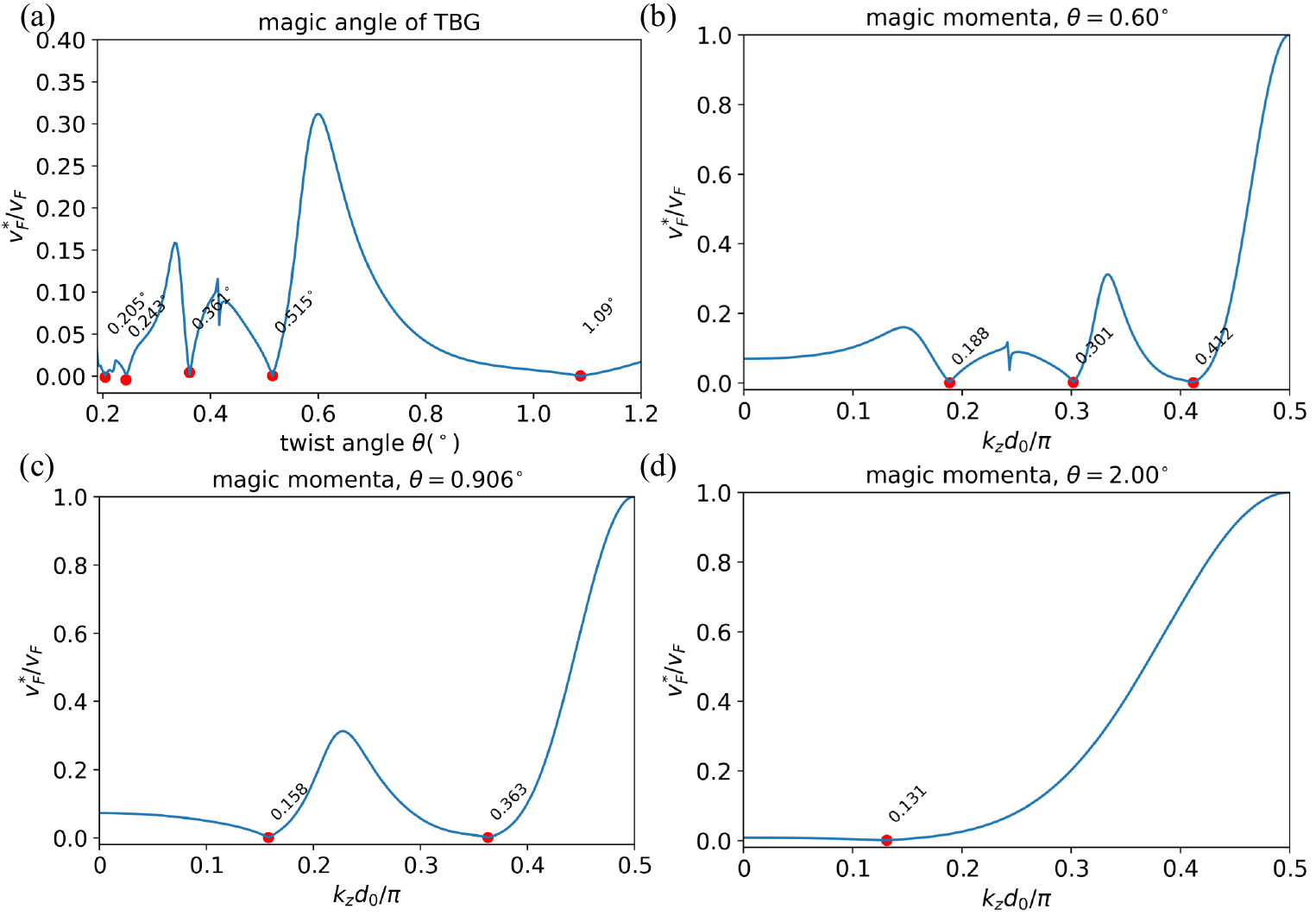}
\caption{(a) The Fermi velocity of TBG near moir\'e $K$ point as a function of twist angle. The red spots mark the magic angles. The $k_z$-dependent in-plane Fermi velocities of bulk ATG with twist angle (b) $0.60\,\degree$, (c) $0.906\,\degree$, and (d) $2.00\,\degree$. The red spots mark the magic momenta at which the in-plane Fermi velocities vanish.}
\label{fig:S1}
\end{figure}

When $u_0=u_0'$, the magic angle of TBG is solely determined by a dimensionless coupling constant $\alpha_0(\theta)=u_0'/(\hbar v_{F} k_\theta)$. In bulk ATG, at each $k_z$ one can define a renormalized coupling constant $\alpha(\theta,k_z)=u_0'\vert 1+ e^{i2k_z d_0}\vert/(\hbar v_{F} k_\theta)$, such that at twist angle $\theta$, when $k_z=k_z^{*}$ satisfies
\begin{equation}
\alpha(\theta,k_z^*)=\alpha_0(\theta_m^{(1)})\;,
\end{equation}
one would obtain  the flat bands at ``magic momentum" $k_z^{*}$  which are exactly the same as those of TBG at the first magic angle $\theta_m^{(1)}$.  Equivalently, the magic momentum condition is simply written as 
\begin{equation}
2\cos(k_{z}^{*} d_{0})=\frac{\sin{\theta/2}}{\sin{\theta_{m}^{(1)}/2}}
\label{eq:magic-kz}
\end{equation}
For small twist angle $\theta$, $\sin{\theta/2}\approx \theta /2$. In order to have real valued solution of $k_z^{*}$ of Eq.~\eqref{eq:magic-kz}, the twist angle of bulk ATG needs to satisfy $\theta < 2\theta_m^{(1)}$.   More generally, when $\theta$ is smaller than twice of the $n$th magic angle of TBG, i.e., $\theta< 2\theta_m^{(n)}$, one can always find the $s$th  magic momentum $k_z^{(s),*}$ which satisfies
\begin{equation}
2\cos(k_{z}^{(s),*} d_{0})=\frac{\sin{\theta/2}}{\sin{\theta_{m}^{(s)}/2}}
\label{eq:magic-kz-s}
\end{equation}
where $\theta^{(s)}_m$ is the $s$th magic angle of TBG with $s=1, 2,..., n$. Technically, we define magic angles of TBG as the twist angles at which the Fermi velocity of Dirac cone vanishes. 
In Fig.~\ref{fig:S1}(a), we present the Fermi velocity of TBG near moir\'e $\mathbf{K}$ point as a function of $\theta$. The red spot represents the magic angles from the first to the fifth.

In Fig.~2(a) in the main text, we plot the values of the magic momenta $\{k_{z}^{(s),*}\}$ (for $s=1,...,5$) as a function of twist angle $\theta$. We also calculate the $k_z$-dependent in-plane Fermi velocity of bulk ATG with three specific twist angles $\theta=0.60\,\degree$, $0.906\,\degree$ and $2.00\,\degree$ in Fig.~\ref{fig:S1}(b), (c) and (d), respectively. When $\theta=0.60\,\degree$, from Fig.~2(a) in the main text, we can learn that there will be three $k_z$ values  which satisfy the magic-momenta condition (Eq.~\eqref{eq:magic-kz-s}), which are marked by red dots in Fig.~\ref{fig:S1}(b). Same in Fig.~\ref{fig:S1}(c), there are two spots representing the two magic momenta at which the in-plane Fermi velocities vanish. In Fig.~\ref{fig:S1}(d), there is only one spot representing one magic momentum. 

\subsection{Band structures for $\theta=1.05 \degree$ at different $k_z$}

\begin{figure}
	\centering
	\includegraphics[width=16cm]{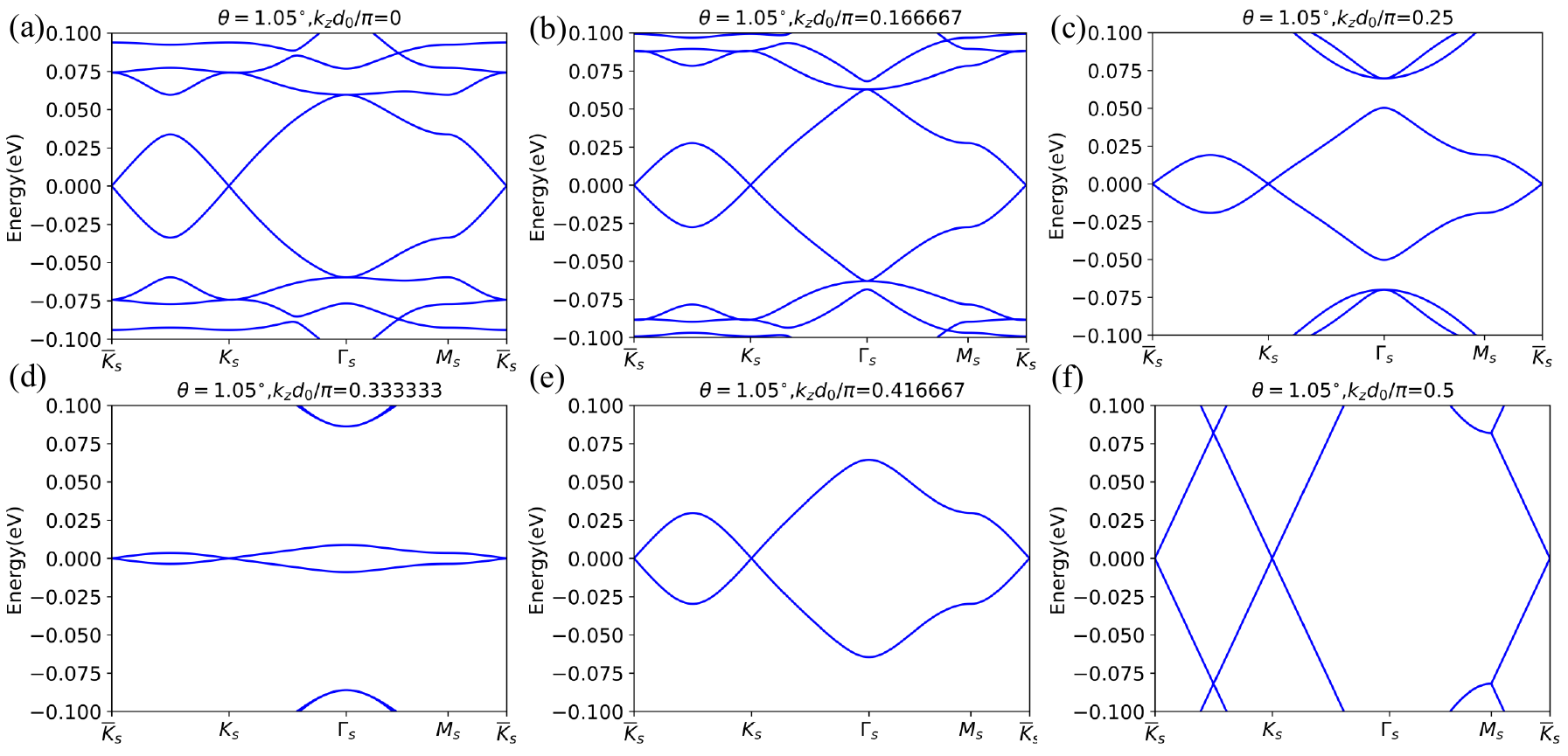}
	\caption{Band structures in bulk ATG with twist angle $\theta=1.05 \degree$ in the $(k_x,k_y)$-plane for (a) $k_{z}d_{0}\!=\!0$, (b) $\pi/6$, (c) $\pi/4$, (d) $\pi/3$, (e) $5\pi/12$ and (f) $\pi/2$.}
	\label{fig:S2}
\end{figure}

In Fig.~\ref{fig:S2}(a)-(f), we explicitly show the evolution of 2D band structures in the $(k_x,k_y)$-plane for $k_z d_0$ varying from 0 to $\pi/2$ in bulk ATG with twist angle $\theta=\theta_{m}^{(1)}=1.05\,\degree$. The previous argument shows that the flat bands of TBG at the first magic angle emerge at $k_z^{(1),*} d_0= \pi/3$, which is confirmed by numerical calculations as shown in Fig.~\ref{fig:S2}(d). At $k_z d_0= \pi/2$, the phase factor $(1+e^{2ik_z d_0})$ vanishes (see Eq.~(1) in the main text), such that the band structure is just those of two decoupled single layer graphene folded into the moir\'e Brillouin zone, as shown in Fig.~\ref{fig:S2}(f).

\begin{figure}
	\centering
	\includegraphics[width=0.95\textwidth]{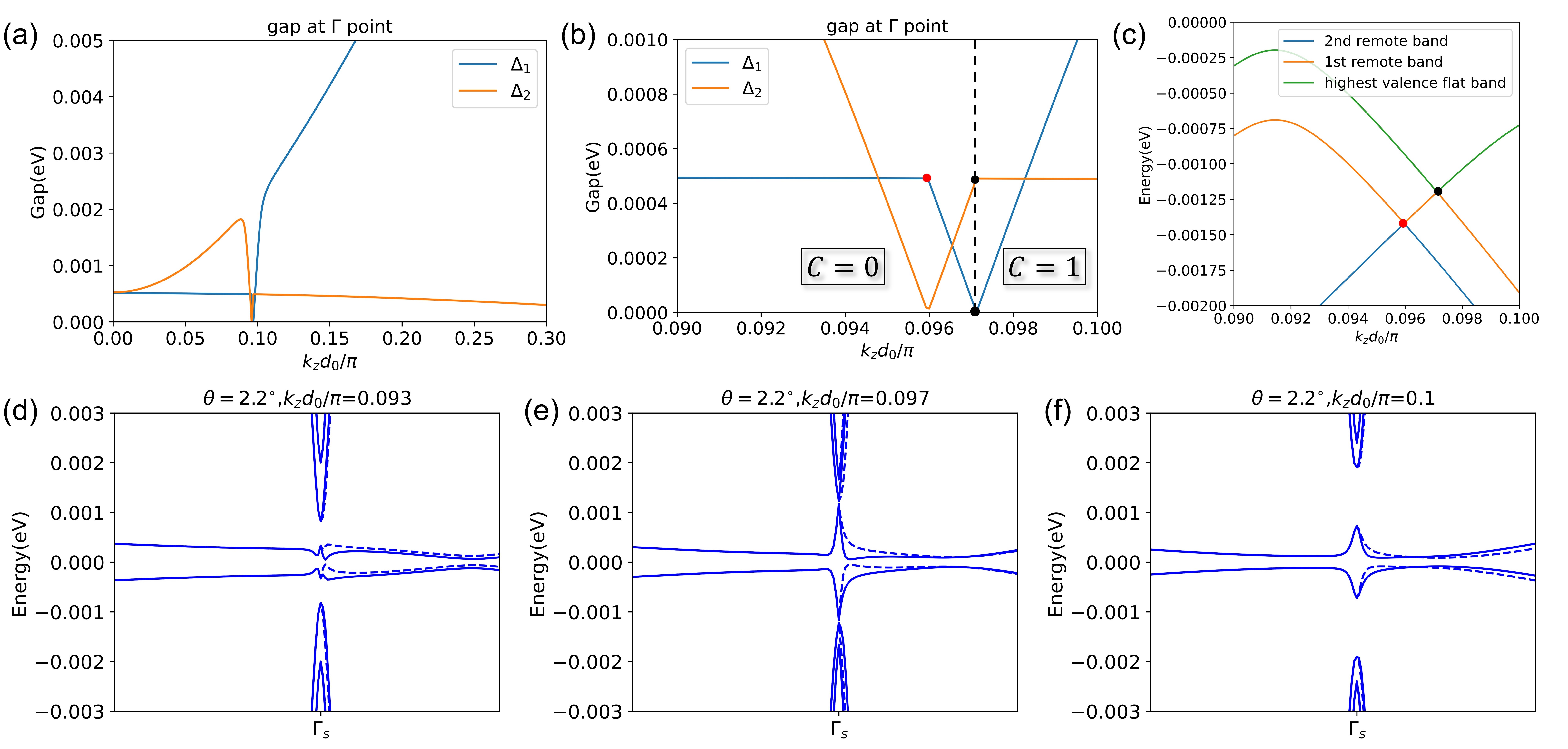}
	\caption{(a) Band gaps $\Delta_{1,2}$ at $\Gamma$ point for bulk ATG with twist angle $2.2\,\degree$ at different $k_{z}$. The blue line represents the band gap at $\Gamma$ point between the flat bands and the remote bands. The orange line represent the band gap at $\Gamma$ point between two remote bands. In (b), we show a zoomed-in view near the critical point. The dashed line and the black dots mark the topological transition point of the flat band. The Chern numbers are labeled as $C$. We also mark with red dots the kink spot in the evolution of $\Delta_1$, which is due to band order switching. This is clearly shown in (c) the evolution of the energy of three relevant bands at $\Gamma$ point, where the band inversion points are annotated by red and black dot, one-to-one corresponding with those in (b). The flat bands and remotes bands near $\Gamma$ point for bulk ATG with (d) $k_{z}d_{0}/\pi=0.093$, (e) $0.097$ and (f) $0.1$. The solid lines represent energy band from the $\mu=+$ valley and the dashed lines represent energy band from the other valley.}
	\label{fig:S4}
\end{figure}

\subsection{Gap closing and topological transition at the vicinity of magic momentum}

\begin{figure}
	\centering
	\includegraphics[width=0.95\textwidth]{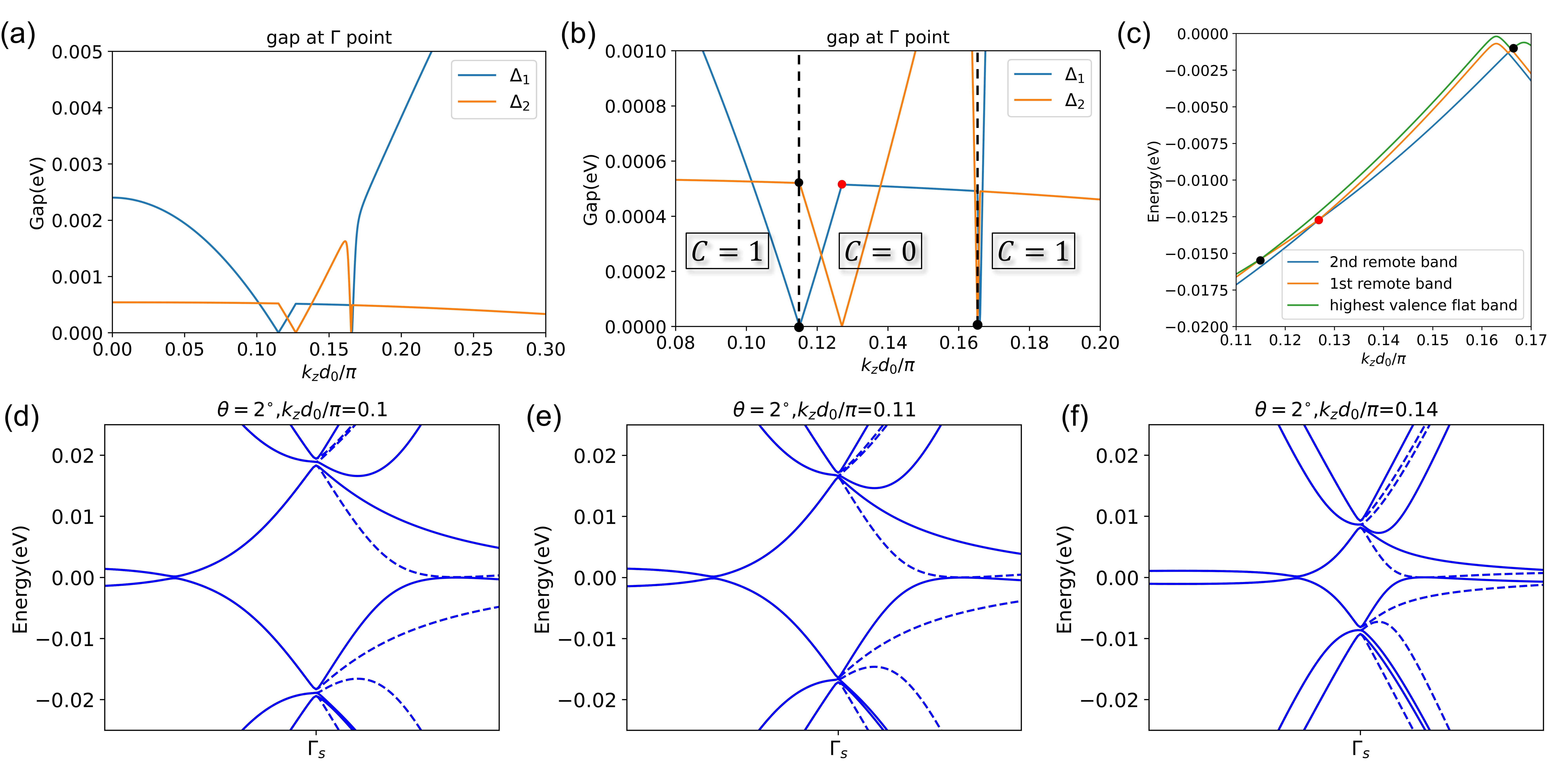}
	\caption{(a) {Band gaps $\Delta_{1,2}$} at $\Gamma$ point for bulk ATG with twist angle $2\,\degree$ at different $k_{z}$. The blue line represents the band gap at $\Gamma$ point between the flat bands and the remote bands. The orange line represent the band gap at $\Gamma$ point between two remote bands. In (b), we show a zoomed-in view near the two critical points. The dashed lines and the black dots mark the topological transition points of the flat band. The Chern numbers are labeled as $C$. We also mark with red dots one of the kink spots in the evolution of $\Delta_1$, which is due to band order switching. This is shown in (c) the evolution of the energy of three relevant bands at $\Gamma$ point, where three band inversion points are annotated by red and black dot, one-to-one corresponding with those in (b). The flat bands and remotes bands near $\Gamma$ point for bulk ATG with (c) $k_{z}d_{0}/\pi=0.10$, (d) $0.11$ and (e) $0.14$. The solid lines represent energy band from the $\mu=+$ valley and the dashed lines represent energy band from the other valley.}
	\label{fig:g2}
\end{figure}

In this section, we provide the detailed information of the band closings at the phase boundaries of the topological transitions. As we mentioned in the main text, we establish an intriguing topological phase diagram by varying both twist angle $\theta$ and $k_{z}$. When the twist angle is fixed between $\theta_{c,0}\!=\!2.3^{\circ}$ and $\theta_{c,1}\!=\!2.15^{\circ}$, the Brillouin zone can be divided into two topologically different parts. When the twist angle is smaller than $\theta_{c,1}$, there will be more topological transitions while varying $k_{z}$ inside the 3D Brillouin zone. Here, we provide the detailed electronic band structures at the near the topological transitions.

For bulk ATG with a twist angle of $\theta=2.2^{\circ}$, the Brillouin zone can be partitioned  into two distinct regions: the Chern number of the highest valence band of $K$ valley equals zero if $k_z \in [-k^{(1),*}_{z},k^{(1),*}_{z}]$ and equals one otherwise. We introduce a small mass term with a value of $m_{z}\!=\!0.001\,$eV and obtain the band structures of bulk ATG at the vicinity of $k_{z}^{(1),*}$. In Fig.~\ref{fig:S4}(a), we present the direct band gaps at $\Gamma$ point for bulk ATG at a function of $k_{z}$. The blue line represents the  gap between the highest valence band  and the band immediately below it (dubbed as  the first remote band) at $\Gamma$ point, denoted as $\Delta_1$ in Fig.~\ref{fig:S4}(a); while the orange line represents the  gap between the first remote band and the second remote band at $\Gamma$ point, denoted as $\Delta_2$ in Fig.~\ref{fig:S4}(a). Fig.~\ref{fig:S4}(b) is a zoomed-in view of the two direct gaps at $\Gamma$ near the gap closing points. The dashed line marks the critical $k_{z}$ value marking the topological transition where $\Delta_1$ vanishes. The corresponding Chern number $C$ of the highest valence band of valley $K$ aside the transition point is also shown. We see that $\Delta_1$ vanishes at $k_{z}\!\approx\!0.097\pi/d_{0}$, across which the Chern number changes by unity. Furthermore, near the phase boundary, another gap closure occurs between two remote bands, but the Chern number of the highest valence band remains unaffected as expected. In Fig.~\ref{fig:S4}(d-f), we explicitly present the band structures in the $(k_x,k_y)$-plane of the bulk ATG with $k_{z}d_{0}/\pi\!=\!0.093$, $0.097$ and $0.1$, respectively, which explicitly demonstrates the gap closure at $k_{z}d_{0}/\pi\!=\!0.097$. 

Furthermore, in the case of bulk ATG with a twist angle of $\theta=2.0^{\circ}$, there are four topologically different regions. In Fig.~\ref{fig:g2}(a), we show the behavior of the direct band gap at $\Gamma$ point. Again, the blue line represents the  gap between the highest valence band and the first remote band (denoted as $\Delta_1$), while the orange line represents the band gap between the two remote bands (denoted by $\Delta_2$). Fig.~\ref{fig:g2}(b) is a magnified view near two critical points, with the values of $k^{(2),*}_{z}d_{0}/\pi=0.115$ and of $k^{(1),*}_{z}d_{0}/\pi=0.166$ at which $\Delta_1$ vanishes. While increasing $k_{z}$ from zero, the Chern number undergoes a transition from one to zero, and subsequently form zero back to one. In Fig.~\ref{fig:g2}(d-f), we present the band structures in the $(k_{x},k_{y})$-plane with $k_{z}d_{0}/\pi\!=\!0.10$, $0.11$ and $0.14$, respectively.

\section{S2. Lattice relaxation of graphite spiral}
\label{sec:relaxation}

\begin{figure}
\centering
\includegraphics[width=16cm]{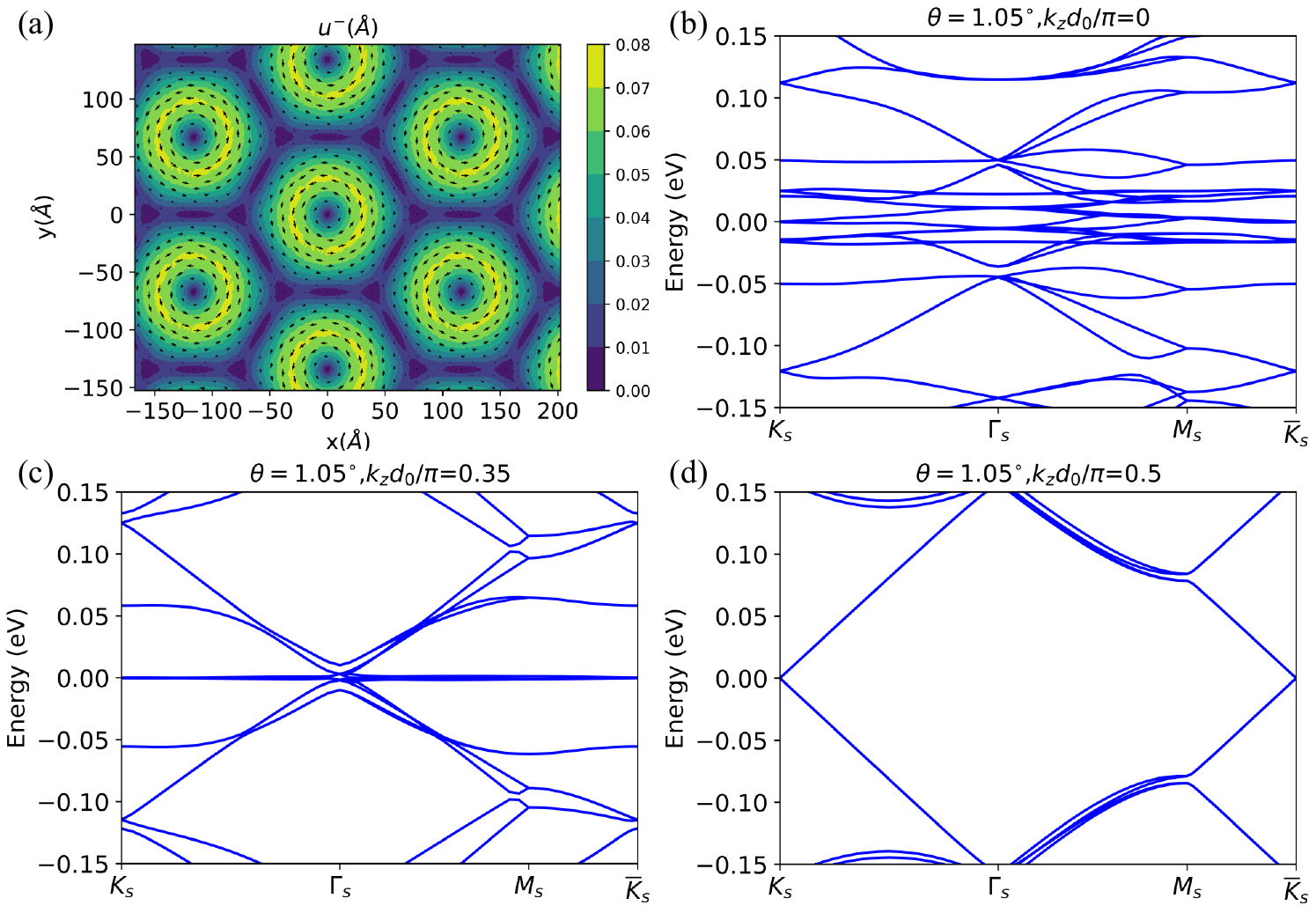}
\caption{
	(a) The relative interlayer in-plane distortion field of bulk ATG with twist angle $\theta = 1.05 \degree$. The colorbar represents the amplitude of the distortion fields. The arrows represent the direction of the distortion fields. We present the band structure of bulk ATG at $\theta = 1.05 \degree$ with fully relaxed lattice structure at (b) $k_{z}d_{0}/\pi=0$, (c) $0.35$ (c) and (d) $0.5$.}
\label{fig:relaxation}
\end{figure}

In this section we study the lattice distortions of bulk ATG using direct molecular dynamics simulations.
Specifically, the structural relaxation is calculated utilizing Large-scale Atomic-Molecular Massively Parallel Simulation (LAMMPS) \cite{LAMMPS}, employing Long-range Carbon Bond-Order Potential (LCBOP) \cite{LCBOP}. The initial structure is a moir\'e supercell with periodic boundary condition applied in both the in-plane and out-of-plane directions. The twist angles between two adjacent layers, according to the commensurate condition, are set to $\pm 1.05\,\degree$, $\pm 1.31\,\degree$, $\pm 1.58\,\degree$ and $\pm1.83\,\degree$. The lattice relaxation calculations are carried out using the steepest descent algorithm. In Fig.~\ref{fig:relaxation}(a), we present the relative interlayer shift in the moir\'e supercell. The lateral lattice distortions in bulk ATG are qualitatively the same as those in TBG \cite{koshino-tbg-prb17,angeli-tbg-prb18}. Due to the presence of the translational symmetry along the $z$-direction and mirror symmetry, it turns out that there is no out-of-plane distortion in bulk ATG. The system under different twist angle exhibits the lowest total energy always at an interlayer distance of $d_0=3.35$\,\AA\ so that the vertical lattice constant is $2d_0=6.7\,$\AA. We have not observed any tendency of dimerization between neighboring layers, consistent with the TEM results. 

Based on the relaxed lattice structure, we further calculate the electronic band structures of bulk ATG at the first magic angle $\theta=1.05\,\degree$ using an atomistic Slater-Koster tight-binding model.
To be specific, the hopping amplitude between two $p_{z}$ orbitals at different sites is expressed as \cite{moon-tbg-prb13}
\begin{equation}\label{eq:tightbinding}
	t(\textbf{d})=V_{\sigma}\left(\frac{\textbf{d}\cdot\hat{\textbf{z}}}{d}\right)+V_{\pi}\left[1-\left(\frac{\textbf{d}\cdot\hat{\textbf{z}}}{d}\right)^{2}\right],
\end{equation}
where $V_{\sigma}\!=\!V^{0}_{\sigma}e^{-(r-d_{0})/\delta_{0}}$ and $V_{\pi}\!=\!V^{0}_{\pi}e^{-(r-a_{0})/\delta_{0}}$. $\textbf{d}=(d_{x},d_{y},d_{z})$ is the displacement vector between two sites. $d_{c} = 3.35\,$\AA\  is the interlayer distance. $a_{0}=a/\sqrt{3} =1.42\,$\AA\  is the distance between two nearest neighbor carbon atoms. $\delta_{0}=0.184 a$, with $a=2.46\,$\AA. $V^{0}_{\sigma} = 0.48\,$eV and $V^{0}_{\pi}=-2.7\,$eV. The band structures of bulk ATG at $\theta = 1.05\,\degree$ at several fixed values of $k_{z}$ are showed in Fig.~\ref{fig:relaxation}(b)-(d). The band width of flat bands varies with $k_{z}$, and reaches a minimum at $k_{z}d_{0}\approx 0.35\pi$. This is slightly different from the value obtained from the magic-momentum condition $k_{z}d_{0}\approx \pi/3$ (Eq.~(1) in the main text) because the criteria for the magic angle is slightly different, i. e., the vanishing Fermi velocity around Dirac points does not exactly coincide with the condition of minimal overall bandwidth.

\section{S3. Landau quantization of continuum model}
\label{sec:landau}
We consider the Landau quantization of the continuum model for bulk ATG in the presence of an out-of-plane magnetic field $B \hat{\textbf{z}}$. The associated vector potential can always be chosen to be in-plane such that $k_z$ remains a good quantum number. Therefore, the prefactor $1+\exp (i\phi)$ with $\phi=2 k_z d_0$ is totally inert in the following derivations. Therefore, the derivations of the Landau quantization in bulk ATG are exactly the same as those of TBG but with $k_z$-dependent moir\'e potential. Technically, we follow the procedures given in \cite{moire-butterfly-macdonald-prb11}, which are also adopted in \cite{hejazi-ll_tbg-prb-2019,zhang-ll_tbg-prb-2019}. The only difference is the gauge choice in the initial Hamiltonian.

\subsection{Hamiltonian in Landau level basis}
In our set-up, we start with the continuum Hamiltonian
\begin{align}
	\H^{\mu}_{\theta}=\left[\begin{array}{cc}
		-\hbar v_{F} (\k-\K_{1,\mu})\cdot\bm{\sigma}^{\mu} & T (1 + e^{i\phi}) \\
		T^{\dagger}(1 + e^{-i\phi}) &-\hbar v_{F} (\k-\K_{2,\mu})\cdot\bm{\sigma}^{\mu}
	\end{array}\right]
	\label{eq:ham3d_full}
\end{align}
where $\k = (k_x,k_y)$ and $k_z$ appearing in $\phi$ is irrelevant in the following derivations. We restore all the units and expand the moir\'e potential
\begin{equation}
	\begin{split}
		\bm{\sigma}^{\mu=\pm} &= (\mp\sigma_{x},\sigma_{y}) \\
		\K_{l,\mu} &= k_\theta \left(\frac{\sqrt{3}}{2}, \frac{(-1)^l \mu}{2} \right) \\
		T_{\mu} &= \sum^{3}_{j=1} T_{j,\mu} e^{i \mu \Q_j \cdot\vr} \\
		T_{j,\mu} &= \left[\begin{array}{cc}
			u_0 & u_0' w^{\mu(j-1)} \\
			u_0' w^{-\mu(j-1)} & u_0 
		\end{array}\right] \\
		\Q_{1} &=(0,0), \quad \Q_{2}=\sqrt{3} k_\theta (-1/2,-\sqrt{3}/2), \quad \Q_{3}=\sqrt{3} k_\theta (1/2,-\sqrt{3}/2)
	\end{split}
\end{equation}
with $w = \exp (i 2\pi/3)$. In the following, we only show the details of calculations for $\mu=+$ while that of $\mu=-$ is exactly the same. We will give the final results for both valleys at the end of this section.

Now we introduce an out-of-plane magnetic field via vector potential in the Landau gauge $\mathbf{A} = (-By,0,0)$. The diagonal part Eq.~\eqref{eq:ham3d_full} can be easily written in the Landau level basis while the off-diagonal term cannot. Let us first focus on the diagonal part. For $\mu=+$, we Peierls substitute $k_x$ by $k_x - eBy/\hbar$ then 
\begin{align}
	- \hbar v_F (\k-\K_{l,+})\cdot\bm{\sigma}^{+} = \frac{\hbar \sqrt{2} v_F}{\lb} \left[\begin{array}{cc}
		0 & \had (k_x) \\
		 \ha (k_x) & 0
	\end{array}\right] + \hbar v_F \left[\begin{array}{cc}
	0 & -K_{l,+,x} - i K_{l,+,y} \\
	-K_{l,+,x}+ i K_{l,+,y} & 0
	\end{array}\right]
\end{align}
where the magnetic length $\lb = \sqrt{\hbar/eB}$ and the ladder operators
\begin{equation}
	\begin{split}
			\ha (k_x) &= \frac{\lb}{\sqrt{2}} \left(k_x - \frac{y}{\lb^2} - i k_y\right) \\
		\had (k_x) &= \frac{\lb}{\sqrt{2}} \left(k_x - \frac{y}{\lb^2} + i k_y\right)\\
		\rm{and} &\quad [\ha (k_x), \had (k_x)] = 1 
	\end{split}.
\end{equation}
We introduce the Landau level basis
\begin{equation}
	\begin{split}
		\ket{l,\alpha; n, k_x} &= \hfd_{l,\alpha; n,k_x} \ket{0} \\ \rm{with} &\quad \ket{l,\alpha; n, k_x} = \frac{(\had (k_x))^n}{\sqrt{n !}}\ket{l,\alpha;0, k_x}\\
		&\quad \ha (k_x)\ket{l,\alpha;0, k_x} = 0
	\end{split}
\label{eq:LLbasis}
\end{equation}

where $\alpha = A,B$ is the sublattice and $l=1,2$ is the layer index. Then, the diagonal part can be suitably written as
\begin{equation}
	\begin{split}
		\hat{H}^{\mu=+}_{D} = & \sum_{l,n,k_x} \hbar v_F (-K_{l,+,x} - i K_{l,+,y}) \  \hfd_{l,A; n,k_x} \hf_{l,B; n,k_x} + \rm{H.c.} \\
		&+ \sum_{l,n,k_x} \frac{\sqrt{2(n+1)}\hbar v_F}{\lb} \   \hfd_{l,A; n+1,k_x} \hf_{l,B; n,k_x} + \rm{H.c.}
	\end{split}.
\label{eq:hamD_LL1}
\end{equation}

Now we write the off-diagonal part $\hat{H}^{\mu=+}_{M}$ in the Landau level basis. To prepare the basis transformation, it is worthwhile to rewrite $\hat{H}^{\mu=+}_{M}$ in the second-quantized form explicitly showing the plane-wave basis, which is defined as
\begin{equation}
	\begin{split}
		\ket{l,\alpha; \k} &= \hcd_{l,\alpha}(\k)\ket{0} \\
		\bracket{\vr}{l,\alpha; \k} &= \frac{1}{\sqrt{S}} e^{i \k \cdot \vr} \chi_{l,\alpha}
	\end{split}
\end{equation}
with $\chi_{l,\alpha}$ represents the spinor part of the wavefunction and $S$ in the total area. In this basis, $\hat{H}^{\mu=+}_{M}$ is written as:
\begin{equation}
	\hat{H}^{\mu=+}_{M} = \sum_{\k,\alpha \beta} \sum_{j=1}^{3}  (1 + e^{i\phi}) (T_{j,+})_{\alpha \beta}\  \hcd_{1,\alpha}(\k+\Q_j) \hc_{2,\alpha}(\k)  + \rm{H.c.}.
\label{eq:hamM_planewave}
\end{equation}

By the completeness of the Landau level basis,
\begin{equation}
	\ket{l,\alpha;\k} = \sum_{n} \ket{l,\alpha;n,k_x} \bracket{l,\alpha;n,k_x}{l,\alpha;\k}.
\end{equation}
The formula for the basis transformation is thus
\begin{equation}
	\hc_{l,\alpha}(\k) = \sum_{n} \hf_{l,\alpha; n,k_x} \tilde{H}_n (k_y) e^{-i k_x k_y \lb^2}
\end{equation}
with $\tilde{H}_n (k_y)$ is the Fourier transform of the $n$-th Hermite function $H_n (y)$, i.e., the wavefunction of 1D quantum harmonic oscillator
\begin{equation}
	\tilde{H}_n (k_y) = \int \frac{dy}{\sqrt{L_y}} e^{-ik_y y} H_n (y).
\end{equation}
So, the moir\'e potential in the Landau level basis reads
\begin{equation}
	\hat{H}^{\mu=+}_M = \sum_{j=1}^{3} \sum_{m,n,k_x} (1 + e^{i\phi}) F_{mn,k_x} (\Q_j) \sum_{\alpha \beta}  \left(T_{j,+}\right)_{\alpha \beta} \  \hfd_{1,\alpha; m,k_x+Q_{j,x}} \hf_{2,\alpha; n,k_x} + \rm{H.c.}
\label{eq:hamM_LL0}
\end{equation}
where we use a shorthand notation
\begin{equation}
	F_{mn,k_x} (\Q_j) = e^{i (k_x + Q_{j,x}) Q_{j,y} \lb^2} \  \sum_{k_y} e^{i  k_y Q_{j,x} \lb^2} \ \tilde{H}_m^{*} (k_y + Q_{j,y}) \tilde{H}_n (k_y).
\end{equation}

\subsection{Displacement operators}
We now calculate the phase factor $F_{mn,k_x} (\Q_j)$. Transforming back $\tilde{H}_n (k_y) $ to $H_n (y)$, the factor $F_{mn,k_x} (\Q_j)$ becomes
\begin{equation}
	F_{mn,k_x} (\Q_j) = e^{i (k_x + Q_{j,x}) Q_{j,y} \lb^2} \  \bra{m,0} e^{i Q_{j,y}} \ket{n,-Q_{j,x}}
\label{eq:phasefac}
\end{equation} 
where we use the notation in Eq.~\eqref{eq:LLbasis} omitting irrelevant the $(l,\alpha)$ spinor part. Here, $k_x$ in $\ket{n,k_x}$ plays the role of the cyclotron center, which sits at $y = k_x \lb^2$. One can move the cyclotron center back to the origin $y=0$ by applying a so-called \textit{displacement operator} on the state. It turns out the phase factor itself can be written in terms of displacement operator such that one can calculate it analytically.

We provide here a brief self-contained introduction on how displacement operator enters in our calculations. More details can be found in the classic paper \cite{cahill-glauber-pr-1969}. For the basis $\ket{n,k_x}$, we define
\begin{equation}
	D (\alpha) = e^{\alpha \had (0) - \alpha^* \ha (0)}
\end{equation} 
with $\alpha \in \mathbb{C}$. One can check that 
\begin{equation}
	\ha(0) D (\alpha) \ket{0,0} = \alpha D(\alpha) \ket{0,0}
\end{equation} 
which means that $D (\alpha) \ket{0,0}$ is an eigenstate of $\ha(0)$ with eigenvalue $\alpha$. The two above definitions for displacement operator are equivalent. Other properties of $D$-operator are 
\begin{equation}
	D (\alpha) D (\beta) = D (\alpha + \beta) e^{\frac{\alpha \beta^* - \alpha^* \beta}{2}}
\label{eq:D_product}
\end{equation}
and the commutation relation 
\begin{equation}
	D (\alpha) (\had (0) + \alpha^*) = \had (0) D (\alpha),
\end{equation}
both of which can be proved by direct calculations \cite{cahill-glauber-pr-1969}.

In our problem, we notice that
\begin{equation}
	\ha (-Q_{j,x}) \ket{0,-Q_{j,x}} = 0 \quad \Leftrightarrow \quad \ha (0) \ket{0,-Q_{j,x}} = \frac{Q_{j,x} \lb}{\sqrt{2}} \ket{0,-Q_{j,x}} \quad \Leftrightarrow \quad \ket{0,-Q_{j,x}} = D \left(\frac{Q_{j,x} \lb}{\sqrt{2}} \right) \ket{0,0}.
\end{equation}
So, as implied by the name, $\ket{0,-Q_{j,x}}$ is obtained by moving the cyclotron center of $\ket{0,0}$ from the origin to $y = -Q_{j,x} \lb^2$. Also,
\begin{equation}
	\begin{split}
		D \left(\frac{Q_{j,x} \lb}{\sqrt{2}} \right) &= e^{i k_y Q_{j,x} \lb^2} \\
		D \left( - i\frac{Q_{j,y} \lb}{\sqrt{2}} \right) &= e^{i Q_{j,y} y}.
	\end{split}
\end{equation}
Meanwhile,
\begin{equation}
	D \left(\frac{Q_{j,x} \lb}{\sqrt{2}} \right)  \had (0) = \had \left(\frac{Q_{j,x} \lb}{\sqrt{2}} \right)  D \left(\frac{Q_{j,x} \lb}{\sqrt{2}} \right). 
\label{eq:commuterel}	
\end{equation}
Therefore, the matrix element in Eq.~\eqref{eq:phasefac} can be written in terms of displacement operators
\begin{align*}
	\bra{m,0} e^{i Q_{j,y}} \ket{n,-Q_{j,x}} &= \bra{m,0} D \left( - i\frac{Q_{j,y} \lb}{\sqrt{2}} \right) \frac{[\had (-Q_{j,x})]^n}{\sqrt{n !}} D \left(\frac{Q_{j,x} \lb}{\sqrt{2}} \right) \ket{0,0} \\
	& = \bra{m,0} D \left( - i\frac{Q_{j,y} \lb}{\sqrt{2}} \right) D \left(\frac{Q_{j,x} \lb}{\sqrt{2}} \right) \ket{n,0}  \\
	& = \bra{m,0} D \left(\frac{(Q_{j,x} - i Q_{j,y}) \lb}{\sqrt{2}} \right) \ket{n,0} e^{-i \frac{Q_{j,x} Q_{j,y} \lb^2}{2}}.
\end{align*}
So, the phase factor becomes 
\begin{equation}
	F_{mn,k_x} (\Q_j) = e^{i \frac{(2 k_x + Q_{j,x}) Q_{j,y} \lb^2}{2}}  \  \bra{m,0} D (z_j) \ket{n,0} \equiv e^{i \frac{(2 k_x + Q_{j,x}) Q_{j,y} \lb^2}{2}} F^0_{mn} (\Q_j)
\end{equation} 
with $z_j = \lb / \sqrt{2} (Q_{j,x} - Q_{j,y})$. The matrix element $F^0_{mn} (\Q_j)$ has an analytical form involving generalized Laguerre polynomial $\{ L^{(\alpha)}_n \}$ with $\alpha > -1$:
\begin{equation}
	F^0_{mn} (\Q_j) = \left\{
	\begin{array}{ll}
		e^{-\frac{|z_j|^2}{2}} z_j^{m-n} \sqrt{\frac{n !}{m !}} L^{(m-n)}_n (|z_j|^2) & \mbox{if $m \geq n$, } \\
		e^{-\frac{|z_j|^2}{2}} (-z_j^*)^{n-m} \sqrt{\frac{m !}{n !}} L^{(n-m)}_m (|z_j|^2) & \mbox{if $m < n$.}
	\end{array}
\right.
\end{equation}
The moir\'e potential Eq.~\eqref{eq:hamM_LL0} becomes
\begin{equation}
	\hat{H}^{\mu=+}_M = \sum_{j=1}^{3} \sum_{mn,\alpha \beta} (1 + e^{i\phi}) \left(T_{j,+}\right)_{\alpha \beta} F^0_{mn} (\Q_j) e^{i \frac{Q_{j,x} Q_{j,y} \lb^2}{2}}  \sum_{k_x} e^{i k_x Q_{j,y} \lb^2} \  \hfd_{1,\alpha; m,k_x+Q_{j,x}} \hf_{2,\alpha; n,k_x} + \rm{H.c.}
	\label{eq:hamM_LL1}
\end{equation}

\subsection{Commensurate condition}
Omitting layer, sublattice and Laudau level indices, the sum over $k_x$ in Eq.~\eqref{eq:hamM_LL1} is reminiscent of 1D tight-binding model if we identify in the following way:
\begin{itemize}
\item The position of each repetition motif is $k_x \lb^2$.
\item The lattice spacing is given by $Q^m_x \lb^2 > 0$, which is the greatest common divisor of $\{\forall j, Q_{j,x} \lb^2\}$.
\item The nearest hopping is given by an oscillating phase $e^{i k_x Q_{j,y} \lb^2}$.
\end{itemize}
Such 1D chain has lattice translational symmetry if, given $Q_{j,y}$,
\begin{equation}
	\exists p \in \mathbb{N} \quad \mbox{s.t.} \quad e^{i (k_x + p Q^m_x) Q_{j,y} \lb^2} = e^{i k_x Q_{j,y} \lb^2} \quad \Leftrightarrow \quad \exists (p,q) \in \mathbb{N}^2 \quad \mbox{s.t.} \quad Q^m_x Q_{j,y} \lb^2 = 2 \pi \frac{q}{p}
\label{eq:commensurate_0}.
\end{equation} 
To fulfill this commensurate condition for all $Q_{j,y}$, it suffice that $Q^m_y > 0$, i.e., the greatest common divisor of $\{\forall j, Q_{j,y} \}$, satisfies Eq.~\eqref{eq:commensurate_0}. So, we obtain the version of the commensurate condition in reciprocal space
\begin{equation}
	Q^m_x Q^m_y \lb^2 = 2 \pi \frac{q}{p} \quad \mbox{with $p,q$ coprime}.
\label{eq:commensurate_k}
\end{equation} 
Accordingly, the commensurate condition in real space with $Q^m_x = 2 \pi / \sqrt{3} L_s$ and $Q^m_y = 2 \pi / L_s$ is given by the magnetic flux $\Phi$ in the Moir\'e pattern unit-cell of size $\Omega_0 = \sqrt{3} L_s^2 / 2$ 
\begin{equation}
	\frac{\Phi}{\Phi_0} = \frac{p}{2q} \quad \mbox{with $p,q$ coprime}
	\label{eq:commensurate_r}
\end{equation} 
where $\Phi_0$ is the magnetic flux quantum $h/e$. 

Once the commensurate condition is satisfied, the unit-cell of the effective 1D chain in our problem has $p$ motifs. So, we can write
\begin{equation}
	k_x = \tilde{k}_x + (s p + r) Q^m_x
\end{equation}
with $\tilde{k}_x \in [0, Q^m_x[$ and $(s,r)$ are integers such that $r = 0,\dots,p-1$. Then, we define $Q_{j,x} \equiv \Delta r_j Q^m_x$ and
\begin{equation}
	\hf_{i,\alpha; n, sp+r} (\tilde{k}_x) \  \equiv \ \hf_{i,\alpha; n, k_x}.
\end{equation}
So, the sum over $k_x$ in Eq.~\eqref{eq:hamM_LL1} becomes
\begin{align*}
	\sum_{k_x} e^{i k_x Q_{j,y} \lb^2} \  \hfd_{1,\alpha; m,k_x+Q_{j,x}} \hf_{2,\alpha; n,k_x} &= \sum_{\tilde{k}_x} e^{i \tilde{k}_x Q_{j,y} \lb^2} \sum_{s,r} e^{i r Q^m_x Q_{j,y} \lb^2} \ \hfd_{1,\alpha; m,sp+r + \Delta r_j} (\tilde{k}_x) \hf_{2,\alpha; n,sp+r} (\tilde{k}_x) \\
	& = \sum_{\tilde{k}_x} e^{i \tilde{k}_x Q_{j,y} \lb^2} \sum_{r=0}^{p-1} e^{i r Q^m_x Q_{j,y} \lb^2} \sum_{ \tilde{k}_y}\ e^{-i \tilde{k}_y \Delta r_j Q^m_x \lb^2} \hfd_{1,\alpha; m,r + \Delta r_j} (\kt) \hf_{2,\alpha; n,r} (\kt)
\end{align*}
where $\kt = (\tilde{k}_x,\tilde{k}_y)$. In the last line, we use periodic boundary conditions and the following Fourier transformation
\begin{align}
	\hf_{l,\alpha; n,sp+r} (\tilde{k}_x) &= \frac{1}{\sqrt{N}} \sum_{\tilde{k}_y} \hf_{l,\alpha; n, r} (\kt) e^{i \tilde{k}_y (sp+r) Q^m_x \lb^2} \quad \mbox{with} \quad \tilde{k}_y Q^m_x \lb^2 \in [0,\frac{2\pi}{p}[\\
	\hf_{l,\alpha; n,r+p} (\kt) &= \hf_{l,\alpha; n,r} (\kt)
\end{align}
The moir\'e potential in this new basis becomes
\begin{equation}
	\begin{split}
			\hat{H}^{\mu=+}_M = \sum_{j=1}^{3} &\sum_{mn,\alpha \beta} (1 + e^{i\phi}) \left(T_{j,+}\right)_{\alpha \beta} F^0_{mn} (\Q_j) e^{i \frac{Q_{j,x} Q_{j,y} \lb^2}{2}}  \\
			& \times \sum_{\kt} \sum_{r=0}^{p-1} e^{i \tilde{k}_x Q_{j,y} \lb^2} e^{i (r Q_{j,y} - \Delta r_j \tilde{k}_y ) Q^m_x \lb^2} \  \hfd_{1,\alpha; m,r + \Delta r_j} (\kt) \hf_{2,\alpha; n,r} (\kt) + \rm{H.c.}
	\end{split}
	\label{eq:hamM_LL2}.
\end{equation}
Therefore, the Dirac Hamiltonian part Eq.~\eqref{eq:hamD_LL1} becomes
\begin{equation}
	\begin{split}
		\hat{H}^{\mu=+}_{D} = & \sum_{l,n,\kt} \sum_{r=0}^{p-1} \hbar v_F (-K_{l,+,x} - i K_{l,+,y}) \  \hfd_{l,A; n,r} (\kt) \hf_{l,B; n,r} (\kt) + \rm{H.c.} \\
		&+ \sum_{l,n,\kt} \sum_{r=0}^{p-1} \frac{\sqrt{2(n+1)}\hbar v_F}{\lb} \   \hfd_{l,A; n+1,r} (\kt) \hf_{l,B; n,r}(\kt) + \rm{H.c.}
	\end{split}.
	\label{eq:hamD_LL2}
\end{equation}

The calculations for $\mu=-$ is exactly parallel to what we have shown for $\mu=+$. So, we give directly the results. The moir\'e potential is 
\begin{equation}
	\begin{split}
		\hat{H}^{\mu=-}_M = \sum_{j=1}^{3} &\sum_{mn,\alpha \beta} (1 + e^{i\phi}) \left(T_{j,-}\right)_{\alpha \beta} F^0_{mn} (-\Q_j) e^{i \frac{Q_{j,x} Q_{j,y} \lb^2}{2}} \\
		& \times \sum_{\kt} \sum_{r=0}^{p-1} e^{-i \tilde{k}_x Q_{j,y} \lb^2} e^{-i (r Q_{j,y} - \Delta r_j \tilde{k}_y ) Q^m_x \lb^2} \  \hfd_{1,\alpha; m,r + \Delta r_j} (\kt) \hf_{2,\alpha; n,r} (\kt) + \rm{H.c.}
	\end{split}
\end{equation}
and the Dirac Hamiltonian part reads
\begin{equation}
	\begin{split}
		\hat{H}^{\mu=-}_{D} = & \sum_{l,n,\kt} \sum_{r=0}^{p-1} \hbar v_F (K_{l,-,x} - i K_{l,-,y}) \  \hfd_{l,A; n,r} (\kt) \hf_{l,B; n,r} (\kt) + \rm{H.c.} \\
		&+ \sum_{l,n,\kt} \sum_{r=0}^{p-1} -\frac{\sqrt{2(n+1)} \hbar v_F}{\lb} \   \hfd_{l,B; n+1,r} (\kt) \hf_{l,A; n,r}(\kt) + \rm{H.c.}
	\end{split}.
\end{equation}

\subsection{Numerical implementation}
The commensurate condition gives rise to Hofstadter butterfly in the Landau level spectrum as a function of magnetic field. In numerical implementation, we have to introduce a cut-off on the number of Landau levels. As the Dirac part is (almost) diagonalized in the Landau level basis, the coupling between Landau levels comes from two places: one is $v_F k_\theta$ in the diagonal term and the other is the moir\'e potential, which has an amplitude $u_0'$. So, denoting $E_c =  \max (v_F k_\theta,u_0')$, a viable cut-off needs to encounter the coupling between $n=0$ to $n_0$ such that
\begin{equation}
	E_c \sim \frac{\sqrt{2n_0}\hbar v_F}{\lb} \Leftrightarrow n_0 \sim \frac{\sqrt{3} E_c^2 a^2}{8\pi \hbar^2 v_F^2 \theta^2} \cdot \frac{\Phi_0}{\Phi }.
\end{equation} 
Note that $E_c = u_0'$ only if $\theta < 0.64 \degree$. In our plots, the twist angle is larger than $1 \degree$, so $E_c = \hbar v_F k_\theta$ and
\begin{equation}
	n_0 \sim \frac{2 \sqrt{3} \pi}{9} \cdot \frac{\Phi_0}{\Phi } = \frac{1.21}{\frac{\Phi}{\Phi_0}},
\end{equation}
which is independent of $\theta$. Therefore, to achieve the numerical convergence of the Hofstadter butterfly, one needs a cut-off $N_c$, which is several times larger than $n_0$. Since $n_0 \sim 100$ for $\Phi/\Phi_0 = 0.01$, we choose in our numerical calculations $N_c = 500$. For even smaller $\Phi/\Phi_0$, we use $N_c=1600$.

Another important point in the numerical calculations is that one should choose different cut-off, differed by one, for two sublattices. For example, in valley $\mu=+$, the zeroth Landau level is purely $B$-sublattice polarized and topologically protected. The higher $n$-th Landau level involves the $n$-th state of sublattice $B$ and the $(n-1)$-th state of sublattice $B$. If we choose the same cut-off $N_c$ for two sublattices, then the $N_c$-th state of sublattice $A$ has no partner to couple and induces a floating zero energy state. This would lead to non-physical numerical error hindering the convergence of Hofstadter butterfly. 

\subsection{Wannier plot}
Once we obtain the Laudau levels of a given $k_z$-indexed TBG, we can extract the Chern number of gaps between Landau levels following the Streda formula. This information is precisely provided by Wannier plots. In this subsection, we explain how to draw the Wannier plots shown in Fig.~3 of the main text, inspired from \cite{hejazi-ll_tbg-prb-2019}.

In Wannier plot, we actually show the density of states $\rho$ as a function of filling factor $n$ and magnetic field $B$. For illustrative purposes, we have to broaden the Dirac $\delta$-like peaks in $\rho$ at a given energy $E$ using Lorentz distributions, namely:
\begin{equation}
	\rho (E) = \sum_{i} \delta (E-E_i) \to \sum_{i} \frac{1}{\pi} \frac{\gamma}{ (E-E_i)^2 + \gamma^2}
\end{equation}
where the parameter $\gamma$ mimics Landau level broadening due to thermal or disorder effects. The filling factor $n$ at energy $E$ is obtained by counting the number of Landau levels below this energy, namely:
\begin{equation}
	n(E) = \frac{1}{2q} \sum_{i} \theta (E-E_i) \to \sum_{i} \frac{1}{2q\pi} \arctan \left( \frac{E-E_i}{\gamma} \right)
\end{equation}
where we add a normalization factor $1/2q$ since the number of magnetic bands is proportional to $q$, where $p$ and $q$ are co-prime numbers satisfying $\Phi/\Phi_0 = p/2q$. In this way, we can achieve the conventional definition of filling factor, for example, $n \in [-2,2]$ in Wannier plot if one consider the two flat bands in two valleys only in one spin sector.

Technically, the empirical parameter $\gamma$ has been thus adjusted for each specific case to achieve best resolution of Wannier plots (usually $\gamma<10^{-4}$ eV). In our plots, we also rescale the density of states by its maximum $\rho_{\rm{max}}$. Then, we only plot the points having $\rho/\rho_{\rm{max}} \le 1 \%$ to focus on the topological nature of large gaps.

\section{S4. Hofstadter butterfly spectra}

\begin{figure}
	\centering
	\includegraphics[width=0.45\textwidth]{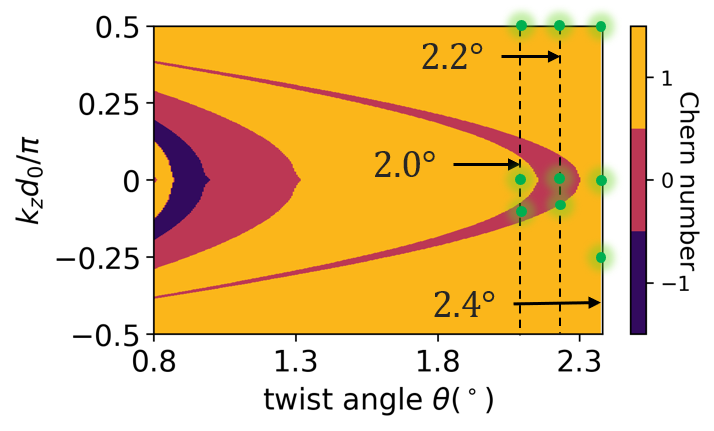}
	\caption{Topological phase diagram. The green dots indicate the momenta $k_z$ that we discuss. Their associated twist angle are marked by dashed lines.}
	\label{fig:chern_kz}
\end{figure}

In this section, we provide a detailed analysis on the evolution of Hofstadter butterfly spectra of $k_z$-indexed TBG in bulk ATG of three twist angles $\theta = 2.0 \degree, \  2.2 \degree$ and $2.4 \degree$. These three twist angles are the representative of the three topological phases, respectively, as shown in Fig.~\ref{fig:chern_kz}. For each chosen twist angle, we select three representative $k_z$ momenta (annotated by green dots in Fig.~\ref{fig:chern_kz}) and account for the Hofstadter butterfly spectrum of their corresponding $k_z$-index TBG. In the following discussions, we always consider only one spin sector and neglecting Zeeman effect on Landau levels. The magnetic flux ratio $\Phi_0/\Phi$ is defined with respect to the physical twist angle and the corresponding 2D moir\'e superlattice.

\begin{figure}
	\centering
	\includegraphics[width=0.95\textwidth]{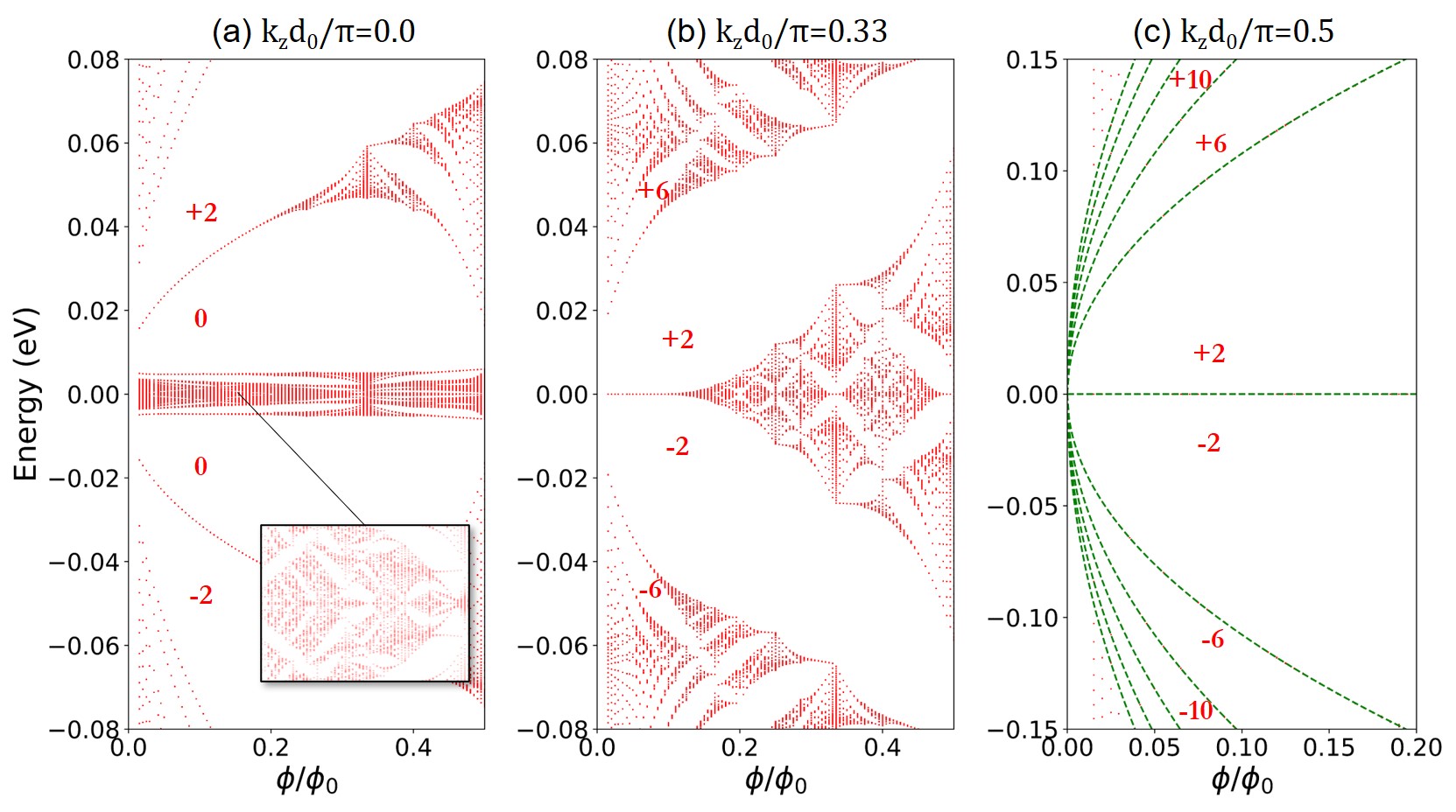}
	\caption{Hofstadter butterfly spectra for $\theta=2.4 \degree$ at $k_z d_0/\pi = 0$, 1/3 and 1/2. Landau level dispersion in monolayer graphene is marked with green dashed lines. The Chern number of gap are indicated by red.}
	\label{fig:Hofst_theta2.40}
\end{figure}

Let us first study the case of $\theta=2.40 \degree$. According to the topological phase diagram Fig.~\ref{fig:chern_kz}, all the $k_z$-indexed TBG in bulk ATG is adiabatically connected to TBG of twist angle larger than the first magic angle, i.e., there is no magic momentum for $\theta=2.40 \degree$. The three representative $k_z$ are $k_z d_0/\pi =0, 1/3, 1/2$ (see Fig.~\ref{fig:Hofst_theta2.40}(a-c), respectively), which effectively correspond to TBG of twist angle $\theta=1.2 \degree, \ 2.4 \degree$ and TBG with vanishing moir\'e potential, respectively. Here, ``effectively'' means that $k_z$-indexed TBG in bulk ATG of $\theta=2.40 \degree$ share the same dimensionless ration $\alpha$ (defined in the main text) with 2D TBG of certain twist angle. 

For $k_z d_0/\pi =1/2$, the interlayer moir\'e coupling is completely suppressed so that the $k_z$-indexed TBG is basically a trivial addition of two layers of graphene forbidding any interlayer coupling. As shown in Fig.~\ref{fig:Hofst_theta2.40}(c), Landau levels follow exactly the same $\sqrt{B}$-dispersion as in graphene (green dashed line). The fractal pattern cannot be seen even for $\Phi/\Phi_0 = 0.2$ in the given energy scale because the corresponding magnetic flux in monolayer graphene's small unit-cell is only $10^{-4}|$. Because of topological pinning of zeroth Landau level around Dirac cones at zero energy, the Chern number of the gaps follows the $(2n+1)$-rule in monolayer graphene if considering only one spin sector. Therefore, the Chern number of two decoupled graphene monolayers, which both follows the $(2n+1)$-rule, exhibits the $(4n+2)$-rule, as indicated in Fig.~\ref{fig:Hofst_theta2.40}(c). 

For $k_z d_0/\pi =1/3$, we effectively study a TBG of twist angle $2.40 \degree$. Although this angle is twice larger than the first magic angle $\sim 1.1 \degree$, the moir\'e potential already substantially renormalize the Fermi velocity around Dirac cone. Combining with the large moir\'e superlattice, the zeroth Landau level starts to split at $\phi/\phi_0=0.15$, i.e., $B=20.8$T, as shown in Fig.~\ref{fig:Hofst_theta2.40}(b). When $B<20.8$T, Landau levels still follow the $\sqrt{B}$-dispersion except that the Fermi velocity is strongly suppressed with $v_F^*/v_F\sim 0.4$. So, the Chern number of gaps still follow the $(4n+2)$-rule, as for $k_z d_0/\pi =1/2$. 

When $k_z \to 0$, the effective TBG under study is twisted by $1.2 \degree$, which is close to the first magic angle. The bandwidth and the Fermi velocity are then radically reduced so that the Landau levels are hard to resolve in the given energy scale as shown in the Hofstadter spectrum Fig.~\ref{fig:Hofst_theta2.40}(a). In the inset, we zoom in on the low-energy Landau levels to energy scale of $\sim 1$ meV. A rich and complex fractal pattern of Landau levels appears at sub Tesla magnetic field. The cluster of these tightly distributed Landau levels stems from the two (nearly-)flat bands of TBG. The big gap between this cluster and other Landau levels turns out to have a vanishing Chern number. The cluster preserves its energy range while varying magnetic field, as stipulated by Streda's formula \cite{streda-1982}. 

\begin{figure}
	\centering
	\includegraphics[width=0.5\textwidth]{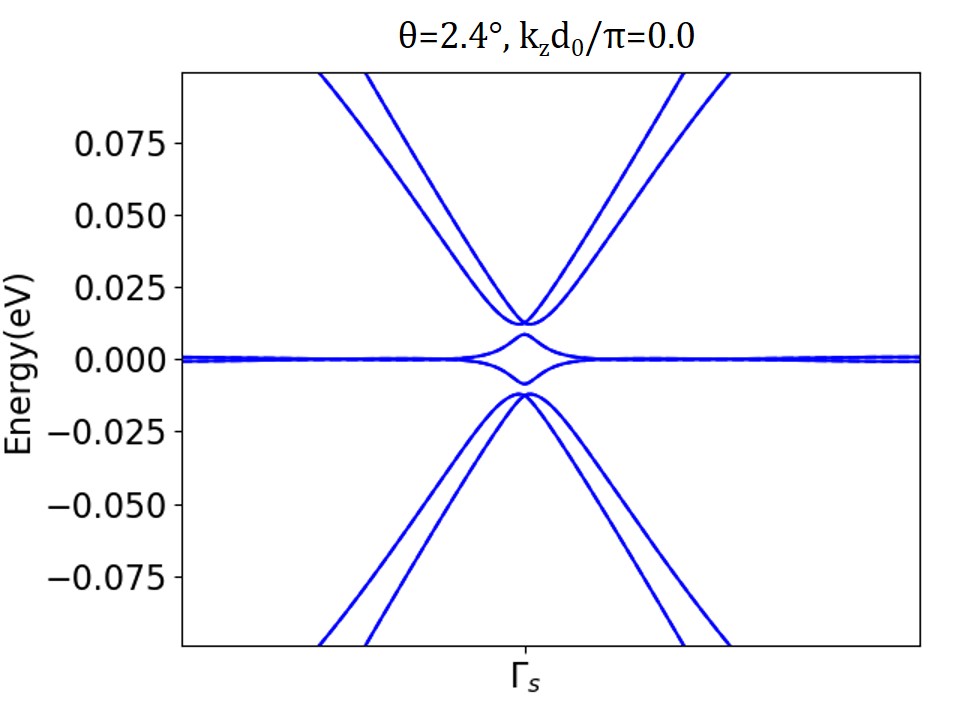}
	\caption{Band structure zoomed at $\Gamma_s$ for $\theta=2.40 \degree$ at $k_z=0$.}
	\label{fig:zoomband_theta2.40}
\end{figure}

The Hofstadter spectrum at low field can be understood by looking into the 2D band structure at $k_z=0$ of bulk ATG, as shown in Fig.~\ref{fig:zoomband_theta2.40}. The two flat bands are separated from the four remote bands by a gap at $\Gamma_s$ point, where the remote bands in the same valley are twice degenerate thanks to the $C_3$ and $M_y$ symmetries \cite{hejazi-prb19}. Most saliently, while extrapolating the gap to zero magnetic field, the separation between the first high-energy Landau level and the low-energy cluster is precisely the gap between the flat bands and the two high-energy bands at $\Gamma_s$ as shown in Fig.~\ref{fig:zoomband_theta2.40}. Therefore, the Landau levels aside the big gap with zero Chern number have to stem from the $\Gamma_s$ point of the flat bands and the remote bands. At the vicinity of $\Gamma_s$, the six bands in study can be described by a six-band $k \cdot p$ Hamiltonian \cite{hejazi-prb19}, which is reminiscent of that found in semiconductors with spin-orbit coupling. The off-diagonal part of the Hamiltonian attributes non-zero Berry curvature to the bands. Therefore, the Landau levels of these bands cannot naively classified into topologically trivial Schr\"odinger fermion with quadratic dispersion or conventional massive Dirac fermion with $\pi$-Berry phase (e.g., that found in hBN). What we have here is something in between such that:
\begin{itemize}
\item Landau levels from the remote bands follow the $\sqrt{B}$ dispersion while the zeroth Landau level does not keep constant but vary with magnetic field. 
\item The Landau levels from $\Gamma_s$ of the flat bands follows the linear $B$ dispersion. 
\end{itemize}

Furthermore, the Chern number of gaps follows the $2n$-rule. This can be understood by looking into Fig.~\ref{fig:zoomband_theta2.40}. The factor 2 comes from the valley degeneracy. Although the first high energy band is twice degenerate at $\Gamma_s$, their distinct dispersion away from $\Gamma_s$ induces different Landau level splitting so that the increment of the Chern number is 2 instead of 4.

\begin{figure}
	\centering
	\includegraphics[width=0.95\textwidth]{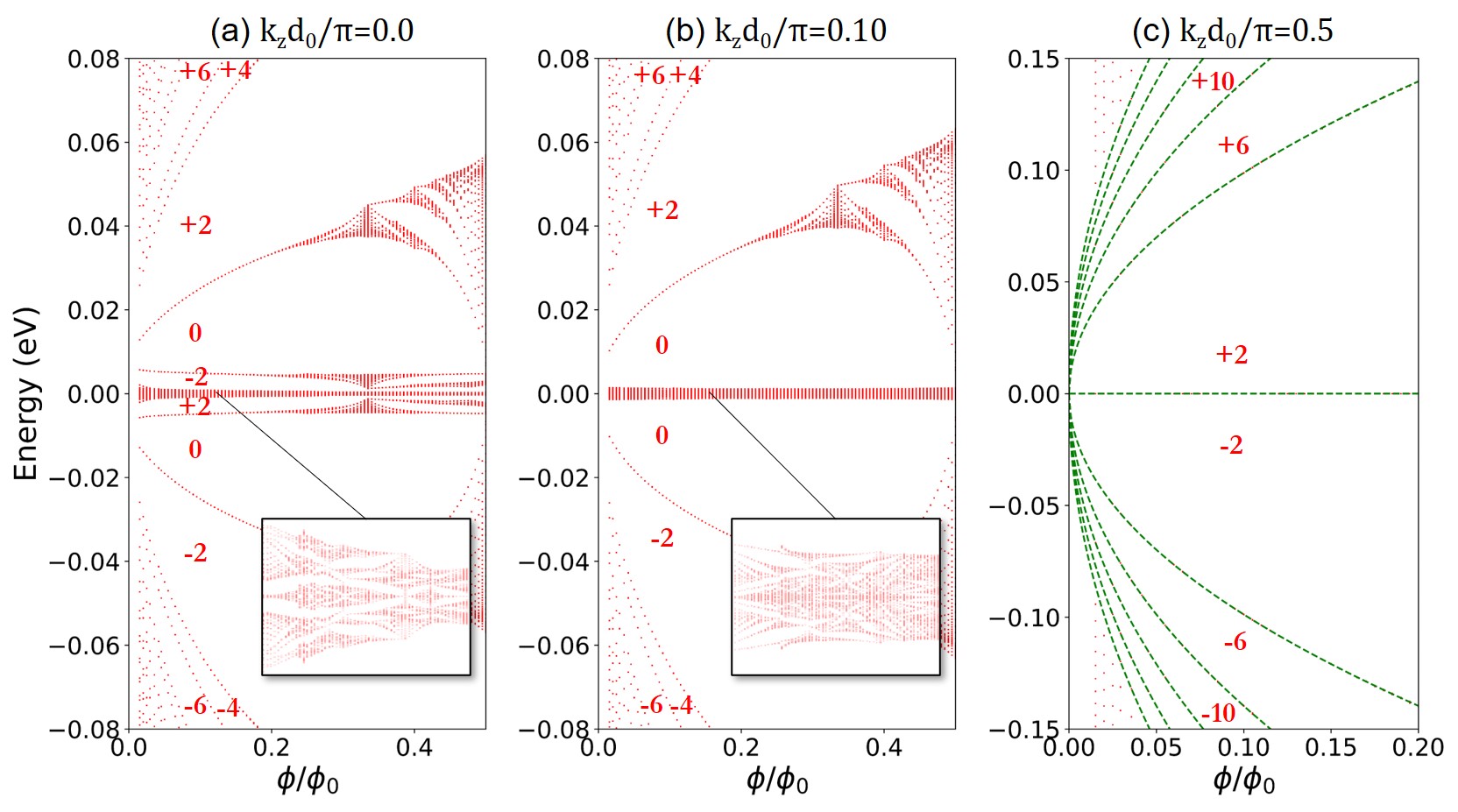}
	\caption{Hofstadter butterfly spectra for $\theta=2.2 \degree$ at $k_z d_0/\pi = 0$, 0.10 and 1/2. Landau level dispersion in monolayer graphene is marked with green dashed lines. The Chern number of gap are indicated by red.}
	\label{fig:Hofst_theta2.20}
\end{figure}

In the case of $\theta=2.20 \degree$ and $2.00 \degree$, the Hofstadter spectra at $k_z d_0/\pi = 1/2$ [see Fig.~\ref{fig:Hofst_theta2.20}(c) and Fig.~\ref{fig:Hofst_theta2.00}(c), respectively] are exactly the same as for $\theta=2.40 \degree$ because they correspond to the same decoupled bilayer graphene with vanishing interlayer coupling. 

\begin{figure}
	\centering
	\includegraphics[width=0.8\textwidth]{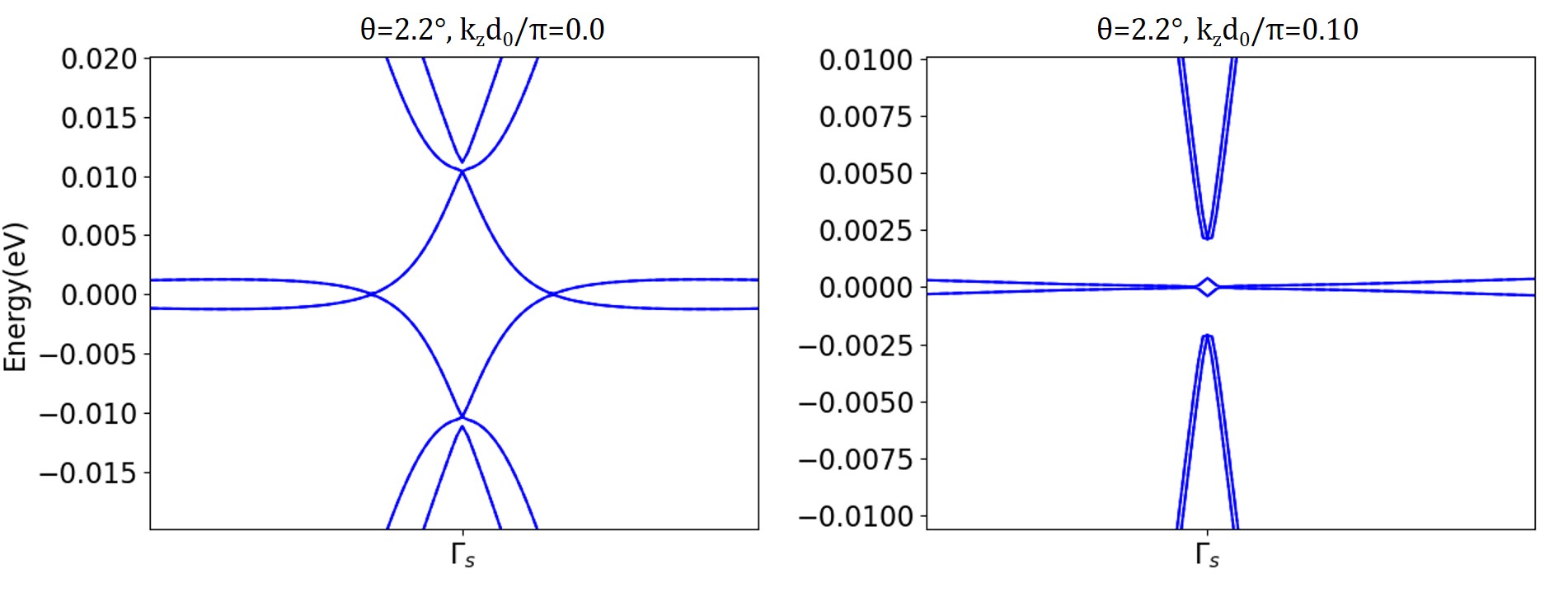}
	\caption{Band structure zoomed at $\Gamma_s$ for $\theta=2.20 \degree$ at $k_z d_0/\pi=0$ and 0.10.}
	\label{fig:zoomband_theta2.20}
\end{figure}

\begin{figure}
	\centering
	\includegraphics[width=0.95\textwidth]{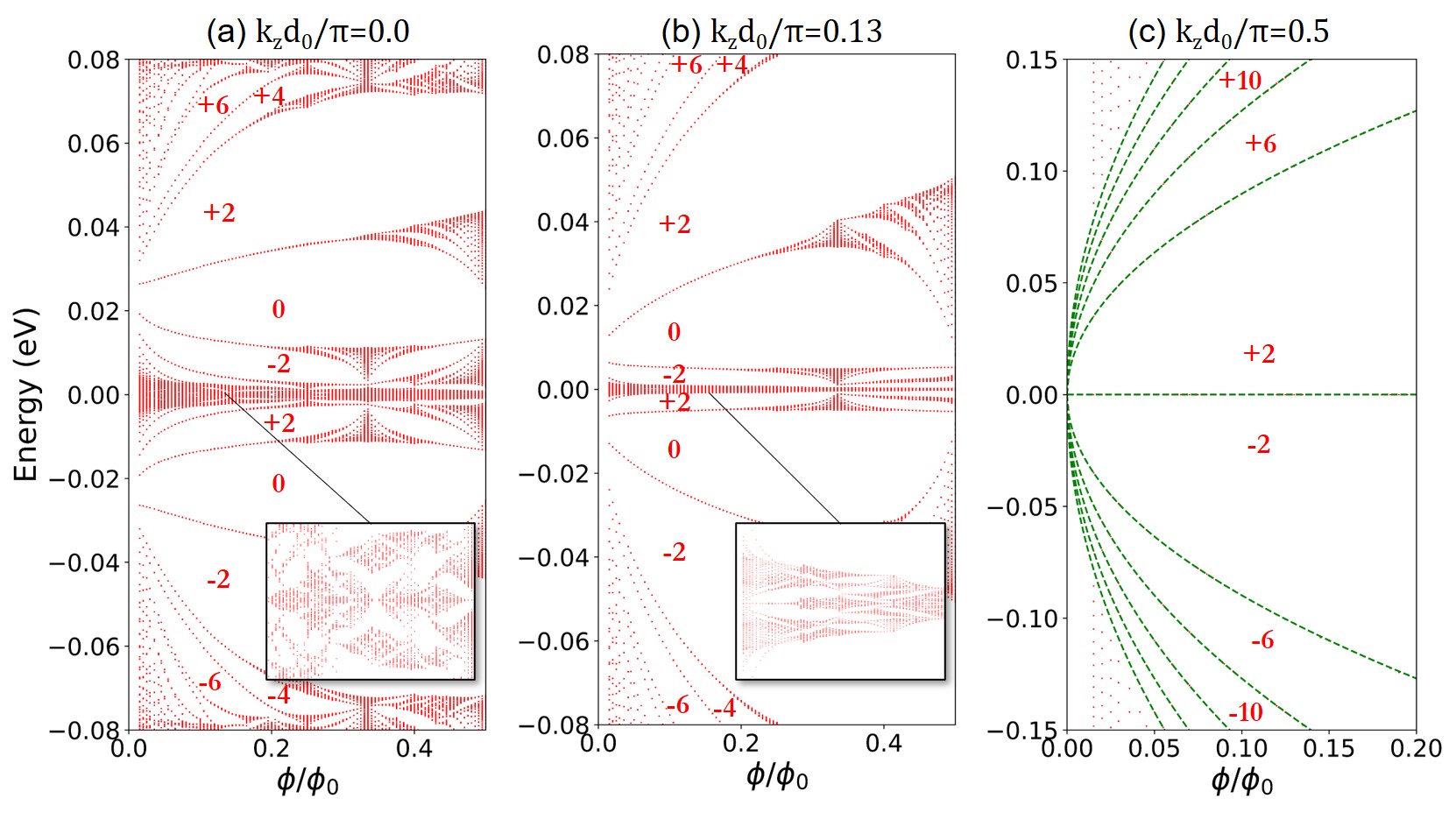}
	\caption{Hofstadter butterfly spectra for $\theta=2.0 \degree$ at $k_z d_0/\pi = 0$, 0.13 and 1/2. Landau level dispersion in monolayer graphene is marked with green dashed lines. The Chern number of gap are indicated by red.}
	\label{fig:Hofst_theta2.00}
\end{figure}

When $\theta=2.20 \degree$, we choose two other $k_z$ as $k_z d_0 /\pi=0$ and 0.10. According to the topological phase diagram, the former is in the $C=0$ region and the latter is on the phase boundary. The Landau levels of the flat bands are zoomed in on the energy scale, shown in the inset of Fig.~\ref{fig:Hofst_theta2.20}(a,b). When $k_z d_0 /\pi=0$, the Chern number and the energy dispersion of Landau levels originated from remote bands have the same behavior as for $\theta=2.40 \degree$ at $k_z=0$ for the same reasons. However, in contrast to $\theta = 2.40 \degree$, the energy of Landau level of the flat band edges has the $\sqrt{B}$ dispersion, which may be seen as a signature of topological transition by exchange of Berry curvature at band touching points [see Fig.~\ref{fig:zoomband_theta2.20}(left)]. When $k_z d_0 /\pi=0.10$, gap reopens at $\Gamma_s$ but the Landau levels are hard to resolve because of completely flattened bands [see Fig.~\ref{fig:zoomband_theta2.20}(right)] close to the topological transition boundary.

\begin{figure}
	\centering
	\includegraphics[width=0.8\textwidth]{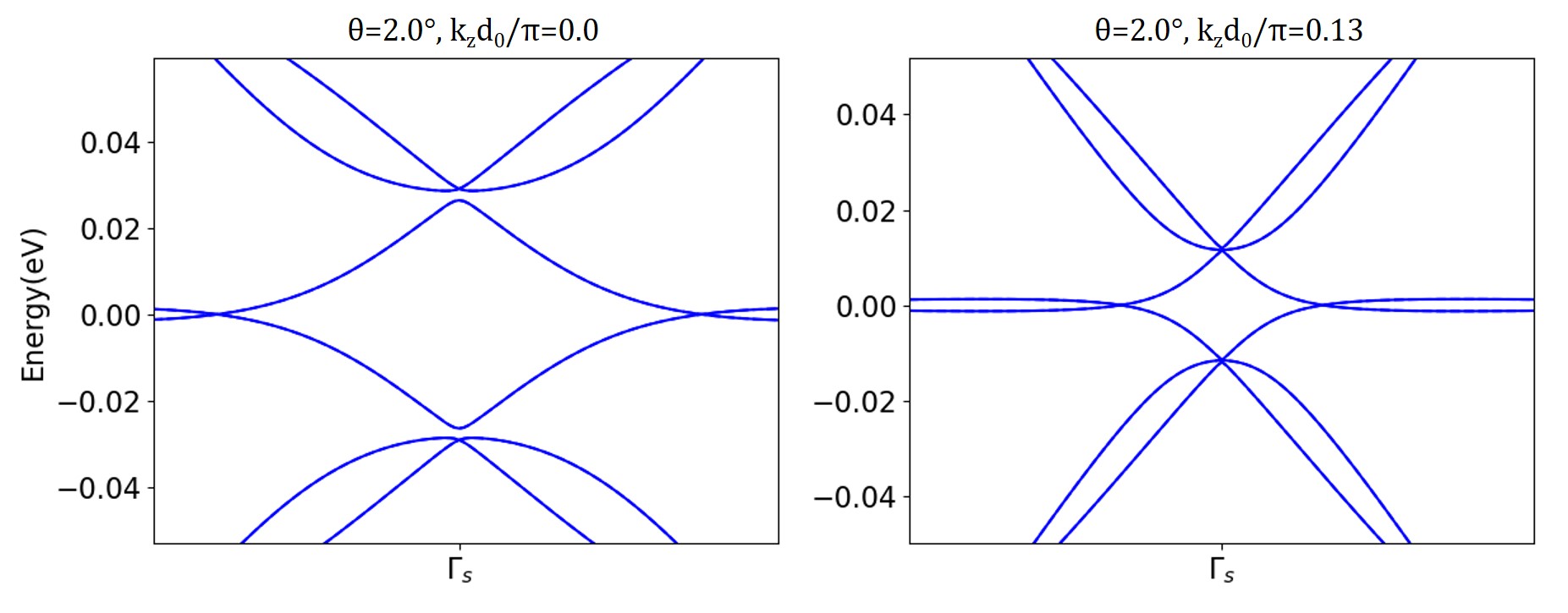}
	\caption{Band structure zoomed at $\Gamma_s$ for $\theta=2.40 \degree$ at $k_z d_0 /\pi = 0$ and 0.13.}
	\label{fig:zoomband_theta2.00}
\end{figure}

When $\theta=2.00 \degree$, we choose two other $kz$ as $k_z d_0 /\pi=0$ and 0.13. According to the topological phase diagram, the former is in the $C=1$ region but smaller than the first magic momentum and the latter is in the small region of $C=0$. The Landau levels of the flat bands are zoomed in on the energy scale, shown in the inset of Fig.~\ref{fig:Hofst_theta2.00}(a,b). It turns out that the Hofstadter spectra Fig.~\ref{fig:Hofst_theta2.00}(b) and the band structure Fig.~\ref{fig:zoomband_theta2.00} for $\theta=2.00 \degree$ at $k_z d_0 /\pi=0.13$ are similar to the case for $\theta=2.20 \degree$ at $k_z d_0 /\pi=0$. Therefore, the understanding of them should be also the same since they belongs to the same topological phase. However, the case for $\theta=2.00 \degree$ at $k_z d_0 /\pi=0$ is different. Although the band structure for $k_z d_0 /\pi=0$ (left panel of Fig.~\ref{fig:zoomband_theta2.00}) is similar to $k_z d_0 /\pi=0.13$ (right panel of Fig.~\ref{fig:zoomband_theta2.00}), they belongs to distinct topological phase, according to the phase diagram Fig.~\ref{fig:chern_kz}. This can be seen from the zeroth Landau level of the remote band in Fig.~\ref{fig:Hofst_theta2.00}(a), which follows a dispersion linear in $B$ instead of $\sqrt{B}$, i.e., a hint of change in Berry curvature.

\section{S5. Bound states of the spiral dislocation line}
\begin{figure}
	\centering
	\includegraphics[width=16cm]{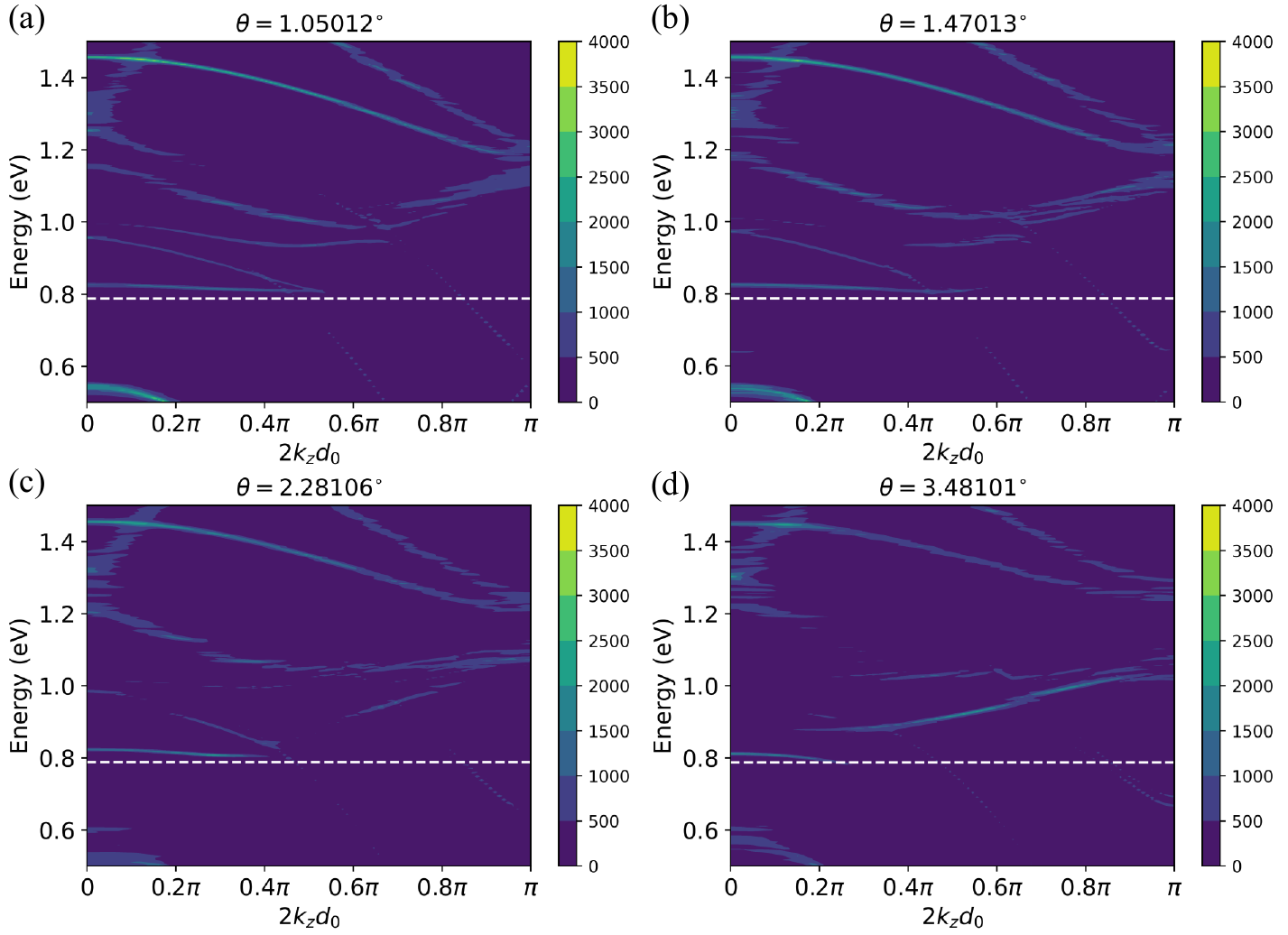}
	\caption{The spectral function of the bounded states near the spiral dislocation line for twisted graphene spiral with (a) $2\times2$ supercell and twist angle $\theta=1.05\degree$; (b) $2\times2$ supercell and twist angle $\theta=1.47\degree$; (c) $4\times4$ supercell and twist angle $\theta=2.28\degree$; and (d) $6\times6$ supercell and twist angle $\theta=3.48\degree$.}
	\label{fig:dislocation}
	
\end{figure}

In the main text, we have studied the physical properties of twisted graphene spiral in regions far from the dislocation line, which forms a bulk alternating twisted graphite. We now investigate the properties of the dislocation line. We start with the lattice structure of twisted graphene spiral. The dislocation line is located at the $AA$ point in the moir\'e supercell.  We consider the lattice distortion in the $z$ direction at position $\textbf{r}$, while neglecting the in-plane displacement, i.e. $u_{x}=u_{y}=0$. The lattice distortion in the $z$ direction at a lateral position $\textbf{r}$ (by setting the dislocation line as the origin) can be expressed as:
\begin{equation}
u_{z}(\textbf{r})=\frac{2d_{0}}{2\pi}\arg(r_{x}+\,i\,r_{y}),
\end{equation}
where the function $\text{arg}$ denotes the argument of the complex number. The presence of the dislocation line breaks the moir\'e superlattice translational symmetry. Thus, we set up an open boundary condition in the $x$-$y$ direction and a periodic boundary in the $z$ direction. We perform the lattice relaxation calculation utilizing Large-scale Atomic-Molecular Massively Parallel Simulation (LAMMPS) \cite{LAMMPS}, employing long-rang bond-order potential for carbon (LCBOP) \cite{LCBOP}. The initial structure is an $n\times\,n$ ($n=2, 4, 6$) supercell with the dislocation line at the center of the sample.  Based on the relaxed lattice structure, we calculate the band structure of twisted graphene spiral including the screw dislocation line  using the atomistic tight binding model described by Eq.~(\ref{eq:tightbinding}). Then, we project the energy band to the nearest neighbor carbon atoms around the dislocation line, in order to evaluate the single-particle spectral function of the bounded states associated with the spiral dislocation line.
In Fig.~\ref{fig:dislocation}, we present the spectral function of the bounded states near the dislocation line for twisted graphene spiral with different twist angles. Specifically, We construct a $2\times2$ supercell for twisted graphene spiral with twist angle $\theta=1.05\degree$, a $2\times2$ supercell for GS with $\theta=1.47\degree$, a $4\time4$ supercell with $\theta=2.28\degree$ and a $6\times6$ supercell with $\theta=3.48\degree$. The dashed lines mark the energy of the Dirac point. We observe the presence of several bounded states that cross the energy level of the Dirac point. These bound states do not show any qualitative difference at different twist angles.

\end{document}